\documentclass[a4paper,11pt]{article}
\pdfoutput=1
\usepackage{jheppub}
\usepackage{bbold}
\usepackage{amsmath,mathtools}
\usepackage{empheq}
\usepackage{booktabs}
\usepackage{multirow}
\usepackage{xspace}
\usepackage{calc}
\usepackage{placeins}
\usepackage{subcaption}
\usepackage{todonotes}
\usepackage{tabu}
\DeclareRobustCommand{\ensuremathrm}[1]{\ensuremath{\mathrm{#1}}\xspace}

\DeclareRobustCommand{\GeV}{\ensuremathrm{GeV}}

\DeclareRobustCommand{\NNNLO}{\ensuremathrm{N^3LO}}
\DeclareRobustCommand{\NNLO}{\ensuremathrm{N^2LO}}
\DeclareRobustCommand{\NLO}{\ensuremathrm{NLO}}
\DeclareRobustCommand{\LO}{\ensuremathrm{LO}}
\DeclareRobustCommand{\HEJ}{\ensuremathrm{HEJ}}
\DeclareRobustCommand{\HIGHEJ}{\emph{High Energy Jets}}
\DeclareRobustCommand{\Sherpa}{\ensuremathrm{Sherpa}}
\DeclareRobustCommand{\Comix}{\textsc{Comix}}
\DeclareRobustCommand{\openloops}{\ensuremathrm{Open Loops}}

\DeclareRobustCommand{\as}{\ensuremath{\alpha_s}\xspace}

\DeclareRobustCommand{\scalemjj}{\ensuremath{\mu_r\!=\!\mu_f\!=\!\max(m_H,m_{12})}\xspace}
\DeclareRobustCommand{\scaleht}{\ensuremath{\mu_r\!=\!\mu_f\!=\!H_T/2}\xspace}
\def\spa#1.#2{\left\langle#1\,#2\right\rangle}
\def\spb#1.#2{\left[#1\,#2\right]} \def\spaa#1.#2.#3{\langle\mskip-1mu{#1} |
  #2 | {#3}\mskip-1mu\rangle} \def\spbb#1.#2.#3{[\mskip-1mu{#1} | #2 |
  {#3}\mskip-1mu]} \def\spab#1.#2.#3{\langle\mskip-1mu{#1} | #2 |
  {#3}\mskip-1mu\rangle} \def\spba#1.#2.#3{\langle\mskip-1mu{#1}^+ | #2 |
  {#3}^+\mskip-1mu\rangle} \def\spav#1.#2.#3{\|\mskip-1mu{#1} | #2 |
  {#3}\mskip-1mu\|^2} \def\jc#1.#2.#3{j^{#1}_{#2#3}}
\DeclareRobustCommand{\mathgraphics}[1]{\vcenter{\hbox{\includegraphics{#1}}}}
\title{Finite Quark-Mass Effects in Higgs Boson Production With Dijets at Large Energies}
\author[a]{Jeppe~R.~Andersen,}
\author[b]{James~D.~Cockburn,}
\author[a]{Marian Heil,}
\author[c]{Andreas Maier}
\author[b]{and Jennifer~M.~Smillie}
\affiliation[a]{Institute for Particle Physics Phenomenology,\\University of
  Durham, South Road, Durham DH1 3LE, UK}
\affiliation[b]{Higgs Centre for Theoretical Physics, University of Edinburgh,\\
  Peter Guthrie Tait Road, Edinburgh EH9 3FD, UK.}
\affiliation[c]{Deutsches Elektronen-Synchrotron, DESY, Platanenallee 6,
  15738 Zeuthen, Germany}
\emailAdd{jeppe.andersen@durham.ac.uk}
\emailAdd{j.d.cockburn@ed.ac.uk}
\emailAdd{marian.heil@durham.ac.uk}
\emailAdd{andreas.martin.maier@desy.de}
\emailAdd{j.m.smillie@ed.ac.uk}

\abstract{ The production of a Higgs boson in association with at least two jets receives contributions both from the fusion
of weak vector bosons (VBF) and from QCD processes, especially gluon fusion (GF).
The former process is important for measuring the coupling of the Higgs boson to
weak bosons, whereas the latter process plays an important role in
determining any $CP$-admixtures in the Higgs sector.
In this paper we go beyond the current state-of-the-art for fixed order
calculations of the GF process (i.e.~one loop $H+2j$ including full quark mass
effects) by including the all-order effects in leading $\log(\hat s/p_t^2)$, together with full quark mass and loop-propagator kinematic
effects.  We calculate the mass-dependent components and implement the
resummation within the framework of High Energy Jets.

The high-energy effects
suppress the prediction compared to fixed order at large $\Delta y_{12}$ and $m_{jj}$
(and therefore within the usual VBF cuts of widely separated jets), just as
found in the limit of $m_t\to \infty$.
The mass dependence is more significant than at fixed order, because the
systematic inclusion of the leading logarithms in $\hat s/p_t^2$ results in a hardening of
the transverse momentum of the Higgs boson, which in turn probes in more
detail the loop-structure of
the coupling. In particular, the full mass dependence reduces the
cross section within VBF cuts by 11\% compared to a calculation based just on
the infinite top mass limit, but the impact of the bottom quark remains small. This all implies that the gluon-fusion contribution within
VBF-cuts is less severe than current estimates suggest.
}

\preprint{\begin{minipage}[t]{\widthof{DCPT/18/XX, DESY
        18-217,}}DCPT/18/204, DESY 18-217,\\IPPP/18/102, MCnet-18-33
  \end{minipage}
}
\begin{document}
\maketitle
\flushbottom

% \newpage
\section{Introduction}
\label{sec:Introduction}

In this paper we calculate the gluon fusion component of Higgs boson production
in association with two jets supplementing the fixed order results with the
leading logarithmic corrections in $\hat s/p_t^2$
to all orders in the coupling \emph{and} including full quark mass
effects. The necessary mass-dependent components for the resummation are
derived, and allow for a first
calculation of the interference of top and bottom quark mass contributions in
$H+2j$.  Within VBF cuts, NLO predictions exhibit an instability which is cured by the
inclusion of the all-order high-energy logarithms.
 Precise, stable
predictions of the gluon fusion component in $H+2j$ are essential for the
analysis of the Higgs boson couplings both to weak bosons and to gluons mediated
by heavy quarks.

Let us first review the current status of the calculation of the quark-mass
effects in Higgs Boson production. Higgs bosons are most copiously produced at the CERN LHC through heavy-quark
mediated gluon fusion (GF), where the Born-level process is at one-loop and at
order $\as^2$. Even after the discovery~\cite{Aad:2012tfa,Chatrchyan:2012ufa}
of the Higgs boson, an accurate prediction of the production mechanism is
needed to reveal any possible effects on the production rate from the
much sought-after physics beyond the Standard Model. Since the coupling of
the Higgs boson to a quark is proportional to the quark mass, the
gluon fusion process is dominated by the contribution from a top-quark
loop. This process is known to \NNNLO (to order $\as^5$) in the limit
of infinite top-mass~\cite{Anastasiou:2014vaa,Anastasiou:2015ema,Anastasiou:2016cez,Cieri:2018oms}.
%% accuracy needed to a) disseminate any possible BSM effects (couple to top
%% loop) and b) to disentangle VBF and GF production. Not just to measure the
%% production accurately and individually - GF as background, GF as signal to
%% check for CP admixtures
Finite top-mass effects in the inclusive cross section can be taken into
account at one order lower (to \NNLO in \as) by a formal expansion in
$(m_h/m_t)$. The effects on the total cross section of the finite top-mass
are found to be very small
indeed~\cite{Harlander:2009mq,Harlander:2009my,Harlander:2012hf}. An explicit
calculation of the loop contribution using the full propagator dependence
allows the inclusion also for the contribution from bottom-quarks. It is here
found that the bottom-top interference effects are of the order of
-5\%~\cite{Anastasiou:2016cez}.

Higgs boson production in association with \emph{one} jet obviously forms a
subset of the higher order corrections to the calculation of inclusive
Higgs-boson production. As such, it is known to \NNLO in \as (to order
$\as^5$) in the limit of infinite top-mass. It was very recently calculated
to \NLO with full dependence on the heavy-quark
propagator~\cite{Jones:2018hbb}. This explicit calculation of the quark loops
can be used to check the earlier reported top-bottom interference effects,
which were approximated using a small-mass expansion for the amplitudes
involving bottom quarks, and the infinite mass limit for the top-quark
contribution~\cite{Lindert:2017pky}. For $H+j$-production at \NLO, the effect of
the full dependence on the heavy-quark propagator momentum is a 9\%
\emph{increase} in the overall cross section over the result obtained in the
infinite top-mass limit. The quality of the approximations obtained using the
infinite top mass limit in $H+j$-production is therefore much worse than for
inclusive Higgs boson production. Furthermore, there is a strong phase-space
dependence: for transverse momenta of the Higgs boson larger than 800~GeV, the
effects of the full dependence on the heavy-quark propagators leads to a
\emph{suppression} over the result for an infinite top mass of more than an
order of magnitude. For processes with more than just the Higgs boson in the
final state, the limit of infinite top-mass loses not only the dependence on
the mass of the propagating quarks, but also the full kinematic dependence on
the propagators in the loop-diagrams.

Higgs boson production in association with \emph{two} jets can proceed
through both of the processes of weak boson fusion (VBF) and gluon
fusion. The VBF process reveals the direct coupling between the Higgs boson
and the weak bosons, whereas the GF $H+2j$-process allow for studies of possible
$CP$-admixtures in the Higgs sector~\cite{Klamke:2007cu,Andersen:2010zx}. A
precise study of either of these effects requires a separation of the
contribution from the two processes, which has to be guided by detailed
calculations. Luckily, these indicate that the interference between the two
processes is
negligible~\cite{Andersen:2007mp,Bredenstein:2008tm,Dixon:2009uk}, so the two
processes can in principle be studied independently.

The VBF process is known fully differentially at
\NNLO~\cite{Cacciari:2015jma,Cruz-Martinez:2018rod} (i.e.~to order $\as^2\alpha^2$) and the
inclusive cross section is known in the effective structure function approach
to \NNNLO~\cite{Dreyer:2016oyx} (i.e.~to order $\as^3\alpha^2$). One
important lesson from these calculations is that while the higher order
perturbative corrections to the inclusive cross sections are very small
indeed, within typical VBF-cuts the \NNLO-corrections to the \NLO-result
can be 3-4 times larger, and reduce the cross section by 4\%, with effects of
up to 7\% on distributions.

The contribution through GF to $H+2j$ is known with full dependence on the
heavy-quark propagator just at LO~\cite{DelDuca:2001eu,DelDuca:2001fn}, and at
NLO in the infinite top-mass limit~\cite{Campbell:2006xx,Campbell:2010cz}. The
situation is the same for $H+3j$~\cite{Cullen:2013saa,Greiner:2016awe}.  In the
current paper we present a calculation of higher-order perturbative corrections
to the GF component of $H+2j$-production, maintaining the full dependence on the
heavy-quark propagator in the heavy-quark mediated coupling to the Higgs boson,
and including the effects of propagating both bottom and top quarks. The results
obtained are exact in the limit of large dijet invariant mass, which is relevant
for the VBF- and gluon-fusion CP-studies, and are furthermore matched to the
highest-order fixed-order perturbative result which could be produced with
available tools, in this case \Sherpa~\cite{Gleisberg:2008ta} with the extension
of \openloops~\cite{Cascioli:2011va}\footnote{This gives $pp\to H+2j-$processes
  with full quark mass dependence.  The corresponding results for $pp\to H+3j$
  in \cite{Greiner:2016awe} could have been included directly if the
  implementation was readily available or if the results
  were available as event files.}. The
results rely on the observation that the high-energy limit
commutes with any limit taken on the masses of the propagating quarks in the
coupling to the Higgs boson~\cite{DelDuca:2003ba}, and the results are obtained within the framework
of \HIGHEJ\
(\HEJ)~\cite{Andersen:2009nu,Andersen:2009he,Andersen:2011hs,Andersen:2012gk,Andersen:2016vkp,Andersen:2017kfc,Andersen:2018tnm}.

In section~\ref{sec:QuarkMasses} we explore the structure of the amplitudes for
the different subprocesses of $pp\to H+2j$ with full dependence on finite quark
masses.  We use these results to construct matrix-elements within \HEJ which
contain all finite quark mass effects and maintain accuracy to leading-logarithm
in $s/t$ at all orders in $\as$.  This manifestly includes the
calculation of subprocesses with a high number of high-energy jets, going far
beyond what is possible at fixed-order with finite quark mass effects.  In
section~\ref{sec:matching} we describe the different types of matching to
fixed-order which we employ in the \HEJ predictions.  This is quite involved
owing to the variety of fixed-order samples available.  In
section~\ref{sec:results} we present our results, focussing separately on the
effects of higher perturbative orders and the effects of finite top mass, before
we compare our most-accurate \HEJ prediction with the most-accurate available
fixed-order prediction.  In section~\ref{sec:summary} we summarise our findings.

%%% Local Variables:
%%% mode: latex
%%% TeX-master: "main"
%%% End:

\section{Quark Mass Effects in Higgs Boson Production with \HEJ}
\label{sec:QuarkMasses}

The \HEJ framework is a perturbative framework for QCD processes which achieves
leading logarithmic accuracy in variables which scale as $s/t$.  These are seen
to arise at all orders in $\as$ from
BFKL~\cite{Fadin:1975cb,Kuraev:1976ge,Kuraev:1977fs,Balitsky:1978ic}.  Logarithmic accuracy
at the integrated level is obtained through a power series expansion of the
relevant matrix elements in $s/p_\perp^2$.  This gives the dominant terms in
the multi-Regge kinematic (MRK) limit (or high-energy limit), defined for a $2\to n$ QCD process as
\begin{align}
  \label{eq:helimit}
  s_{ij}\to\infty, \qquad |p_{i\perp}| \; \mathrm{finite} \quad i,j=1,\ldots,n.
\end{align}
In this limit, only a subset of flavour and momentum configurations contribute
at leading power.  From Regge theory, these are the configurations which permit
the maximum number of $t$-channel gluon exchanges in a ladder chain with
particles ordered in rapidity.  We will
call these ``FKL configurations''\footnote{Named after Fadin, Kuraev and
  Lipatov.}.  For example, in pure jet production $qQ\to qggQ$ and $gg\to gg$ are FKL configurations
while $q\bar{q}\to gg$ and $qg\to q Q\bar{Q}g$ are not (all particles written in
rapidity order).

Furthermore, it is known that scattering amplitudes in the high-energy limit
factorise into products of independent scalar factors and emission vertices,
which each depend on a reduced set of momenta~\cite{DelDuca:1999ha,DelDuca:2003ba}.  This is critical for
obtaining simple expressions for large $n$.  In order to achieve this simple form though, many
approximations are required which limit the power of the description away from
the strict limit, i.e.~in physical regions of phase space.  Within \HEJ we
achieved the $t$-channel factorised structure while making fewer approximations
through the use of vector currents in place of scalar
factors~\cite{Andersen:2009nu}.  This immediately renders the \HEJ description
of $2\to 2$ processes exact and preserves the correct position of the
$t$-channel poles, and significantly improves the description of the matrix
elements in the physical region.

The production of a Higgs boson with at least two jets using currents in \HEJ
has been described in the infinite quark mass limit in~\cite{Andersen:2017kfc}.
The $t$-channel factorised structure of the amplitudes lends itself to the
inclusion of finite quark mass effects because the diagrams and loops do not
become any more complicated than those at leading order for arbitrarily large
$n$.  This has been a bottleneck for fixed order calculations where calculations
with finite quark masses have stalled at $H+3$-jets at leading-order.  Finite
quark mass effects in $H+2j$ were studied in the high-energy limit
in~\cite{DelDuca:2003ba}. In this section, we
recap some results and describe the necessary adaptation to incorporate the
results in the current-structure of \HEJ.  We will start our discussion with the
simplest leading-order configurations and gradually move on to more complex
arrangements. We then describe how we supplement these expressions with the
leading-logarithmic real and virtual corrections to all orders in $\alpha_s$
which allows us to resum the logarithms in $s/t$.

\subsection{Finite Mass Dependence in $qQ \to qHQ$}
\label{sec:qQ_to_qQH}

We first consider the gluon-fusion production of a Higgs boson with
two jets originating from two quarks with different flavours $q$ and $Q$. At leading
order, only two diagrams contribute to this production channel for each quark
flavour propagating in the loop, see figure~\ref{fig:qQ_to_qHQ}. As
illustrated, we choose the momentum
assignment $q(p_a) Q(p_b) \to q(p_1) H(p_H) Q(p_2)$ with all momenta
left-to-right.

\begin{figure}
  \centering
  \includegraphics{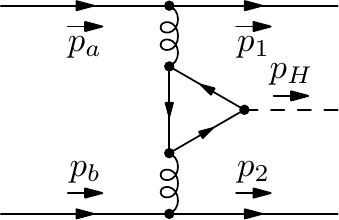} \qquad \includegraphics{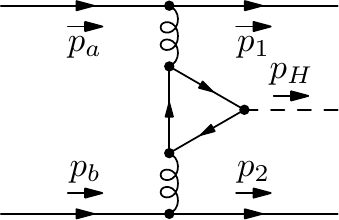}
  \caption{Leading-order diagrams contributing to the process $qQ \to qHQ$.}
  \label{fig:qQ_to_qHQ}
\end{figure}

The quark-loop insertions in the two diagrams are symmetric under charge
conjugation. They can be written as a colour-diagonal vertex of the
form
\begin{equation}
  \label{eq:VH}
  V^{\mu\nu}_H(q_1, q_2) = \mathgraphics{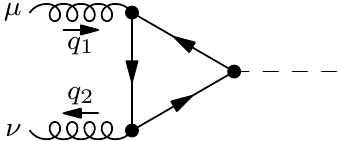} =
  \frac{\alpha_s m^2}{\pi v}\big[
  g^{\mu\nu} T_1(q_1, q_2) - q_2^\mu q_1^\nu T_2(q_1, q_2)
  \big]\,,
\end{equation}
where $m$ is the quark mass, $\alpha_s=\frac{g_s^2}{4\pi}$ is the strong
coupling constant and $v\approx 246\,\GeV$ is the Higgs vacuum expectation
value. For convenience, we list the form factors $T_1$ and $T_2$ in
appendix~\ref{sec:ggH_form_factors}. The expressions are given there for a
single propagating quark flavour; in practice, we sum the contributions from a
top quark and a bottom quark propagating in the loop at the amplitude level.  We
therefore automatically have contributions from both flavours \emph{and} the
interference between them when we square the amplitude.  The colour and
helicity summed and averaged square of the matrix element then takes the
factorised form
\begin{equation}
  \label{eq:Mqq_to_qHQ}
 \overline{ \left|{\cal M}_{qQ \to qHQ}\right|}^2 = \frac{1}{4(N_C^2-1)} ||{\cal S}^m_{qQ
    \to qHQ}||^2\cdot \left(g_s^2 C_F \frac{1}{t_1}\right)\cdot
  \left(\frac{1}{t_1t_2}\right)\cdot \left(g_s^2 C_F \frac{1}{t_2}\right),\,
\end{equation}
where the invariants $t_1 = q_1^2$, $t_2 = q_2^2$ are defined in terms of the
$t$-channel momenta $q_1 = p_a-p_1, q_2 = p_2
- p_b$, and $S^m_{qQ \to qHQ}$ is the current contraction
\begin{equation}
  \label{eq:Sqq_to_qHQ}
  S^m_{qQ \to qHQ} = j_\mu(p_1,p_a) V^{\mu\nu}_H(q_1,q_2)
  j_\nu(p_2,p_b),\,\qquad j_\mu(p_o,p_i) = \bar{u}(p_o)\gamma_\mu u(p_i)\,.
\end{equation}
We use the usual conventions for the QCD colour factors $N_C = 3, C_F =
\frac{4}{3}, C_A = 3$.  Throughout this section we use double-bar notation to
indicate summing over spins.

The amplitude has the same structure as in the limit of an infinite quark
mass~\cite{Andersen:2009nu,Andersen:2008ue,Andersen:2008gc}. Indeed, the
only difference is in the expression for $V_H^{\mu\nu}$, which in the limit is given by
\begin{equation}
  \label{eq:VH_inf_mt}
  V^{\mu\nu}_H(q_1, q_2) \xrightarrow{m \to \infty} \frac{\alpha_s}{3\pi
    v} \left(g^{\mu\nu} q_1\cdot q_2 - q_2^\mu q_1^\nu\right).
\end{equation}
The factorisation of eq.~\eqref{eq:Mqq_to_qHQ} into a current
contraction and a product of $t-$channel propagators is exactly what
enables us to perform \HEJ resummation, as we will discuss in more detail
in section~\ref{sec:resummation}. We stress that up to this point we retain the
exact expression for the amplitude without having to resort to approximations
valid in the high-energy limit.

Here we have referred to initial quarks; the treatment of initial antiquarks is
completely analogous.  The only qualitatively different amplitude is
$q\bar{q} \to qH\bar{q}$, which receives contributions from both $t$-channel
gluon exchange as in figure~\ref{fig:qQ_to_qHQ} and two annihilation diagrams with
$s$-channel gluon exchange. Both sets of diagrams are individually gauge
independent, but the annihilation diagrams are subdominant for a large invariant
mass between the two jets, so that the leading contribution to the amplitude is
the same as the pure-quark amplitude eq.~\eqref{eq:Mqq_to_qHQ}.  One may also
consider the process $q\bar{q} \to gHg$, but this is not an FKL configuration
and hence will not contribute a leading power in $s/t$ to the
matrix-element-squared.

\subsection{Finite Mass Dependence in $gq \to gHq$}
\label{sec:gq_to_gHq}

It was shown in~\cite{Andersen:2009he} that in pure jet production the matrix elements for
gluon-initiated $2\to 2$ processes can also be described exactly as a
contraction of currents where the current for the equivalent quark process
($j_\mu^\pm(p_o,p_i)$ in eq.~\eqref{eq:Sqq_to_qHQ}) is multiplied by a scalar factor $K_g/C_F$.  For a backward-moving
incoming gluon for example, $K_g$ is given by
\begin{equation}
  \label{eq:K_g}
 K_g(p_1^-, p_a^-) = \frac{1}{2}\left(\frac{p_1^-}{p_a^-} + \frac{p_a^-}{p_1^-}\right)\left(C_A -
          \frac{1}{C_A}\right)+\frac{1}{C_A}\,.
\end{equation}
Using this instead of a scalar gluon impact factor or a current multiplied by
$C_A/C_F$ (the limiting value also referred to in~\cite{Andersen:2017kfc}) improves the \HEJ description
of jet processes with incoming gluons
for finite rapidity differences.  The $t$-channel factorisation of an amplitude
implies not only that each factor is independent of the momenta of the rest of
the process, but that it is also independent of the particle-content of the rest of the
process.  This description of incoming gluons in inclusive dijet production is
therefore also valid in $H+2j$ production and we will use it in what follows.

At leading order, the $gq\to gHq$ in the initial state is significantly more
involved than the process with two incoming quarks. Of the 20 diagrams
contributing to the leading-order amplitude, 10 can be obtained from charge
conjugation. The remaining diagrams are depicted in
figure~\ref{fig:gq_to_gHq}.

\begin{figure}[h]
  \centering
    \begin{tabu} to \textwidth {*4{X[c]}}
      \subcaptionbox{\label{fig:gq_to_gHq_a}}{\includegraphics{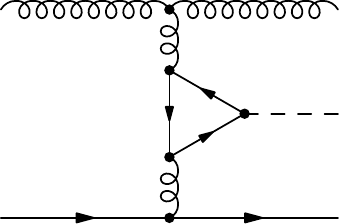}}&
      \subcaptionbox{\label{fig:gq_to_gHq_b}}{\includegraphics{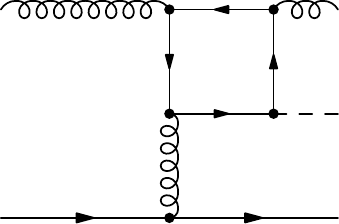}}&
      \subcaptionbox{\label{fig:gq_to_gHq_c}}{\includegraphics{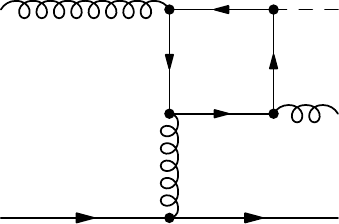}}&
      \subcaptionbox{\label{fig:gq_to_gHq_d}}{\includegraphics{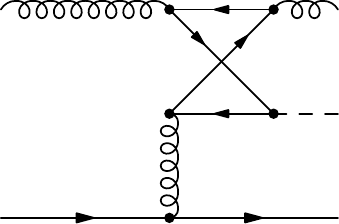}}\\[1em]
    \end{tabu}\\[1em]
    \begin{tabu} to \textwidth {*3{X[c]}}
      \subcaptionbox{\label{fig:gq_to_gHq_e}}{\includegraphics{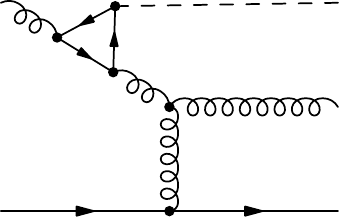}}&
      \subcaptionbox{\label{fig:gq_to_gHq_f}}{\includegraphics{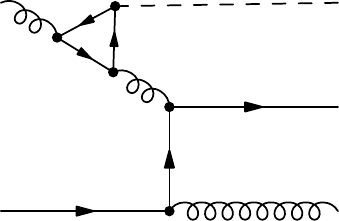}}&
      \subcaptionbox{\label{fig:gq_to_gHq_g}}{\includegraphics{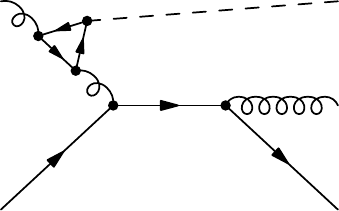}}\\[1em]
      \subcaptionbox{\label{fig:gq_to_gHq_h}}{\includegraphics{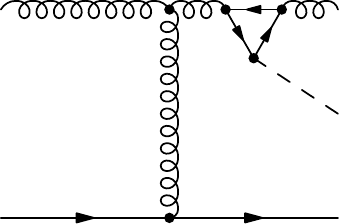}}&
      \subcaptionbox{\label{fig:gq_to_gHq_i}}{\includegraphics{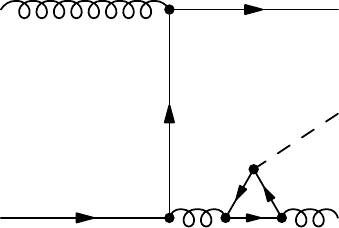}}&
      \subcaptionbox{\label{fig:gq_to_gHq_j}}{\includegraphics{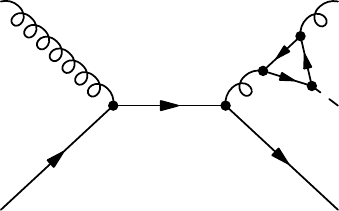}}
    \end{tabu}
    % tabu somehow messes up the figure counter. This is a crude workaround.
    \addtocounter{figure}{1}
   \caption{Leading-order diagrams contributing to the process $gq \to
     gHq$. Diagrams with clock-wise fermion flow in the heavy quark loop can
     be obtained via charge conjugation and are not shown. }
   \label{fig:gq_to_gHq}
 \end{figure}

The amplitude with full quark-mass dependence is known for general
kinematics~\cite{DelDuca:2001eu,DelDuca:2001fn}; it does not have a $t$-channel
factorised form.  In the following subsections, we discuss the different hierarchies
which can exist between invariants in the process and the expressions we will
use to describe this process in the corresponding regions of phase space.

\subsubsection{Central Higgs-Boson Emission}
\label{sec:Higgs_central}

Let us first discuss the case where the Higgs boson is emitted in
between the two jets, with a large rapidity separation from each jet. More
concretely, we consider the momentum assignment $g(p_a) q(p_b) \to
g(p_1) H(p_H) q(p_2)$ with the hierarchy
\begin{equation}
  \label{eq:hierarchy_Higgs_central}
 s_{12} \gg s_{1H}, s_{2H} \gg t_1, t_2, m_H^2\,,
\end{equation}
where $s_{ij} = (p_i + p_j)^2$ are invariant masses of the outgoing
particles, $t_1=(p_a-p_1)^2, t_2=(p_b-p_2)^2$. Assuming the gluon to be emitted backwards, this hierarchy
implies the rapidity ordering $y_1 \ll y_H \ll y_2$. The forward
emission of the gluon is of course completely analogous.

In~\cite{DelDuca:2003ba}, it was shown that the amplitude in this limit
assumes a similar factorised form as in the pure quark case.  This is
also true within the \HEJ formalism~\cite{Andersen:2017kfc}.  As an
example, the colour summed and averaged square of the helicity-conserving amplitude
for a positive-helicity gluon and a negative-helicity quark can
therefore be written as
\begin{equation}
  \label{eq:M_gq_to_gHq}
    \left|{\cal M}_{g^+q^- \to g^+Hq^-}\right|^2 = \frac{1}{N_C^2-1}
    ||{\cal S}^m_{g^+q^- \to g^+Hq^-}||^2
 \left(g_s^2 K_g(p_1^-, p_a^-)
      \frac{1}{t_1}\right)
    \cdot \left(\frac{1}{t_1t_2}\right)\cdot \left(g_s^2 C_F \frac{1}{t_2}\right)
\end{equation}
where $K_g(p_1^-,p_a^-)$ was given in eq.~\eqref{eq:K_g}.  The current contraction is given by
\begin{equation}
  \label{eq:S_gq_to_gHq}
S^m_{g^+q^- \to g^+Hq^-} = j_\mu^+(p_1, p_a) V_H^{\mu\nu}(q_1,q_2)
j_\nu^-(p_2,p_b)\,,\qquad j_\mu^\pm(p_o,p_i) = \bar{u}^\pm(p_o)\gamma_\mu
u^\pm(p_i)\,,
\end{equation}
as in eq.~\eqref{eq:Sqq_to_qHQ} for $qQ\to qHQ$.
% The only non-factorising helicity amplitudes involve a flip of the gluon
% helicity and identical helicities of the incoming
% partons~\cite{Andersen:2009he}. These amplitudes are suppressed in the
% high-energy limit defined by eq.~\eqref{eq:hierarchy_Higgs_central} and
% therefore neglected here.
% We point out that, independently of
% the chosen gauge, the contribution from the box diagrams,
% figure~\ref{fig:gq_to_gHq_b}, \ref{fig:gq_to_gHq_c},
% \ref{fig:gq_to_gHq_d}, is suppressed in the high-energy limit
% defined by eq.~\eqref{eq:hierarchy_Higgs_central}. This observation
% will facilitate the high-energy resummation discussed in
% section~\ref{sec:resummation}.

\subsubsection{Peripheral Higgs-Boson Emission}
\label{sec:Higgs_peripheral}

We now consider the case where we drop the strong-ordering requirement between
the Higgs boson and one of the jets.
% If the Higgs boson is emitted close to one of the jets or outside the
% rapidity range spanned by the two jets the ordering in
% eq.~\eqref{eq:hierarchy_Higgs_central} is no longer fulfilled.
For the case
where the Higgs boson is not strongly separated from a quark, we again invoke
$t$-channel factorisation to treat this as in the $qQ\to qHQ$ process.  There
the result with $V_H^{\mu\nu}$ is exact wherever the Higgs boson is emitted and
hence in this case we again use eq.~\eqref{eq:M_gq_to_gHq}.

We now consider Higgs boson rapidities which are not strongly ordered with respect to
the rapidity of the gluon in $gq\to Hgq$. We will still require a separation from the quark,
i.e. $y_1, y_H \ll y_2$ or
\begin{equation}
  \label{eq:hierarchy_Higgs_peripheral}
s_{12}, s_{2H} \gg s_{1H}, t_1, t_2, m_H^2\,.
\end{equation}
We denote such configurations as $gq \to Hgq$, reserving the notation
$gq \to gHq$ for the central Higgs-boson emission discussed in
section~\ref{sec:Higgs_central}.
The $t$-channel factorisation of the amplitude is only guaranteed where there is
a large rapidity separation between outgoing particles, and hence in this reduced limit we should not
expect to recover a form with two $t$-channel poles as in
eq.~\eqref{eq:M_gq_to_gHq}.  However, as there is still a large rapidity
separation to the quark line, we expect to find a factorised form about the
pole in $t_2$ as follows:
\begin{align}
  \label{eq:M_gq_to_Hgq}
    \overline{\left|{\cal M}_{gq \to Hgq}\right|}^2 ={}& \frac{1}{4(N_C^2-1)}
    ||{\cal S}^m_{gq \to Hgq}||^2\cdot (\alpha_s^2 g_s^2 C_A) \cdot \left(\frac{1}{t_2}\right)\cdot
    \left(g_s^2 C_F \frac{1}{t_2}\right)\,,\\
    S^m_{gq \to Hgq} ={}& j_H^\mu(p_1, p_H, p_a) j_\mu(p_2,p_b)\,,
\end{align}
where the remainder of the amplitude has been written as an effective current
$j_H^\mu$.  This current, dependent on the reduced set of momenta
$(p1,p_H,p_a)$, is derived in appendix~\ref{sec:jH}. The derivation follows
closely the approach in~\cite{Andersen:2009he} for calculating the effective
gluon current.  In contrast to that case, or indeed that of a central Higgs boson,
amplitudes flipping the gluon helicity also contribute. Explicit
expressions are given in appendix~\ref{sec:jH}.

\subsection{Finite Mass Dependence in $gg\to gHg$}
\label{sec:H2j_rest}

As noted above, the $t$-channel factorisation which arises in the limit of large
rapidity separations implies that the building block corresponding to each end
of the chain is independent of the rest of the process.  We can therefore
describe the $gg$-initiated state by taking the expressions for $gq\to gHq$ and
adding the necessary change for an incoming gluon derived from pure jets. We
find for central Higgs boson emission from eq.~\eqref{eq:M_gq_to_gHq}:
\begin{align}
  \label{eq:M_gg_to_gHg}
  \begin{split}
    \left|{\cal M}_{g^+g^- \to g^+Hg^-}\right|^2 =& \frac{1}{N_C^2-1}
    ||{\cal S}^m_{g^+q^- \to g^+Hq^-}||^2 \\ &\cdot
 \left(g_s^2 K_g(p_1^-, p_a^-)
      \frac{1}{t_1}\right)
    \cdot \left(\frac{1}{t_1t_2}\right)\cdot \left(g_s^2 K_g(p_2^+,p_b^+)
      \frac{1}{t_2}\right)
  \end{split} \\
  S^m_{g^+g^- \to g^+Hg^-} =& j_\mu^+(p_1, p_a) V_H^{\mu\nu}(q_1,q_2)
  j_\nu^-(p_2,p_b)\,.
\end{align}
Likewise, for a Higgs boson emitted backward of both gluons where only the
subset of hierarchies applies (eq.~\eqref{eq:hierarchy_Higgs_peripheral}), we find from
eq.~\eqref{eq:M_gq_to_Hgq}:
\begin{align}
  \label{eq:M_gg_to_Hgg}
    \overline{\left|{\cal M}_{gg \to Hgg}\right|}^2 ={}& \frac{1}{4(N_C^2-1)}
    ||{\cal S}^m_{gg \to Hgg}||^2\cdot (\alpha_s^2 g_s^2 C_A) \cdot \left(\frac{1}{t_2}\right)\cdot
    \left(g_s^2 K_g(p_2^+,p_b^+) \frac{1}{t_2}\right)\,,\\
    S^m_{gq \to Hgq} ={}& j_H^\mu(p_1, p_H, p_a) j_\mu(p_2,p_b)\,.
\end{align}
Note that the gluon closest to the Higgs boson will always have the more
complicated treatment derived in the previous subsection (i.e.~it is the momentum of the
gluon closest in rapidity to the Higgs boson which will enter $j_H^\mu$).

\subsection{The First Set of Next-to-Leading Logarithmic Corrections}
\label{sec:first-set-next}

\begin{figure}[btp]
  \centering
  \begin{tabular}{c@{\hspace{2cm}}c}
    \includegraphics{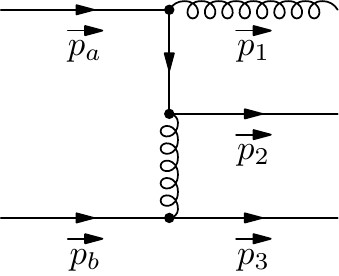} &\includegraphics{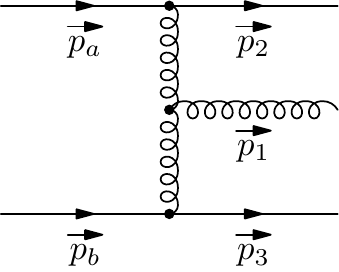}\\
$y_1 < y_2 < y_3$ & $y_2<y_1<y_3$
  \end{tabular}
  \caption{Rapidity-ordered ladder diagrams for $qQ\to gqQ$.  The ordering in
    eq.~\eqref{eq:unoexample} (left) contains just one $t$-channel gluon propagator, whereas the FKL
    configuration (right) contains the maximum number, two.}
  \label{fig:tprops}
\end{figure}

We have now derived the \HEJ description of the scattering amplitudes for all
$2\to H+2$ processes which will contribute at leading power for Higgs-plus-dijets.
At the start of this section, we identified the necessary subprocesses as the
FKL configurations of flavour and momenta.  In \cite{Andersen:2017kfc}, we
extended the \HEJ framework to also describe Born processes where we have
relaxed the requirement of rapidity ordering on exactly one gluon, allowing it
to be emitted outside of the rapidity range defined by an outgoing quark, e.g.:
\begin{align}
  q(p_a) Q(p_b) \to g(p_1) q(p_2) H(p_H) Q(p_3) \qquad y_1 < y_2 \ll y_3\,.
  \label{eq:unoexample}
\end{align}
The addition of the colour-neutral Higgs
boson does not affect the following argument, so for a moment we will consider
only the coloured particles.  When constructed as a rapidity-ordered ladder
diagram, see figure~\ref{fig:tprops},
the ordering above contains one
$t$-channel quark propagator and one $t$-channel gluon propagator as opposed to
the two $t$-channel gluon propagators one would find if $y_1$ and $y_2$ were
reversed.  It is therefore a non-FKL configuration,
and as it is missing one gluon propagator it is suppressed by one power of $s$ at
matrix-element-squared level compared to the FKL configuration.  This is
formally therefore a next-to-leading logarithmic contribution to the dijet cross
section; however one can still construct the leading logarithmic contributions
to each particular subprocess.  We denote these ``unordered'' configurations.
This particular class of processes was chosen as it had been observed after
matching to leading-order that they contributed significantly in regions of
phase space with large transverse momentum for example.  Their inclusion
therefore allows \HEJ to reduce its dependence on fixed-order matching~\cite{Andersen:2017kfc} (see
section~\ref{sec:matching} for a full discussion of all matching included in the
current study).

\begin{figure}[bt]
  \centering
  \includegraphics{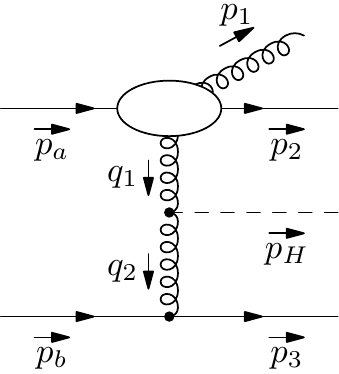}
  \caption{For the rapidity ordering in eq.~\eqref{eq:unoexample} where there is only strong ordering
    between $y_2$ and $y_3$, we should only expect to find factorisation about the
    $t$-channel pole between these particles so the structure of the amplitude
    should be as shown.  We find this in eq.~\eqref{eq:MunocentralH}.}
  \label{fig:strcuture}
\end{figure}

From eq.~\eqref{eq:unoexample}, these subprocesses at Born level have just one
strong rapidity-ordering between the coloured particles and one therefore
constructs an effective current to describe the $q(p_a) \to g(p_1) q(p_2) g^*$
end of the chain, denoted $ j_\mu^{{\rm uno}}(p_2,p_1,p_a)$ (illustrated in
figure~\ref{fig:strcuture}).
This must now
carry two colour indices as it consists of terms with different colour flow.
The matrix element for the case of central Higgs boson emission (where the Higgs
boson is emitted between \emph{the quarks}) is then given by
\begin{align}
    \overline{\left|{\cal M}_{qf_2 \to gqHf_2}\right|}^2 ={}&
    \frac{1}{4(N_C^2-1)}
    ||S^{\rm uno}_{qf_2 \to gqHf_2}(p_1,p_2,p_3,p_a,p_b,q_1,q_2)||^2 \notag\\
    &\cdot \left(g_s^4 K_{\rm uno} \frac{1}{t_1}\right)\cdot
    \left(\frac{1}{t_1t_2}\right)\cdot \left(g_s^2 K_{f_2}
      \frac{1}{t_2}\right)\,,\label{eq:MunocentralH}\\  S^{\rm uno}_{qf_2 \to
      gqHf_2}(p_1,p_2,p_3&,p_a,p_b,q_1,q_2)={} \frac{1}{\sqrt{C_F}} j_\mu^{
      {\rm uno}\ cd}(p_2,p_1,p_a)
    V_H^{\mu\nu}(q_1,q_2)j_\nu(p_3,p_b)T^c_{3b}\,,\notag
\end{align}
where $K_{\rm uno}=-1/2$ and we are using $q_1=p_a-p_1-p_2$, $q_2=p_3-p_b$,
$t_i=q_i^2$ (as in figure~\ref{fig:strcuture}).  It is clear therefore
that the arguments of $S^{\rm uno}_{qQ\to gqHQ}$ are not independent; we have
chosen to display the implicit dependence on $q_1, q_2$  in anticipation of processes with additional
gluons, where the dependence is explicit.  The expression for
$j_\mu^{{\rm uno}}(p_2,p_1,p_a)$ is given in
appendix~\ref{sec:curr-single-unord}.

So far in this section, we have discussed how to construct the \HEJ
approximation to Born-level matrix elements.  In the rest of this
section, we outline how to supplement these with the dominant corrections at all
orders in $\alpha_s$.

\boldmath
\subsection{\HEJ Resummation}
\unboldmath
\label{sec:resummation}

In the previous subsection, we have outlined the \HEJ approximation to the
Born-level matrix element for $f_1 f_2\to f_1 H f_2$, $f_1 f_2\to H f_1 f_2$,
$q f_2\to g q H f_2$ and their symmetric equivalents.  These skeletons will
provide all leading-logarithmic contributions and an important class of
next-to-leading contributions to $H+\ge 2$-jet production.  In order to achieve
an all-order resummation we must add both the dominant real and virtual
corrections.  Our method was described in great detail for the case of an
infinite quark mass in~\cite{Andersen:2017kfc}.  The presence of a finite quark
mass does not affect the resummation procedure as the mass dependence enters
only in the Higgs boson
vertex $V_H^{\mu\nu}$ and the effective current $j_H^\mu$, and so the method
remains unchanged.  In the rest of this subsection we therefore briefly
summarise the real corrections, the virtual corrections and the organisation of
the cancellation of the poles to give an all-order finite result.

\subsubsection{Real Emissions}
\label{sec:real-emissions}

The dominant real corrections in the MRK limit arise in FKL flavour and momentum configurations.
Where there are more than two outgoing coloured particles, the unique FKL
configuration for a given incoming state is the process which has additional
gluons emitted with rapidities in between two outgoing particles of identical
flavour to the incoming particles.  The high-energy limit,
eq.~\eqref{eq:helimit}, implies that these must be well separated from all other
outgoing partons. Therefore, taking central Higgs-boson emission as an example,
for a total of $n$ emitted partons with momenta $p_1,\dots,p_n$, and Higgs boson
emitted between $j$ and $j+1$, we have hierarchies of the form
\begin{equation}
  \label{eq:hierarchy_Higgs_central_all}
  y_1 \ll y_2 \ll \dots \ll y_j \ll y_H \ll y_{j+1} \ll \dots \ll
  y_{n-1} \ll y_n \,.
\end{equation}
In this limit, it has been shown that the emission of the $i$th gluon can
be described by a Lipatov vertex~\cite{Fadin:1996nw}, $V_L$, given by~\cite{Andersen:2009nu}
\begin{equation}
  \label{eq:GenEmissionV}
  \begin{split}
  V_L^\nu(q_i,q_{i+1})=&-(q_i+q_{i+1})^\nu \\
  &+ \frac{p_a^\nu}{2} \left( \frac{q_i^2}{p_{i+1}\cdot p_a} +
  \frac{p_{i+1}\cdot p_b}{p_a\cdot p_b} + \frac{p_{i+1}\cdot p_n}{p_a\cdot p_n}\right) +
p_a \leftrightarrow p_1 \\
  &- \frac{p_b^\nu}{2} \left( \frac{q_{i+1}^2}{p_{i+1} \cdot p_b} + \frac{p_{i+1}\cdot
      p_a}{p_b\cdot p_a} + \frac{p_{i+1}\cdot p_1}{p_b\cdot p_1} \right) - p_b
  \leftrightarrow p_n\,,
  \end{split}
\end{equation}
where $q_i=p_a-p_1-\ldots -p_i$ is the incoming $t$-channel momentum and
$q_{i+1}=q_i-p_{i+1}$ is the outgoing $t$-channel momentum (the $q_i$ will also
have $p_H$ subtracted if the emission is forward of the Higgs boson).  This
vertex is derived from the five possible tree-level diagrams for $qQ\to qgQ$,
and then employing $t$-channel factorisation.  At the matrix-element-squared
level after summing over polarisations each contributes a factor of
\begin{align}
  \left(\frac{-g_s^2 C_A}{t_{i-1} t_i} V_L^{\nu_i}(q_{i-1},q_{i}) V_{L\nu_i}(q_{i-1},q_{i}) \right)
  \label{eq:melipatov}
\end{align}
which multiplies the spinor string function, $S$ (see below).

\begin{figure}
  \centering
  \includegraphics{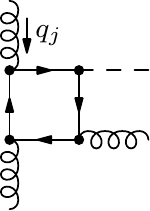}\qquad
  \includegraphics{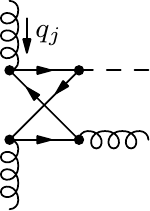}\qquad
  \includegraphics{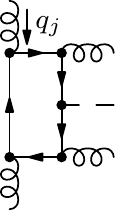}\qquad
  \includegraphics{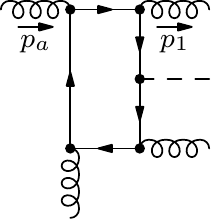}\qquad
  \caption{Examples for additional gluon emission off the heavy quark
    loop. These emissions are suppressed for large rapidity separations.  Here
    vertical lines represent off-shell $t$-channel propagators and horizontal
    lines represent on-shell external particles.}
  \label{fig:real_suppressed}
\end{figure}

This description of real corrections is the same as for the case of pure
tree-level multijet production. A priori, additional gluons could be
emitted off the massive quark loop coupling to the Higgs boson (see
figure~\ref{fig:real_suppressed}). However, in comparison to the emissions
described before, such corrections are suppressed for a
large rapidity separation between the Higgs boson and the gluons and
will be neglected in the following. The absence of $t$-channel
enhancement is obvious in the limit of a large quark mass, where the
quark loop is absorbed into an effective local interaction.

Defining the notation $\cdot g \cdot$ to mean an arbitrary number of gluons
(including zero), we may therefore compactly write the \HEJ approximation to the
Born-level matrix
element for $f_1f_2\to f_1\cdot g \cdot H \cdot g \cdot f_2$ to be
\begin{align}
  \label{eq:MHEJ_realcentral}
    \overline{\left|\mathcal{M}^{\rm HEJ,\ tree}_{\ f_1 f_2 \to f_1\cdot g\cdot
  H\cdot g \cdot f_2}\right|}^2 ={}& \frac1{4(N_C^2-1)} \left\|\mathcal{S}_{f_1f_2\to
                 f_1Hf_2} \right\|^2 \notag\\
  & \cdot\left(\frac{1}{t_j t_{j+1}}\right)\cdot \left(g_s^2 K_{f_1}(p_1^-, p_a^-) \frac{1}{t_1} \right) \cdot \left( g_s^2 K_{f_2}(p_2^+, p_b^+)
    \frac{1}{t_{n}} \right)
  \notag\\ & \cdot \prod_{k=2}^{j} \left(\frac{-g_s^2 C_A}{t_{k-1} t_k}
       V_L^{\nu_k}(q_{k-1},q_{k}) V_{L\nu_k}(q_{k-1},q_{k}) \right)\notag\\ &
        \cdot\prod_{k=j+1}^{n-1} \left(\frac{-g_s^2 C_A}{t_k t_{k+1}}
           V_L^{\nu_k}(q_{k},q_{k+1}) V_{L\nu_k}(q_{k},q_{k+1}) \right),\\
  \label{eq:SHEJ_realcentral}
\mathcal{S}_{f_1f_2\to
                 f_1Hf_2} ={}& j_\mu(p_1, p_a) V_H^{\mu\nu}(q_j,q_{j+1})
j_\nu(p_b,p_n)\,,
\end{align}
where we define $K_q=K_{\bar{q}}=C_F$ for any flavour of quark/antiquark.
Analogous expressions exist for the case of the Higgs boson emitted outside of the
coloured particles in rapidity.

\subsubsection{Virtual Corrections}
\label{sec:virtual-corrections}

The leading-logarithmic terms of the virtual corrections can be obtained via the
\emph{Lipatov Ansatz}~\cite{Kuraev:1976ge}.  This is a prescription where, given
a $t$-channel factorised matrix element as in eq.~\eqref{eq:MHEJ_realcentral}
and a corresponding hierarchy of scales, eq.~\eqref{eq:hierarchy_Higgs_central_all},
each $t$-channel pole is replaced:
\begin{equation}
  \label{eq:virtual}
  \frac{1}{t_i} \to \frac{1}{t_i}\exp[\hat{\alpha}(q_{i\perp})(y_{i+1}-y_i)]\,,
\end{equation}
where $t_i = q_i^2$, $y_{i+1}$ and $y_i$ are the (ordered) rapidities of the emissions
connected by the propagator. Using dimensional regularisation ($D=4-2\epsilon$)
\begin{align}
  \hat{\alpha}(q_i) = - g_s^2 C_A
  \frac{\Gamma(1-\varepsilon)}{(4\pi)^{2+\varepsilon}} \frac2\varepsilon
  \left(\frac{q_{i\perp}^2}{\mu^2} \right)^\varepsilon
  \label{eq:alphahat}
\end{align}
contains divergences as $\varepsilon\to 0$. In the following we describe how to
organise the cancellation of these poles with the poles arising from soft,
real corrections. This prescription correctly describes the leading and even
next-to-leading logarithmic terms, as verified up to \NNLO~\cite{Bogdan:2002sr}.

\subsubsection{Organisation of Cancellation of Poles}
\label{sec:organ-canc-poles}

It is clear that eq.~\eqref{eq:virtual} will generate poles in $\varepsilon$.
The divergences in eq.~\eqref{eq:MHEJ_realcentral} are not immediately obvious;
these appear when the matrix-element-squared is integrated over phase space
where the momentum of the additional gluons described by $V_L$ goes to zero.  The two sources of poles cancel exactly.  In order to organise this
cancellation, we introduce subtraction terms which in real-emission phase space
apply for $0<|p_\perp|<\lambda$.
We use a subtraction term of
\begin{align}
  S=\frac{4}{p_{k\perp}^2}
\end{align}
for $p_{k\perp}<\lambda$ in each of the squares of the Lipatov vertices, such
that
 \begin{align}
   \frac{V^\mu(q_{k-1},q_k)V_\mu(q_{k-1},q_k)}{t_{k-1}t_k} \rightarrow
  \frac{V^\mu(q_{k-1},q_k)V_\mu(q_{k-1},q_k)} {t_{k-1}t_k} -\frac{4}{p_{k\perp}^2} \,.
 \end{align}
for $p_{k\perp}<\lambda$. The subtraction term is integrable in
$D=4-2\epsilon$ and the contribution is added to the virtual corrections,
such that the finite contribution from the virtual corrections is then
described through the prescription
\begin{equation}
  \label{eq:omega0}
  \frac{1}{t_i} \to \frac{1}{t_i}\exp[\omega^0(q_{i\perp})(y_{i+1}-y_i)]\,,
\end{equation}
where the \emph{regularised Regge trajectory}
$\omega^0$ is
\begin{equation}
  \label{eq:w0}
  \omega^0(q_{\perp})=
C_A \frac {\as}{\pi} \ \log \left(\frac{\lambda^2}{q_{\perp}^2} \right).
\end{equation}
It can then be shown analytically (see
e.g.~\cite{Andersen:2008gc,Andersen:2009nu}) that the poles which arise when
integrating over the region $0<|p_\perp|<\lambda$ are equal and opposite to
the virtual poles in eq.~\eqref{eq:virtual}.

 The real radiation is regulated since
\begin{align}
  \frac{V^\mu(q_{k-1},q_k)V_\mu(q_{k-1},q_k)}{t_{k-1}t_k} \xrightarrow{p_{\perp
  k}\to 0} -\frac{4}{p_{k\perp}^2} \,.
\end{align}
For values of $|p_\perp|\lesssim 200$~MeV, the limit is sufficiently good
that the integral of the sum of Lipatov vertices and subtraction term can be
ignored for lower values of $|p_\perp|$.  In that case we define a lower
cut-off $\kappa\simeq200$~MeV of the phase space integrals. The final results are
stable under variation of both $\kappa$ and $\lambda$; the results presented
in this study were obtained with $\lambda=\kappa$.

The final regularised, resummed expressions for the squared matrix element for
the production of a Higgs boson in between partons $j$ and $j+1$
with incoming flavours $f_1,f_2$ is then\footnote{Note that the factor of
  $1/(t_j t_{j+1})$ in the second line was missing in~\cite{Andersen:2017kfc}.}
\begin{align}
  \label{eq:MHEJ_central}
    \overline{\left|\mathcal{M}^{\rm HEJ}_{\ f_1 f_2 \to f_1\cdot g\cdot
  H\cdot g \cdot f_2}\right|}^2 ={}& \frac1{4(N_C^2-1)} \left\|\mathcal{S}_{f_1f_2\to
                 f_1Hf_2} \right\|^2 \notag\\
  & \cdot \left(g_s^2 K_{f_1}(p_1^-,
    p_a^-) \frac{1}{t_1} \right) \cdot\left(\frac{1}{t_j t_{j+1}}\right) \cdot \left( g_s^2 K_{f_2}(p_n^+, p_b^+)
    \frac{1}{t_{n}} \right)
  \notag\\ & \cdot \prod_{k=2}^{j} \left(\frac{-g_s^2 C_A}{t_{k-1} t_k}
       V_L^{\nu_k}(q_{k-1},q_{k}) V_{L\nu_k}(q_{k-1},q_{k}) \right)\notag\\ &
        \cdot\prod_{k=j+1}^{n-1} \left(\frac{-g_s^2 C_A}{t_k t_{k+1}}
           V_L^{\nu_k}(q_{k},q_{k+1}) V_{L\nu_k}(q_{k},q_{k+1}) \right)\notag\\
     & \cdot \prod_{i=1}^{j-1}
     \exp\left[\omega^0(q_{i\perp})(y_{i+1}-y_{i})\right] \cdot \prod_{i=j+2}^n
     \exp\left[\omega^0(q_{i\perp})(y_{i}-y_{i-1})\right] \notag\\ &
     \cdot \exp\left[\omega^0(q_{j\perp})(y_H-y_j) \right] \cdot\exp \left[
     \omega^0(q_{j+1\perp}) (y_{j+1}-y_{H})\right]\,.
\end{align}

The same formula holds for a backward (forward) Higgs-boson emission if $f_1$
($f_2$) is a quark or antiquark with $j=1$ ($j=n$). For a peripheral emission
close to a gluon, there is an equivalent expression but the first two lines
instead mirror eq.~\eqref{eq:M_gq_to_Hgq} and we have one fewer $t$-channel pole
as there is one fewer hierarchy.  Explicitly, the matrix-element-squared is
given by
\begin{align}
  \label{eq:MHEJ_periph}
    \overline{\left|\mathcal{M}^{\rm HEJ}_{\ g f_2 \to Hg \cdot g \cdot f_2}\right|}^2 ={}& \frac1{4(N_C^2-1)} \left\|\mathcal{S}^m_{gf_2\to
                 Hgf_2} \right\|^2 \cdot(\alpha_s^2 g_s^2 C_A) \cdot \left(\frac{1}{t_1}\right)\cdot \left( g_s^2 K_{f_2}(p_n^+, p_b^+)
    \frac{1}{t_{n}} \right)
  \notag\\ & \cdot \prod_{k=2}^{n-1} \left(\frac{-g_s^2 C_A}{t_{k-1} t_k}
       V_L^{\nu_k}(q_{k-1},q_{k}) V_{L\nu_k}(q_{k-1},q_{k}) \right)\notag\\
     & \cdot \prod_{i=1}^{n-1}
     \exp\left[\omega^0(q_{i\perp})(y_{i+1}-y_{i})\right]  \,.
\end{align}
Note here that $q_1=p_a-p_1-p_H$, $q_i=q_{i-1}-p_i$ for $i=2,\ldots,n$ and $t_i=q_i^2$.

The regulated matrix element above
is valid in the phase space of an arbitrary number of extra real gluon emissions each with
$|p_\perp|>\kappa$, provided they are between the extremal partons in
rapidity.  Note that the extremal partons play a special role and are not
allowed to become soft (we do not include the necessary virtual corrections to
regulate the fundamental spinor strings).  In practice we require the extremal
partons to carry a significant fraction of the extremal jet momentum to ensure
that they remain perturbative.

%%% Local Variables:
%%% mode: latex
%%% TeX-master: "main"
%%% End:

\section{Matching to Fixed Order}
\label{sec:matching}
This section describes the matching of results within \HEJ both to the full
leading-order finite quark-mass matrix elements and to the \NLO cross sections
obtained for infinite top mass. We
generate our event weights using the procedure outlined in~\cite{Andersen:2018tnm},
where we begin from fixed-order samples and supplement these with resummation.
The resummation step only applies to the particle and momentum configurations
discussed in section~\ref{sec:QuarkMasses} (namely FKL configurations or FKL
with one unordered gluon).  If a given fixed-order event is not one of these
configurations, it enters our final event sample with its weight unaltered.

As in~\cite{Andersen:2018tnm} we choose the central factorisation and
renormalisation scale for all the
predictions as
$\scalemjj$, where $m_{12}$ is the invariant
mass of the system of the two hardest jets. This scale is chosen because the
study in~\cite{Currie:2017eqf} indicates a better convergence of the
perturbative result for $pp\to 2j$ than traditional $p_t$-based scale choices
such as e.g.~$\scaleht$, in particular for large dijet
rapidity-spans. Since the formalism of the current study is of particular
interest within the VBF-cuts, we choose a central scale which obtains
reasonable uncertainty-estimates for the distributions that the cuts are
based on. We note that, with this choice, the $p_t$-based observables such as
$p_{H\perp}$ show the same pathological scale variance for large values that
the invariant mass-based observable develops for renormalisation
and factorisation scales based on the transverse momenta.

We observed in~\cite{Andersen:2018tnm} that a central scale choice of
$\scaleht$ leads to distributions in rapidity and dijet invariant
mass with values close to the upper edge of the scale variation band obtained when
$\mu_r$ and $\mu_f$ are varied independently by a factor of two around this central
scale choice, keeping their ratio between 0.5 and 2.  The
scale variance obtained with a central scale of $\scaleht$ is
therefore pathologically decreased for distributions at large dijet invariant
mass or large rapidity separations, which are relevant for the VBF
studies. While the scale variations obtained at \NLO with a central scale
choice of $\scalemjj$ are larger, they are also more
reasonable as an indication of the uncertainty due to higher order
corrections within the VBF-cuts.

Section~\ref{sec:matchexcl]} describes the fixed-order samples available which
we use as our starting point and the point-by-point matching applied to the
resummation events.  Section~\ref{sec:match-lead-order} then describes the
matching performed for all events at the level of the total cross section.

\subsection{Matching of Exclusive Amplitudes}
\label{sec:matchexcl]}
\HEJ allows for the perturbative series in each $n$-jet phase space point to
be matched to fixed order~\cite{Andersen:2018tnm}. This, obviously, is
possible only if amplitudes for the $n$-jet phase space point are readily
available. In this study, the fixed-order calculations are performed using
\Sherpa~\cite{Gleisberg:2008ta} with the extension of
\openloops~\cite{Cascioli:2011va} for the evaluation of the
$pp\to H+2j$-processes with full quark-mass dependence. However, this
fixed-order setup
includes just the effects from the top-quark, and not also those of the loops
of bottom-quarks. The effect of both top and bottom quarks is included in the
resummation, and will be discussed later.

Additionally, the $H+3j$-processes are not readily evaluated with the full
quark-mass dependence even at leading order, and even using the infinite
top-mass limit, only the $pp\to H+2j$-process is available at \NLO (and then
obviously $pp\to H+3j$ at tree-level).

The limitations in the fixed-order results mean that the
matching~\cite{Andersen:2018tnm} within \HEJ has to use many more different
components than usual. We will describe them in the following
sections. No point-by-point matching is performed for events with six or more
jets. For such multiplicities the fixed-order results are expensive to compute,
while typically only contributing to less than a percent for all shown observables.

On top of the matching of exclusive events described in the following, the
final predictions for \HEJ will be scaled with the ratio of the inclusive
cross section for $pp\to H+2j$ calculated at infinite top-mass for NLO and
\HEJ.

The described procedure obtains top and bottom mass dependence through the
all-order results, matching to the full top-mass results for $pp\to H+2j$, and
to $pp\to H+3j$ and $pp\to H+4j$ in the limit of infinite top-mass.

\subsubsection{Two-jet Matching with Finite Quark Mass}
\label{sec:2j}
The exclusive two-jet events are matched to full leading order, with finite
quark mass effects. However, as our fixed-order setup allows for just the top-quark
diagrams, technically the matching is performed by multiplying the all-order
results containing both top and bottom mass effects with the ratio of the square of
the full Born-level matrix element evaluated with just the top quark and the
corresponding approximation within \HEJ (using just the propagating
top-quark, with no bottom-quark effects). The final event weights are
therefore proportional to
\begin{equation}
  \label{eq:match_h2j}
  |\mathcal{M}_{\text{HEJ}}^{m_t, m_b}|^2
  \frac{|\mathcal{M}_{\text{LO}}^{m_t}|^2}{|\mathcal{M}_{\text{HEJ, LO}}^{m_t}|^2},\,
\end{equation}
where $\mathcal{M}_{\text{LO}}$ is the leading-order matrix element,
$\mathcal{M}_{\text{HEJ}}$ the all-order \HEJ matrix element,
and $\mathcal{M}_{\text{HEJ, LO}}$ its truncation to leading-order.
The superscript indicates the quark masses that are taken into account.

\subsubsection{Three-, Four- and Five-jet Matching with Infinite Quark Mass }
\label{sec:fininf}
With the method of \cite{Andersen:2018tnm}, the resummation could be
constructed starting from event files from the calculation of Born-level
Higgs-boson production in association with three jets including full momentum
and mass dependence reported in \cite{Greiner:2016awe}.  However, since
these are not available, the three-, four- and five-jet events will be
matched in the infinite top-mass limit, which can be readily evaluated using
\Sherpa~\cite{Gleisberg:2008ta} and
\Comix~\cite{Gleisberg:2008fv}. Technically then, the reweighting of the
event is performed with the ratio of the Born-level evaluation of the
\HEJ-approximation in the infinite top-mass and the full Born-level
expression in the same limit, while the resummation is performed using the
full expressions developed in Section~\ref{sec:resummation} (top and bottom
included). The contribution from the matrix elements to the event
weights is then
\begin{equation}
  \label{eq:match_h3+j}
  |\mathcal{M}_{\text{HEJ}}^{m_t, m_b}|^2
  \frac{|\mathcal{M}_{\text{LO}}^{\text{eff}}|^2}{|\mathcal{M}_{\text{HEJ, LO}}^{\text{eff}}|^2},\,
\end{equation}
where the ``eff'' superscript refers to the limit of an infinite
top-quark mass. In this approximation, the interaction between the Higgs
boson and gluons is described by an operator of dimension five, so that
matrix elements exhibit unphysical scaling in the limit of large
momenta. Since we choose not to include finite top-mass corrections in
the truncated \HEJ matrix element this effect cancels out in the ratio
in eq.~\eqref{eq:match_h3+j}.

The emission of quarks and gluons should resolve the dependence on the loop
momenta only for large energies of the emission (compared to $m_t$). Since
the bulk of each jet multiplicity consists of jet transverse momenta close to
the defined jet threshold, the quark-mass effects should have only a small
effect on the inclusive cross section. The quality of the approximation can
be checked by applying a similar strategy of reweighting in $pp\to H+2j$,
where the full result is known and can be checked against.
\begin{figure}
  \centering
  \begin{subfigure}[t]{0.495\textwidth}
    \includegraphics[width=\linewidth]{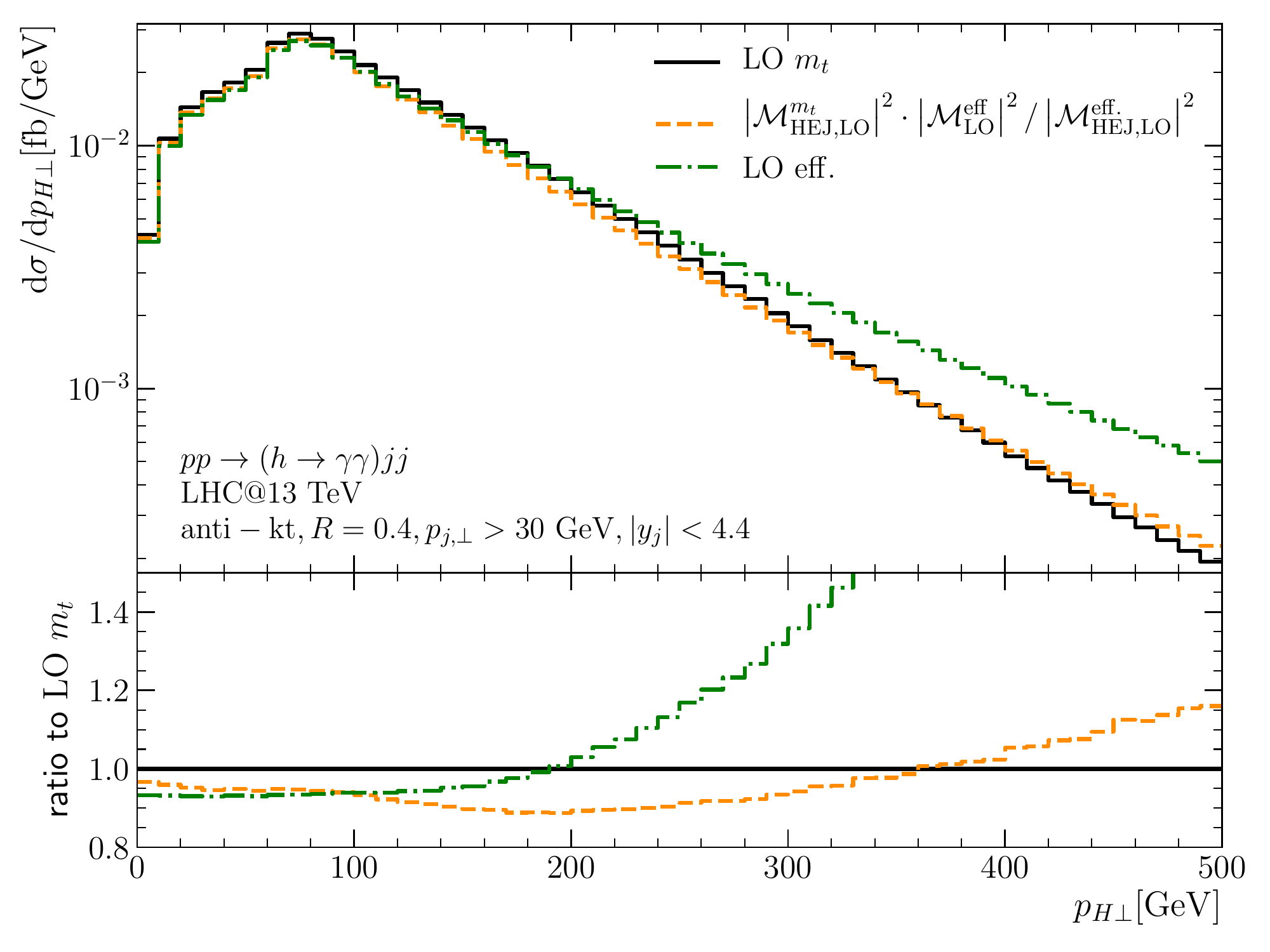}
    \caption{}
  \end{subfigure}
  \begin{subfigure}[t]{0.495\textwidth}
    \includegraphics[width=\linewidth]{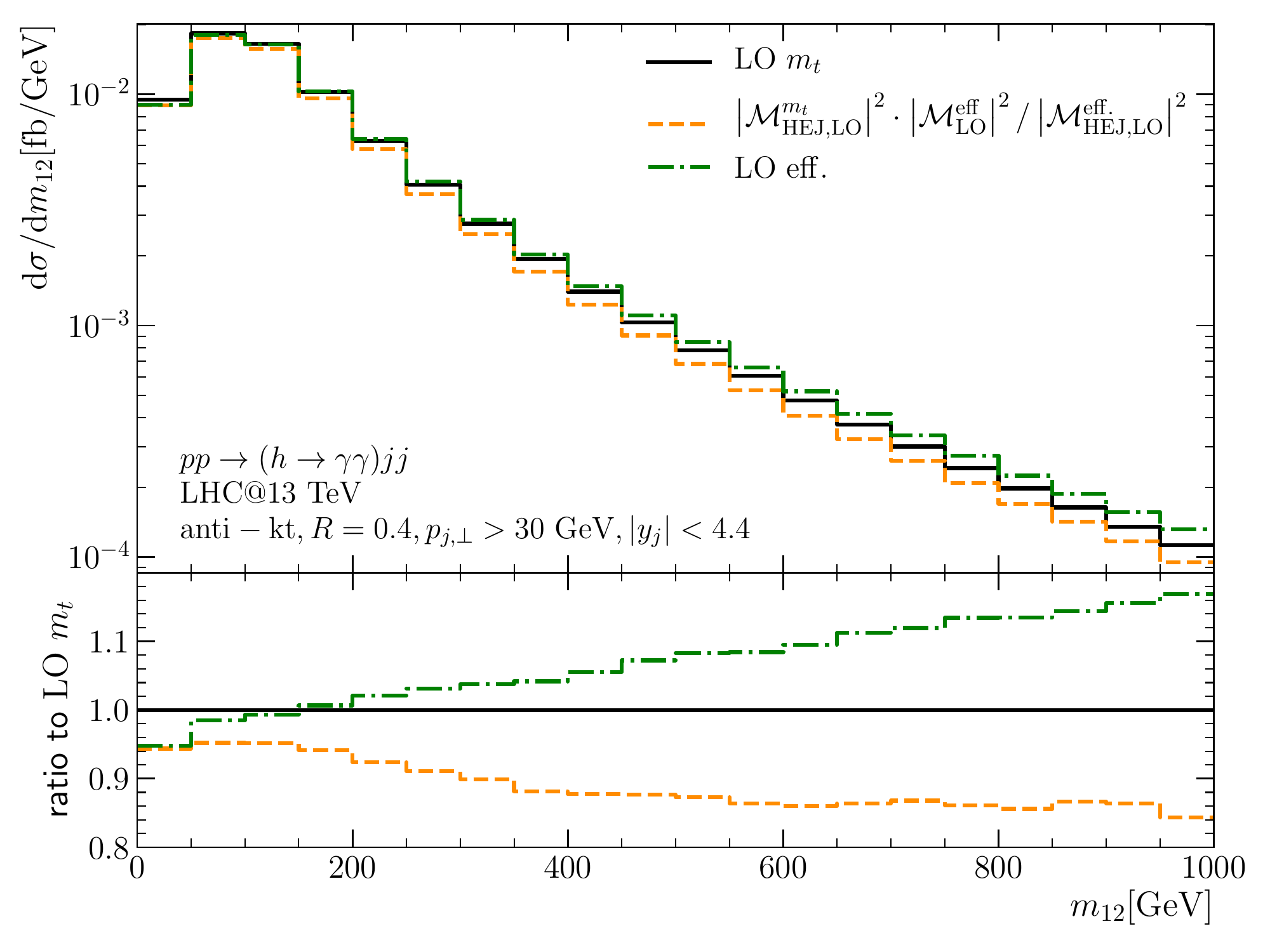}
    \caption{}
  \end{subfigure}
  \begin{subfigure}[t]{0.495\textwidth}
    \includegraphics[width=\linewidth]{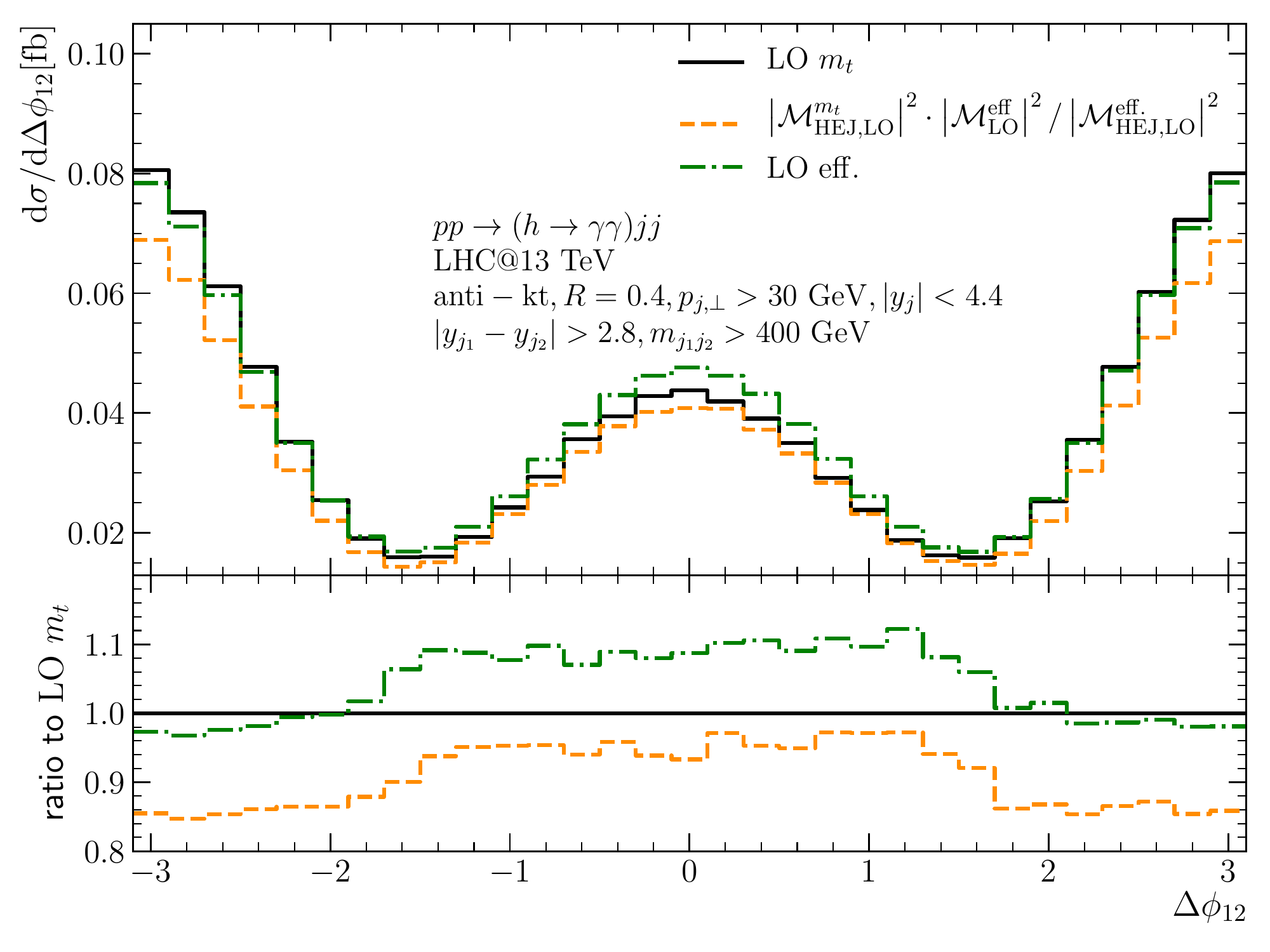}
    \caption{}
  \end{subfigure}
  \begin{subfigure}[t]{0.495\textwidth}
    \includegraphics[width=\linewidth]{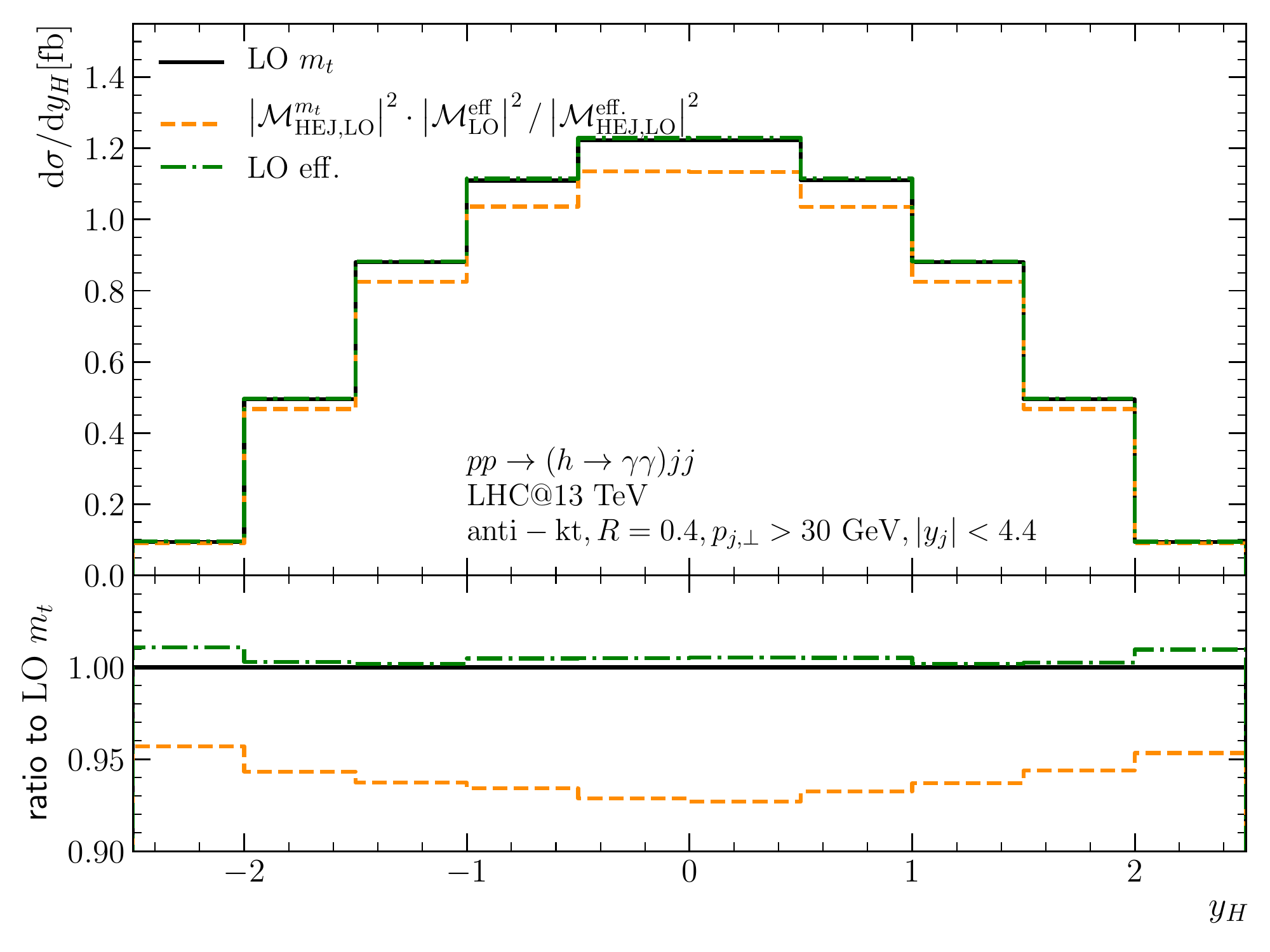}
    \caption{}
  \end{subfigure}
  \caption{In these plots we check the quality of reweighting using a ratio of
    matrix elements with infinite quark mass, using $pp\to H+2j$ where the full
    result is known.  The black solid line shows the LO result with full top
    mass dependence and the green dot-dashed line shows the LO result in the
    effective theory. The orange dashed line shows the HEJ result truncated at
    Born level with full top dependence, reweighted with the ratio of LO to HEJ
  results in the infinite top-mass limit.   The deviation between the orange and
black lines arises from the infinite top-mass limit in the reweighting factor.}
  \label{fig:h2j}
\end{figure}
The result of starting with the \HEJ-approximation of the matrix element
truncated at Born level with full top-mass dependence, multiplied by the
ratio of the full Born-level result to the \HEJ approximation, both evaluated
in the infinite top-mass limit, is shown in figure~\ref{fig:h2j}, and
compared to the Born-level result with both the full top-mass dependence and
in the effective theory of infinite top-mass.  The result obtained from
$m_t\to\infty$ undershoots the full result by 5\% for transverse momenta of
the Higgs boson up to $p_{H\perp}\approx m_t$, at which point it diverges
from, and overestimates, the correct cross section. The net result is that for the tree-level results,
the infinite top-mass limit gives a good approximation to the integrated
cross section obtained with the full top-mass dependence, as observed
in~\cite{DelDuca:2003ba,Greiner:2016awe}, even if it is clear that the
agreement in cross sections is accidental and will depend on the transverse cuts
used.

While the corrections are relatively small and uniform for the differential
cross section with respect to the azimuthal angle between the two
jets and the rapidity of the Higgs boson, there are systematically increasing
corrections to the distribution with respect to the invariant mass between
the two jets, growing to more than 10\% for $m_{j_1 j_2}>700$~GeV.

We now turn the attention to studying the level of approximation to the full
mass-dependent tree-level result by using the full mass-dependence in the
\HEJ-approximation to the full tree-level result followed by matching of the
matrix elements in the infinite top-mass limit (as we have to do for 3, 4 and 5
jets). The results are checked in the case of just two jets. If the reweighting
factor was also evaluated with finite quark masses, the black and orange lines
would be identical, and therefore the difference gives a measure of the quality
of the approximation.  We see that for
rapidity-distributions, exemplified by that of the Higgs-boson, the level of
accuracy obtained is roughly 5\%. The accuracy is better than 12\% in the
distribution of the azimuthal angle between the two jets, and similar for the
invariant mass between the jets, although here, and for the transverse
mass-distribution of the Higgs boson, the corrections increase with increasing
scale.

We emphasise that the seemingly good agreement in the distribution of $y_H$
between the results using the infinite top-mass and full top-mass dependence
is completely accidental, and that the results presented with the dashed
(orange) line and obtained using Eq.~\eqref{eq:match_h3+j} are more
accurate.

We conclude that by using finite quark-masses in the simplified \HEJ
amplitudes, and applying matching in the infinite top-mass limits we can
expect the result with finite top (and bottom) quark mass to be well
approximated for distributions even of 3, 4 and 5 jets.

\subsection{Matching of Leading-Order Results to NLO
  in the Infinite Quark Mass Limit}
\label{sec:match-lead-order}
\begin{figure}
\centering
  \includegraphics[width=0.45\linewidth]{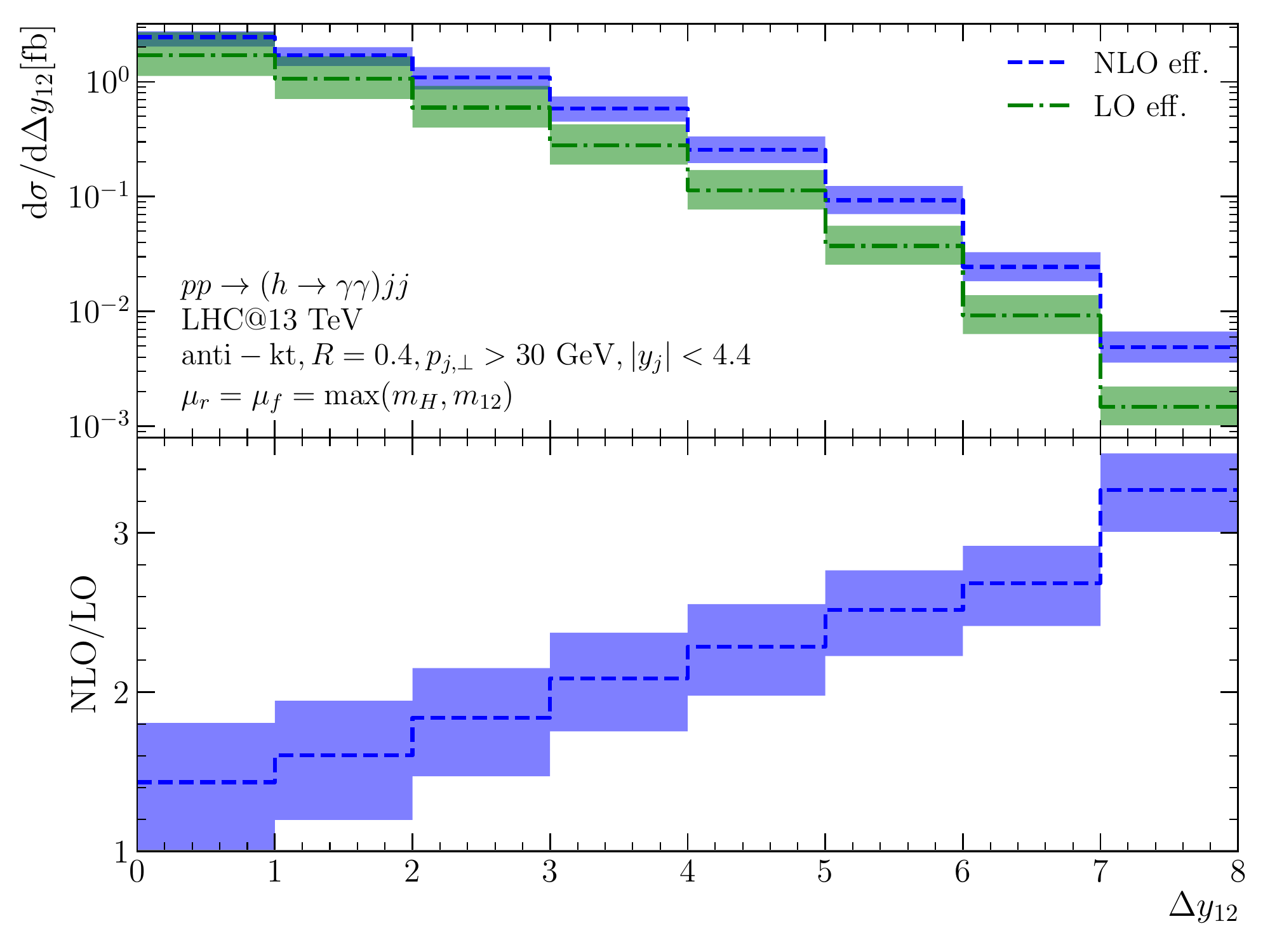}
  \includegraphics[width=0.45\linewidth]{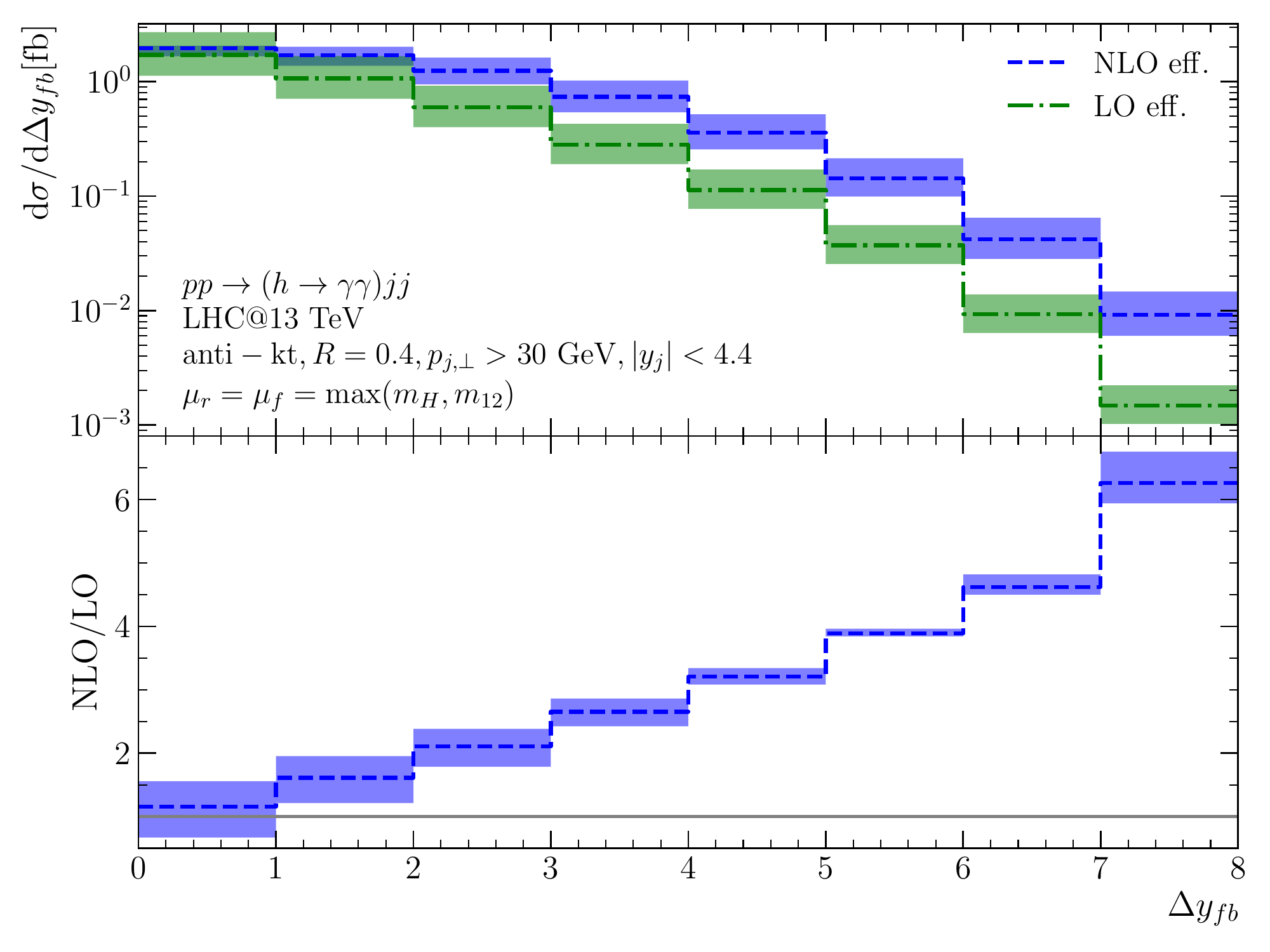}
  \caption{The distribution of the rapidity separation between the hardest jets (left) and the most forward
    and backward jets (right) of Higgs-plus-dijet production for \LO (green, dot-dashed)
    and \NLO (blue, dashed) both in the infinite top mass limit. The bottom
    panel shows the \NLO $K$-factor in each case. The results are obtained with
    $\scalemjj$.}
\label{fig:NLOK-factor}
\end{figure}

The results of the resummation and matching procedure described so far will
be compared to the best possible fixed-order result we can obtain. This
consists of Born-level for full top-mass (but not including the small effect of the
bottom-mass), reweighted bin-by-bin by the differential \NLO $K$-factor calculated
for infinite top-mass. The \LO and \NLO results for the distribution of (left) the
rapidity separation of the hardest two jets and (right) the
maximum rapidity-difference between any two hard jets, $\Delta y_{fb}$ in $pp\to
H+2j$ with infinite top-mass are shown in figure~\ref{fig:NLOK-factor}.
The \NLO $K$-factor is particularly interesting: it has a linear growth in
both cases and is large at large $\Delta y$.  Where it is plotted as function
of $\Delta y_{fb}$, it goes to 1 for $\Delta y_{fb}=0$.  This represents a
region of phase space dominated by exclusive 2-jet events.  For the rapidity
separation of the hardest two jets the $K$-factor reaches a factor of 3 for
rapidity differences of $\Delta y_{12}=8$, and for the most forward/backward
jets reaches a factor of 6 at $\Delta y_{fb}=8$. This obviously brings into
question the validity of \NLO-calculations at such rapidity-differences. The
source of the apparent perturbative instability in the fixed-order result is
treated systematically within \HEJ.  It is here worth mentioning that in the
MRK-limit the all-order \HEJ-results for $pp\to H+2j$ and $pp\to H+3j$ will
contain the same effect from the virtual corrections (and soft emissions) of
a suppressing factor $\propto \exp(\omega(k_\perp^2) \Delta
y_{fb})\propto(\log(\hat s/p_t^2))^{\omega(k_\perp^2)}$ with
$\omega(k_\perp^2)<0$. However, when the perturbative series is terminated at
\NLO-accuracy, the effect of the expansion of the exponential suppression is
included only in the events with Born-level kinematics. The suppression is
missing at \NLO in the corrections from real emissions because of the
fixed-order termination of the perturbative series. At large rapidity-spans
$\Delta y_{fb}$, this will inflate the \NLO-result compared to the all-order
result of \HEJ, irrespectively of the choice of renormalisation and
factorisation scale.

\begin{figure}
\centering
  \includegraphics[width=0.45\linewidth]{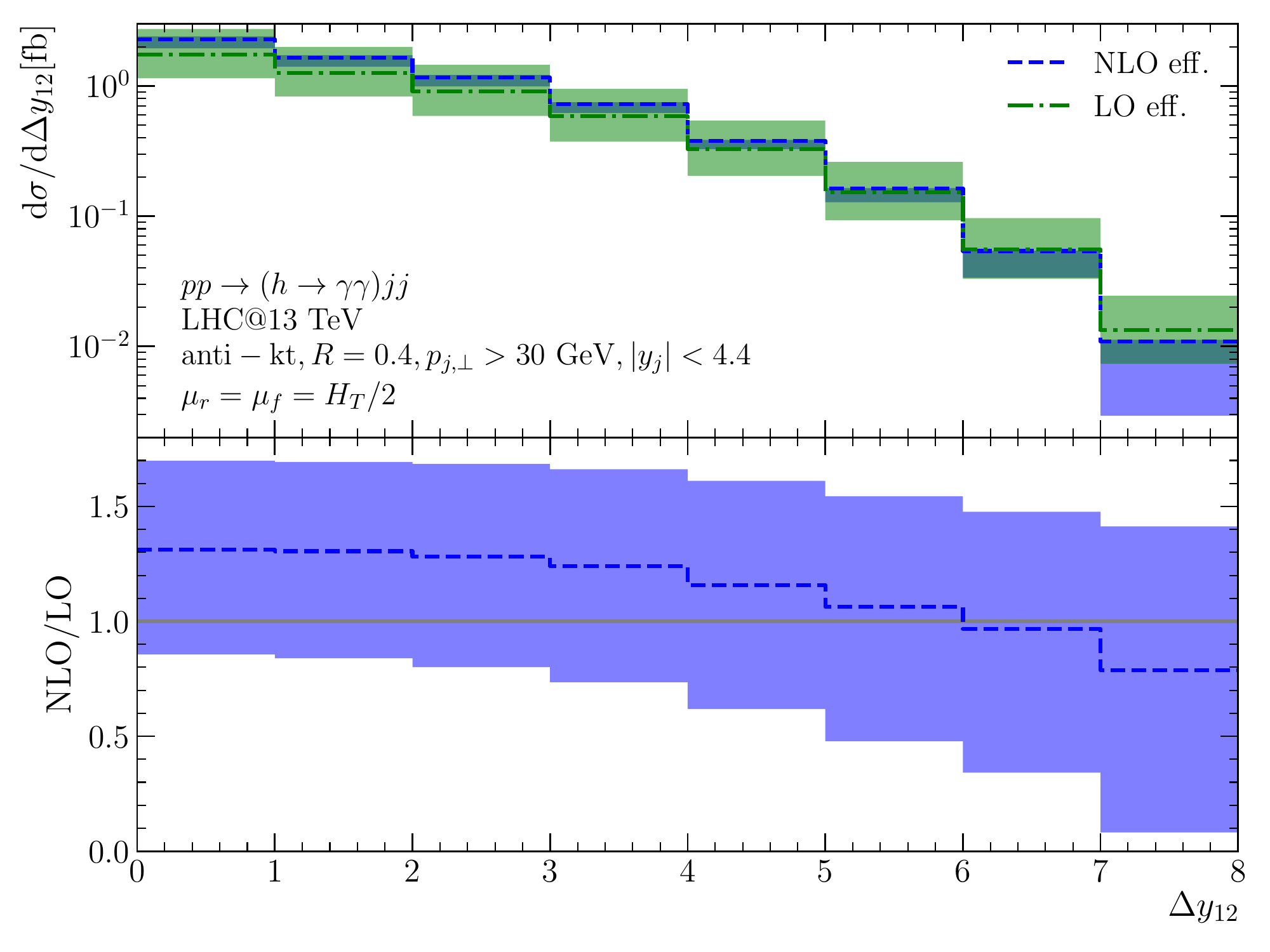}
  \includegraphics[width=0.45\linewidth]{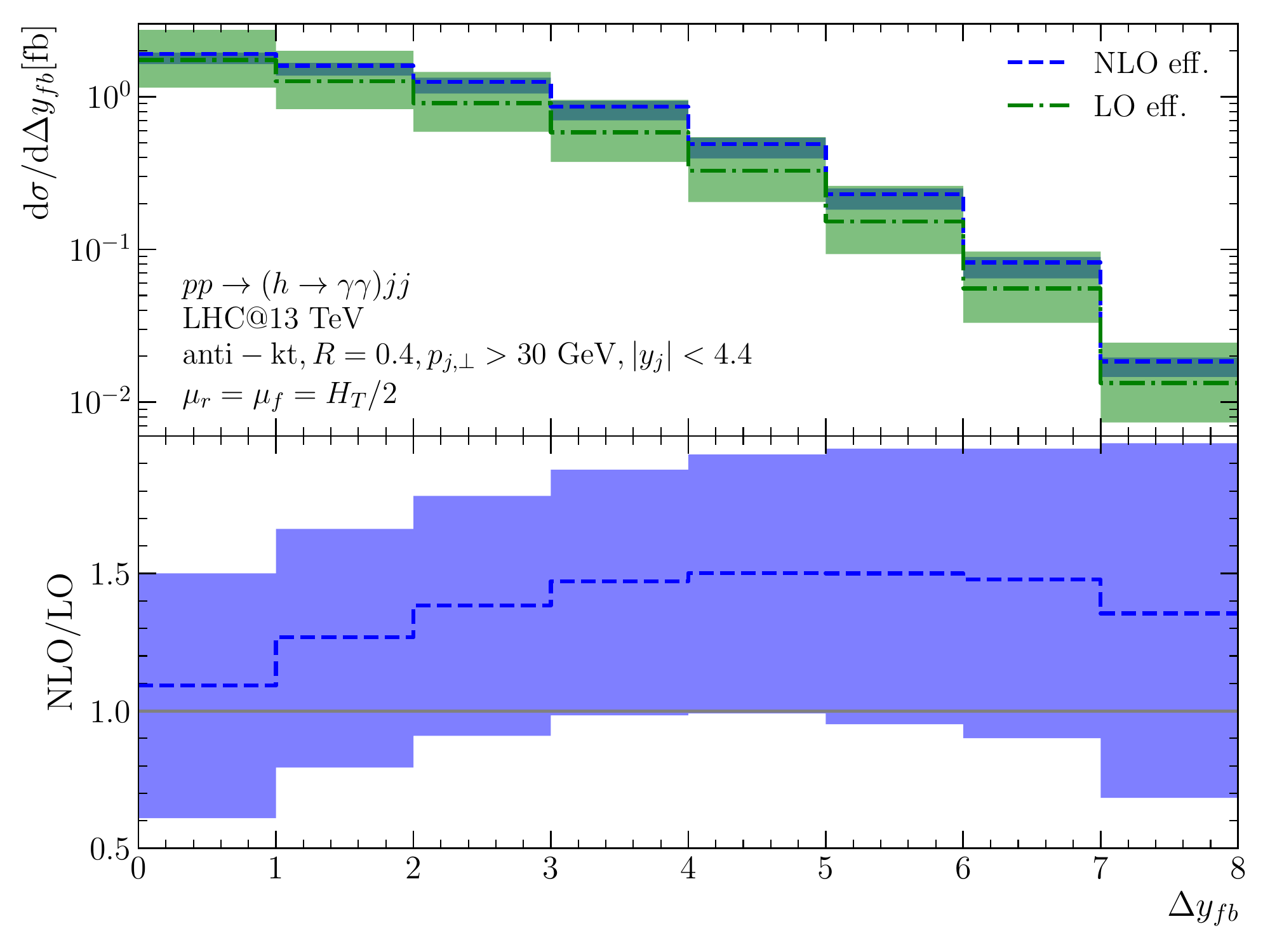}
  \caption{The distribution of the rapidity separation between the hardest jets
    (left) and the most forward and backward jets (right) of Higgs-plus-dijet
    production for \LO (green, dot-dashed) and \NLO (blue, dashed) both in the
    infinite top mass limit. The bottom panel shows the \NLO $K$-factor in each
    case. The results are obtained with $\scaleht$, and should be compared to
    those on  figure~\ref{fig:NLOK-factor}.}
\label{fig:HT2:NLOK-factor}
\end{figure}

The balance between a suppression for $H+2j$ at NLO at large $\Delta y$ of
the two-parton contribution and enhancement of the three-parton contribution
discussed above is obviously influenced by the value of $\alpha_s$ and
therefore the scale choices. Indeed, the effect of choosing instead a central
scale choice of $\scaleht$ is illustrated in
figure~\ref{fig:HT2:NLOK-factor}. As seen in the right plot,
the $K$-factor tends to unity for
$\Delta y_{fb}\to 0$, rises to 1.5 at $\Delta y_{fb}=4$, stabilises and then
starts decreasing at $\Delta y_{fb}\sim 7$. As a function of $\Delta y_{12}$,
the $K$-factor starts at 1.3 for $\Delta y_{12}=0$, and then decreases to 0.7
at $\Delta y_{12}=8$. The smaller $K$-factors observed for the central scale
choice of $\scaleht$ may seem more appealing than the behaviour
observed in figure~\ref{fig:NLOK-factor}; however, the variation obtained
around this central scale will certainly underestimate the uncertainty from
uncalculated higher orders, since the central scale choice leads to a result
close to the edge of the results obtained by the variation. Furthermore, the
scale variation band for \NLO in figure~\ref{fig:HT2:NLOK-factor}~(left)
increases with $\Delta y_{12}$, reaching $-70\%$ in the last bin, above $\Delta
y_{12}>7$. This indicates an instability of the \NLO calculations for
$\scaleht$ at large rapidity differences. All the results presented in the
following with $\scalemjj$ are also presented in appendix~\ref{sec:HT2} for
$\scaleht$. Just as for fixed-order predictions, other processes like $W$+jets
could be used in order to verify which of the scale choices obtains the best
description of data.

%%% Local Variables:
%%% mode: latex
%%% TeX-master: "main"
%%% End:

\section{Results of Finite Quark Masses and All-Order Resummation}
\label{sec:results}
This section will first present results for a separate investigation of the
higher-order effects included with HEJ compared to the fixed-order
approaches. As mentioned before, we employ Sherpa in combination with
OpenLoops to obtain the fixed-order predictions. To evaluate the finite
quark mass corrections within \HEJ, we make use of
QCDLoop~\cite{Carrazza:2016gav}. We adopt our input parameters and cuts
from~\cite{Andersen:2017kfc,Andersen:2018tnm}, following the
experimental analysis of~\cite{Aad:2014lwa}.
Explicitly, we consider the gluon-fusion-induced production of a Higgs
boson decaying into two photons, together with at least two anti-$k_t$
jets with transverse momenta $p_{\perp, j} > 30\,$GeV, rapidities $|y_j|
< 4.4$, and radii $R = 0.4$ at the $13\,$TeV LHC. For the photons, we require
\begin{align}
  \label{eq:photon_cuts}
  |y_{\gamma}| &< 2.37, \qquad 105~\textrm{GeV} < m_{\gamma_1 \gamma_2} < 160~\textrm{GeV},\notag\\
  p_{\perp, \gamma_1} &> 0.35\,m_{\gamma_1 \gamma_2}, \qquad p_{\perp, \gamma_2} > 0.25\,m_{\gamma_1 \gamma_2},
\end{align}
and separations $\Delta R(\gamma, j), \Delta R(\gamma_1, \gamma_2) >
0.4$ from the jets and each other. The Higgs-boson mass is set to
$m_H=125\,$GeV, its width to $\Gamma_H = 4.165\,$MeV and the branching
fraction for the decay into two photons to $0.236\%$. We use the CT14nlo
PDF set~\cite{Dulat:2015mca} as provided by
LHAPDF6~\cite{Buckley:2014ana}. The results presented here are obtained with
the central scale choice of $\scalemjj$; all figures in
this section are reproduced in appendix~\ref{sec:HT2} for the central scale
choice of $\scaleht$.

In addition to inclusive quantities with the basic cuts listed above, we
also consider additional VBF-selection cuts applied to the hardest jets as
in~\cite{Aad:2014lwa}:
\begin{align}
  \label{eq:vbfcuts}
  |y_{j_1}-y_{j_2}| > 2.8, \qquad m_{j_1 j_2} > 400~\textrm{GeV}.
\end{align}

A discussion of the values chosen for the quark masses is in order. In the
gluon-fusion production of a Higgs boson together with light-flavour jets the
heavy quarks only appear in internal loops and are off shell. We therefore do
not use on-shell masses, but instead prefer the $\overline{\text{MS}}$
mass-scheme. The scale $\mu_m$ associated with the $\overline{\text{MS}}$
mass is a priori independent of the renormalisation scale used for the
running coupling. It should be set to a scale characteristic for the heavy
quark loop.

For the bottom quark, the mass is negligible compared to all other
scales in the loop. Since the observables considered in this work depend only
mildly  on the bottom-quark mass, the exact scale choice has
little impact on the prediction. To be definite, we use $\mu_{m_b} =
m_H$ and $m_b(m_H) =
2.8\,$GeV, which can be obtained from input values of $m_b(m_b) =
4.18\,$GeV~\cite{Tanabashi:2018oca}, $\alpha_s(m_Z) = 0.118$ via
renormalisation group evolution at two loops. The effect of higher
orders in the evolution is negligible.

The effect of the top-quark mass is much more important. While there are
ongoing
efforts~\cite{Hoang:2008xm,Kieseler:2015jzh,Butenschoen:2016lpz,Hoang:2017kmk,Nason:2017cxd,Hoang:2018zrp}
to relate the very precise values reported by the LHC and Tevatron
experiments~\cite{Khachatryan:2015hba,Aaboud:2016igd,TevatronElectroweakWorkingGroup:2016lid}
to a well-defined short-distance scheme, the top-quark
$\overline{\text{MS}}$ mass is not known very precisely at the moment. For
this project, the values chosen are $\mu_{m_t} = m_H$ with $m_t(m_H) = 163\,$GeV, in line with direct
determinations of the $\overline{\text{MS}}$
mass~\cite{Langenfeld:2009wd,Abazov:2011pta} and compatible with a pole
mass of $173\,$GeV~\cite{Marquard:2015qpa} within the uncertainties
quoted in~\cite{Abazov:2011pta}.

%%% Local Variables:
%%% mode: latex
%%% TeX-master: "main"
%%% End:

Since the fixed-order setup can take into account the effects only of the
top-quark, all results in section~\ref{sec:effects-high-pert} are for finite
top-mass only (no effects from the bottom-quark
included). Section~\ref{sec:effects-fin-top-mass} investigates the effects on
the results of \HEJ from the finite top and bottom mass compared to the
results obtained for infinite top-mass. Finally,
section~\ref{sec:final-results-hej} compares the most precise predictions
from \HEJ, including both top and bottom mass effects, and the matching to
fixed order discussed in section~\ref{sec:matching}, to that of the
fixed-order finite top-mass results, scaled to NLO accuracy, as described in
section~\ref{sec:match-lead-order}.

\subsection{Effects of Higher Perturbative Orders}
\label{sec:effects-high-pert}
\begin{figure}
  \centering
  \begin{subfigure}[t]{0.495\textwidth}
    \includegraphics[width=\linewidth]{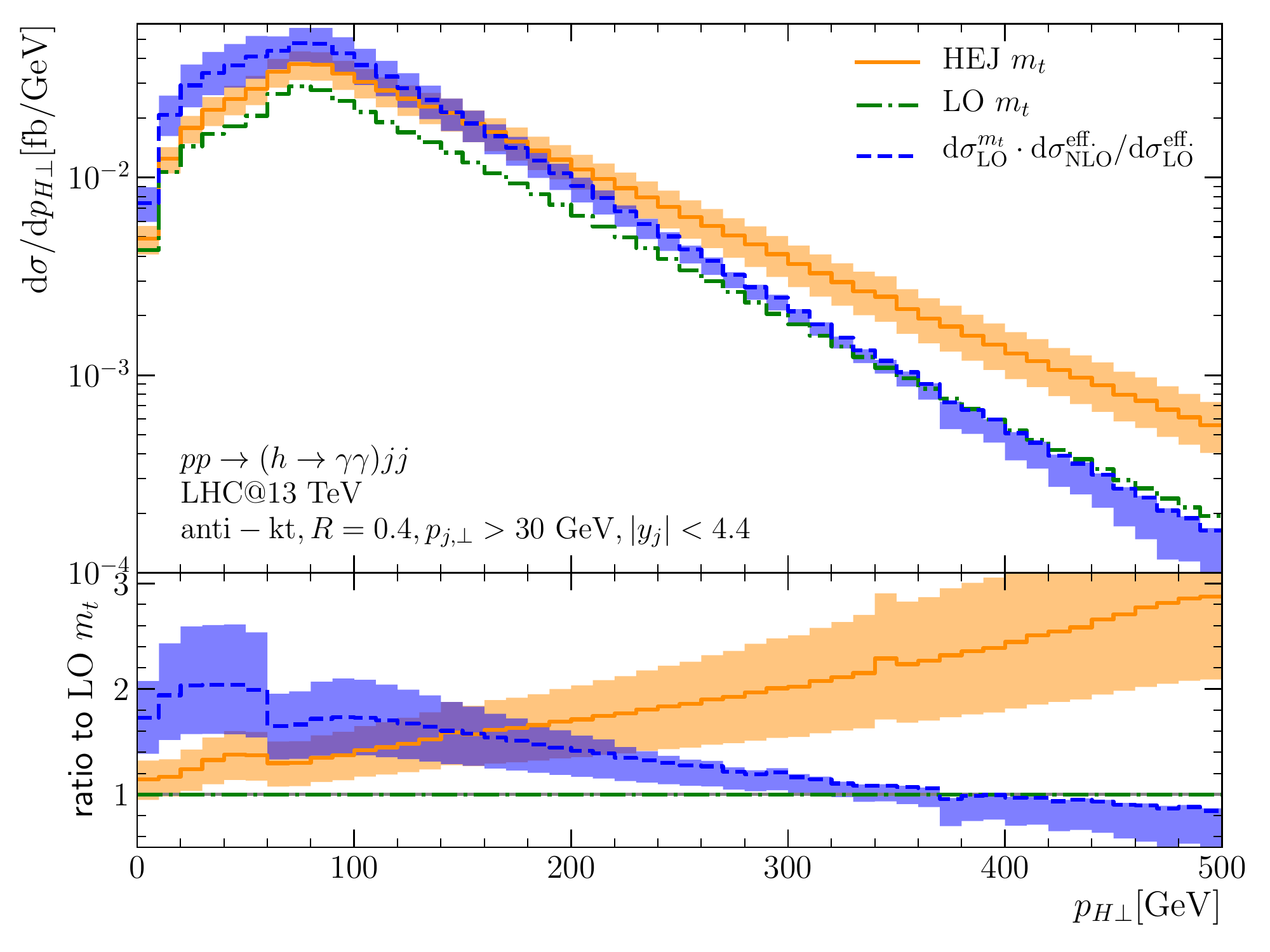}
    \caption{}
    \label{fig:ResultsHigherOrdera}
  \end{subfigure}
  \begin{subfigure}[t]{0.495\textwidth}
    \includegraphics[width=\linewidth]{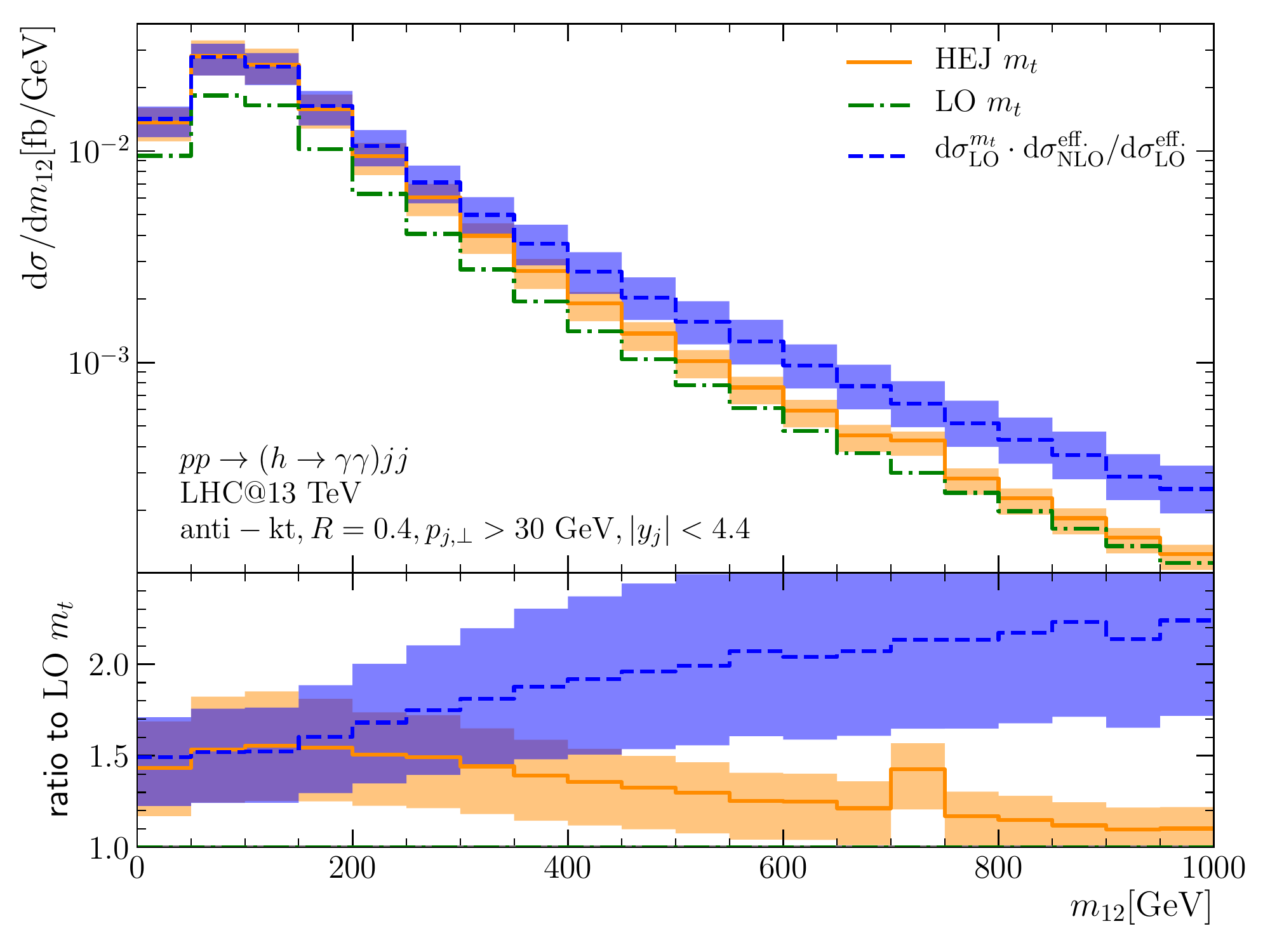}
    \caption{}
    \label{fig:ResultsHigherOrderb}
  \end{subfigure}
  \begin{subfigure}[t]{0.495\textwidth}
    \includegraphics[width=\linewidth]{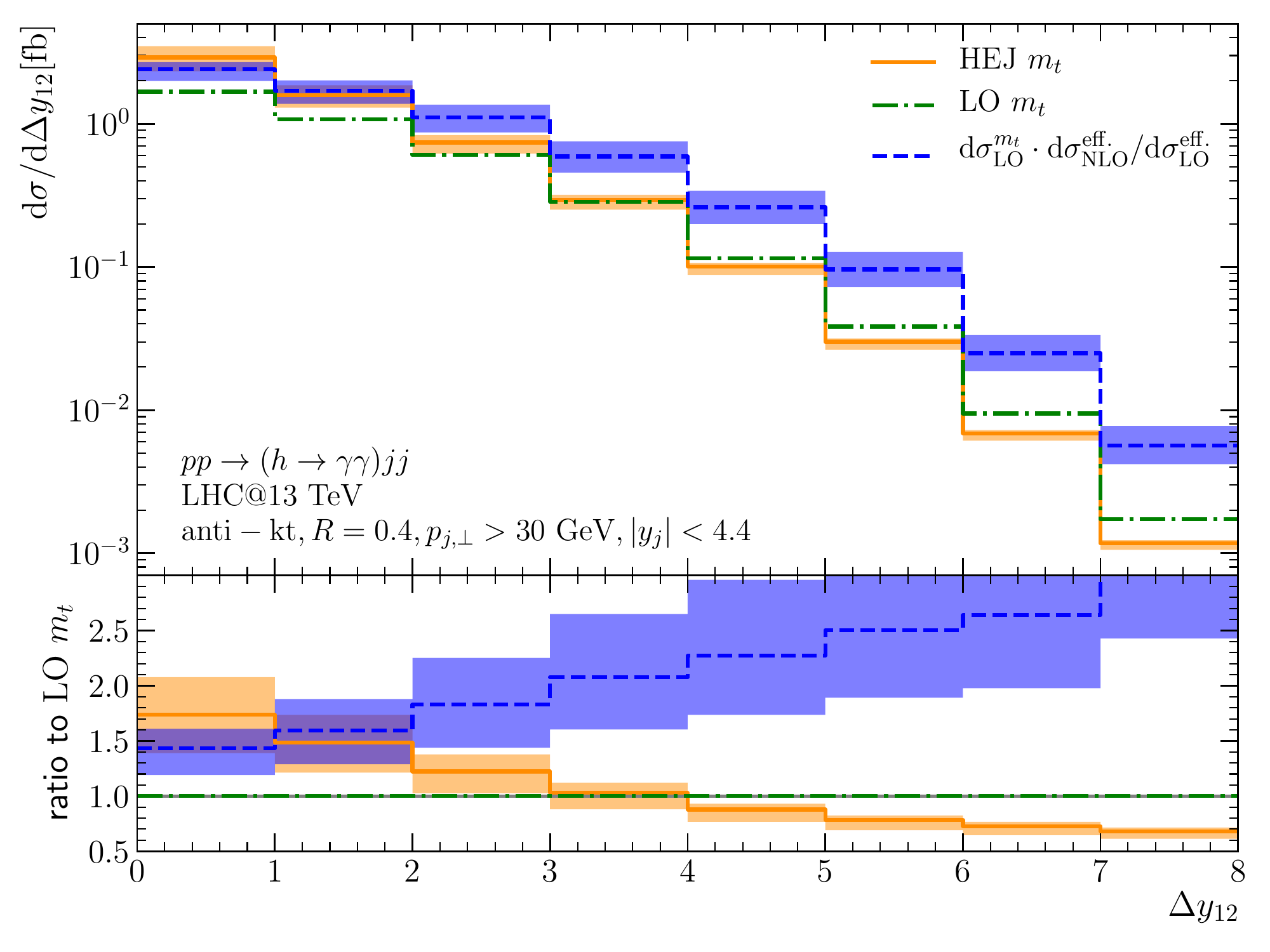}
    \caption{}
    \label{fig:ResultsHigherOrderc}
  \end{subfigure}
  \begin{subfigure}[t]{0.495\textwidth}
    \includegraphics[width=\linewidth]{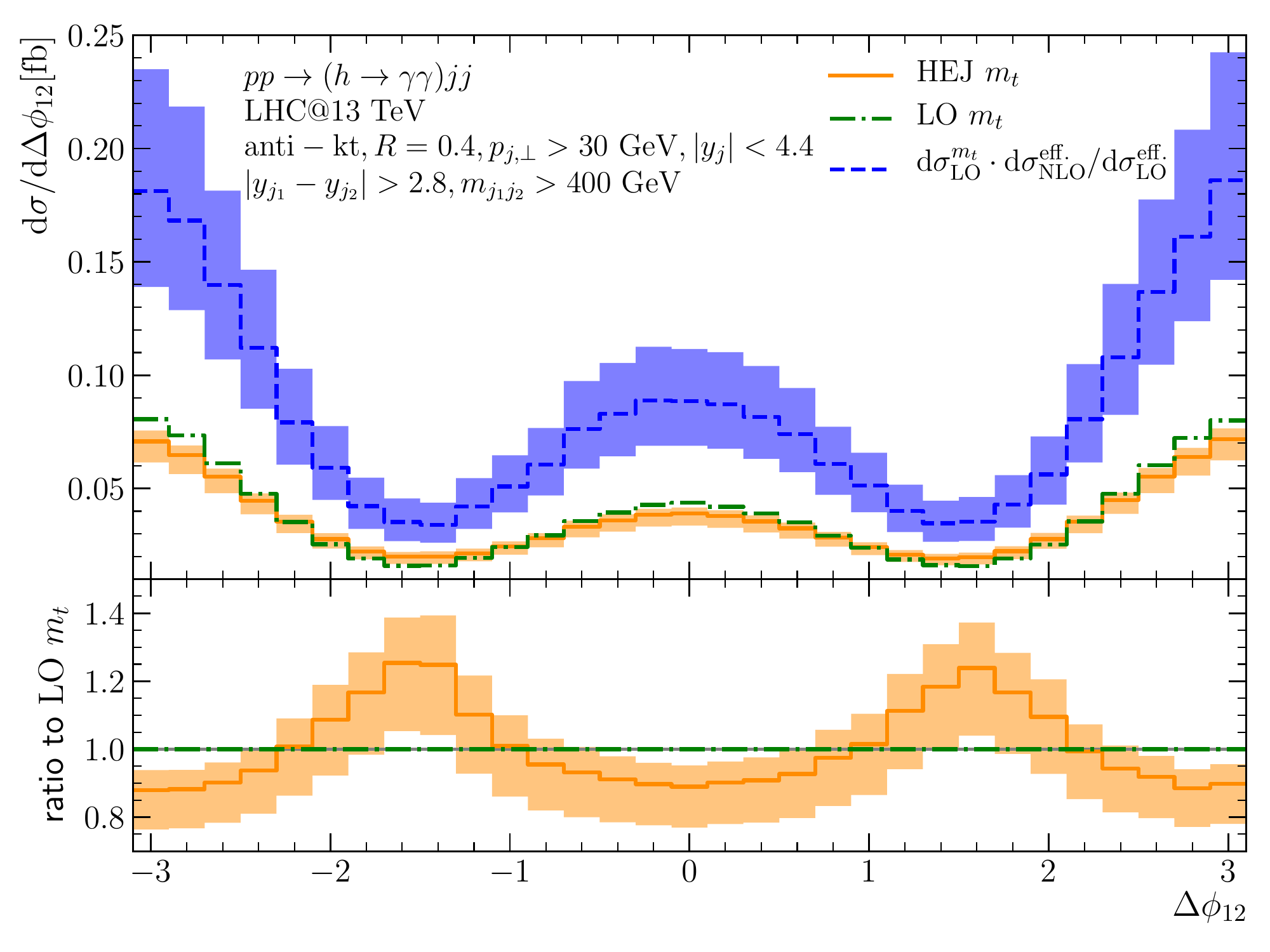}
    \caption{}
    \label{fig:ResultsHigherOrderd}
  \end{subfigure}
  \caption{The results obtained from \HEJ (orange, solid) and \LO (green,
    dot-dashed) both with full top mass. Additionally in blue, dashed is the
    result of scaling the \LO full $m_t$-result bin-by-bin with the \NLO
    $K$-factor in the
    $m_t\to\infty$ limit.  The $K$-factors and their impact
    within the VBF cuts (applied in (d)) are discussed in the text.}
  \label{fig:ResultsHigherOrder}
\end{figure}

Figure~\ref{fig:ResultsHigherOrder} compares the results obtained with finite
top-mass at \LO, the \LO rescaled to \NLO accuracy in the limit of infinite
top-mass, and in all-order \HEJ (using just finite top-mass but no contribution from
the bottom quark). Comparing
the results of $p_{H\perp}$ in figure~\ref{fig:ResultsHigherOrdera} for LO and the rescaling
using the bin-by-bin $K$-factor calculated in the limit of infinite top-mass,
one sees that the \NLO $K$-factor (the ratio between the lines in blue and in
green, indicated by the blue band in the lower plots) varies locally between
0.8 and 2 within ranges of the distributions checked. The \NLO $K$-factor is
decreasing for increasing transverse momentum $p_{H\perp}$, crossing unity at
$p_{H\perp}=340$~GeV.

The \NLO $K$-factors for the distributions in the invariant mass between the
two hardest jets $m_{12}$ (figure~\ref{fig:ResultsHigherOrderb}) and the
rapidity-difference between the two hardest jets
(figure~\ref{fig:ResultsHigherOrderc}) have the same systematic behaviour of
increasing $K$-factor as observed for $\Delta y_{fb}$ in
figure~\ref{fig:NLOK-factor} and discussed there.  The \NLO $K$-factor for $m_{12}$
increases from 1.5 to 2.2 at $m_{12}=1$~TeV, and for $\Delta y_{12}$
the \NLO $K$-factor increases in a straight line from 1.5 to 3 at
$\Delta y_{12}=8$.  This obviously then induces a large $K$-factor when
a large rapidity-separation and invariant mass is required in the VBF-cuts,
as illustrated for the $\phi_{12}$ within these plots seen in
figure~\ref{fig:ResultsHigherOrderd}.  It can also be seen in
figure~\ref{fig:ResultsHigherOrderc} that the ratio between \HEJ and \LO
decreases linearly as a function of $\Delta y_{12}$; this is an illustration of
the logarithmic suppression of events with exactly two jets where $\Delta y =
\log(s/t)$ for large $s$.

The fixed-order matching bin-by-bin (as opposed to phase-space point by
phase-space point employed with \HEJ) does not ensure the same value for the
integrated cross section. The effect of the matching will depend on the
binning width etc. The size of the variation in the cross sections from the
various distributions is one measure of the residual room for improvement in
the matching.
\begin{table}[htbp]
  \centering
  \setlength{\tabcolsep}{12pt}
  \begin{tabular}{lrrr}
    \toprule
    Distribution & $\LO(m_t)~[\text{fb}]$ & $\LO(m_t)*K_{\NLO}(m_t\to\infty)~[\text{fb}]$ & $\HEJ(m_t)~[\text{fb}]$\\
    \midrule
    Inclusive & $3.8^{+2.1}_{-1.3}$&$6.2^{+1.1}_{-1.2}$&
                $5.7^{+1.0}_{-1.1}$\\
    $p_{H\perp}$&&$6.3^{+1.2}_{-1.3}$&\\
    $m_{12}$&&$6.2^{+1.0}_{-1.1}$&\\
    $y_{12}$&&$6.2^{+1.1}_{-1.2}$&\\[1em]
    VBF & $0.24^{+0.12}_{-0.08}$& $0.53^{+0.15}_{-0.13}$& $0.23^{+0.02}_{-0.03}$\\
    VBF, $\phi_{12}$ &&$0.53^{+0.13}_{-0.10}$&\\
    \bottomrule
  \end{tabular}

  \caption{Cross sections obtained at \LO, LO scaled with bin-by-bin
    $K$-factor for various distributions, and the \HEJ with the inclusive
    cross sections scaled to \NLO}
  \label{tab:intcross}
\end{table}
The integrated cross sections obtained from various distributions using the
method of differential $K$-factors are listed in
Table~\ref{tab:intcross}. There is found to be very little variation in the
integrated cross section of just $0.1$~fb, well within the scale variation on
the \NLO-rescaled cross section of $6.2^{+1.1}_{-1.2}$~fb, and an overall
$K$-factor of 1.6. Within the VBF-cuts, the overall \NLO $K$-factor is 2.2,
and the \NLO-rescaled cross section is found to be
$0.53^{+0.15}_{-0.13}$~fb.

Table~\ref{tab:intcross} also contains the result for \HEJ, rescaling the
all-order results with finite top-quark mass with the ratio between the
results obtained for infinite top-quark mass at \NLO and at all orders with
\HEJ. Thus the matching is different to that applied at lowest order and
normalises to the \NLO cross section in the infinite top-quark mass
limit. The inclusive cross section for \HEJ matched as described is found to
be $5.7^{+1.0}_{-1.1}$fb, slightly lower than the \LO-result for finite
top-quark mass multiplied by the \NLO $K$-factor from the infinite top-quark
mass. While the results for the inclusive cross sections are similar at \NLO
and in \HEJ, the distributions differ significantly. As is evident from
figure~\ref{fig:ResultsHigherOrder}, the differential distribution from \HEJ
is harder in $p_{H\perp}$ compared to the scaled \LO-result, while the
spectrum is decreasing significantly faster for both $m_{12}$ and
$\Delta y_{12}$. This means that even though the total cross section for
\HEJ is matched to \NLO (in the infinite top-mass limit) with a scale-dependent $K$-factor of $1.4^{+0.4}_{-0.4}$, within the VBF-cuts the result of
$0.23^{+0.02}_{-0.03}$~fb happens to be closer (but with a reduced scale
dependence) to the \LO cross section of $0.24^{+0.12}_{-0.08}$ . It is just a
numerical coincidence of the cuts applied that the cross sections agree.  As
seen already in the discussion of the \NLO corrections, the perturbative
corrections are large in the VBF region. There is no reason to believe the
perturbative series has converged already at \NLO.

\subsection{Effects of the Finite Top Mass}
\label{sec:effects-fin-top-mass}
The impact of the full top-quark mass-dependence on the Born-level result for
$pp\to H+2j$ was already investigated in figure~\ref{fig:h2j}. While the
effect on the integrated cross section is very small, the effect on the
differential distribution in $p_{\perp H}$ is enormous. The infinite top-mass
approximation undershoots the full-top mass result by 5\% for $p_{\perp H}$ up
to 200~GeV and then increasingly overshoots for
increasing transverse momentum, reaching 40\% error already at
$p_{\perp H}=340$~GeV. Similarly, for the invariant mass between the two
hardest jets, the distribution for the infinite top-mass result starts off
undershooting the true result by 5\%, crossing at $m_{12}=150$~GeV and
increasing to 16\% by $m_{12}=1$~TeV. The error due to the infinite top-quark mass
approximation is very small and uniform in the rapidity of the Higgs boson.

We now turn our attention to the impact of both the finite top-quark and
bottom-quark mass on the results of \HEJ. First, we list in
table~\ref{tab:crosssections} the result for the cross section with
inclusive- and the VBF-cuts for infinite top-quark mass and finite top-quark
mass for fixed order (\LO scaled with \NLO in the limit of infinite top-quark
mass). For \HEJ we also list the result using both finite masses for the
top-quark and bottom-quark.
\begin{table}
  \centering
  \begin{tabular}{lrrrr}
    \toprule
    &\multicolumn{2}{c}{Fixed Order}&\multicolumn{2}{c}{\HEJ}\\
    \cmidrule(r){2-3}\cmidrule(l){4-5}
    &Inclusive $H+2j$&VBF cuts&Inclusive $H+2j$&VBF cuts\\\midrule
    $m_t\to\infty$   & $6.2^{+1.1}_{-1.2}$ fb& $0.54^{+0.16}_{-0.12}$ fb& $6.2^{+1.1}_{-1.2}$ fb& $0.26^{+0.02}_{-0.04}$ fb\\[.5em]
    $m_t=163$ GeV     & $6.2^{+1.1}_{-1.2}$ fb &$0.53^{+0.15}_{-0.13}$ fb& $5.7^{+1.0}_{-1.1}$ fb& $0.23^{+0.02}_{-0.03}$ fb\\[.5em]

    $m_t=163$ GeV & \multirow{2}{*}{-} & \multirow{2}{*}{-}
    &\multirow{2}{*}{$5.7^{+1.0}_{-1.1}$ fb}&
                                              \multirow{2}{*}{$0.23^{+0.02}_{-0.03}$ fb}\\
    $m_b = 2.8$ GeV\\
\bottomrule
  \end{tabular}
  \caption{Cross sections obtained in fixed order perturbation theory
      (either full \NLO for the results using infinite top-quark mass or \LO
      scaled bin-by-bin with the $K$-factor obtained in the infinite
      top-quark mass limit) and in \HEJ for $pp\to H+2j$ with inclusive and
      VBF-cuts. See text for further comments.}
  \label{tab:crosssections}
\end{table}
The finite top-quark mass has a much larger impact on the results of \HEJ than at
fixed order, which might at first seem surprising, since the results of \HEJ
are matched to the fixed-order
results. The larger impact of the top-mass effects are therefore not a result
of the approximations in \HEJ. Instead, as is evident in the distributions of
figure~\ref{fig:ResultsHigherOrder}, the higher-order corrections of \HEJ emphasise the distribution at
larger $p_{H\perp}$, where the corrections from the finite quark-mass are
large. Therefore, the top-quark mass corrections of the \HEJ-results amount
to a 9\% reduction within the inclusive and 11\% within the VBF-cuts. We do
not observe any effect  of the non-zero bottom mass beyond 1\% for any of the observables
studied. The impact obviously increases, if the bottom mass is chosen larger~\cite{Greiner:2016awe}.
%much in line with $m_b/m_t$,
%the expected relative size of the interference term, and the results
%of~\cite{Greiner:2016awe}.

\begin{figure}
\centering
\begin{subfigure}[t]{0.495\textwidth}
  \includegraphics[width=\linewidth]{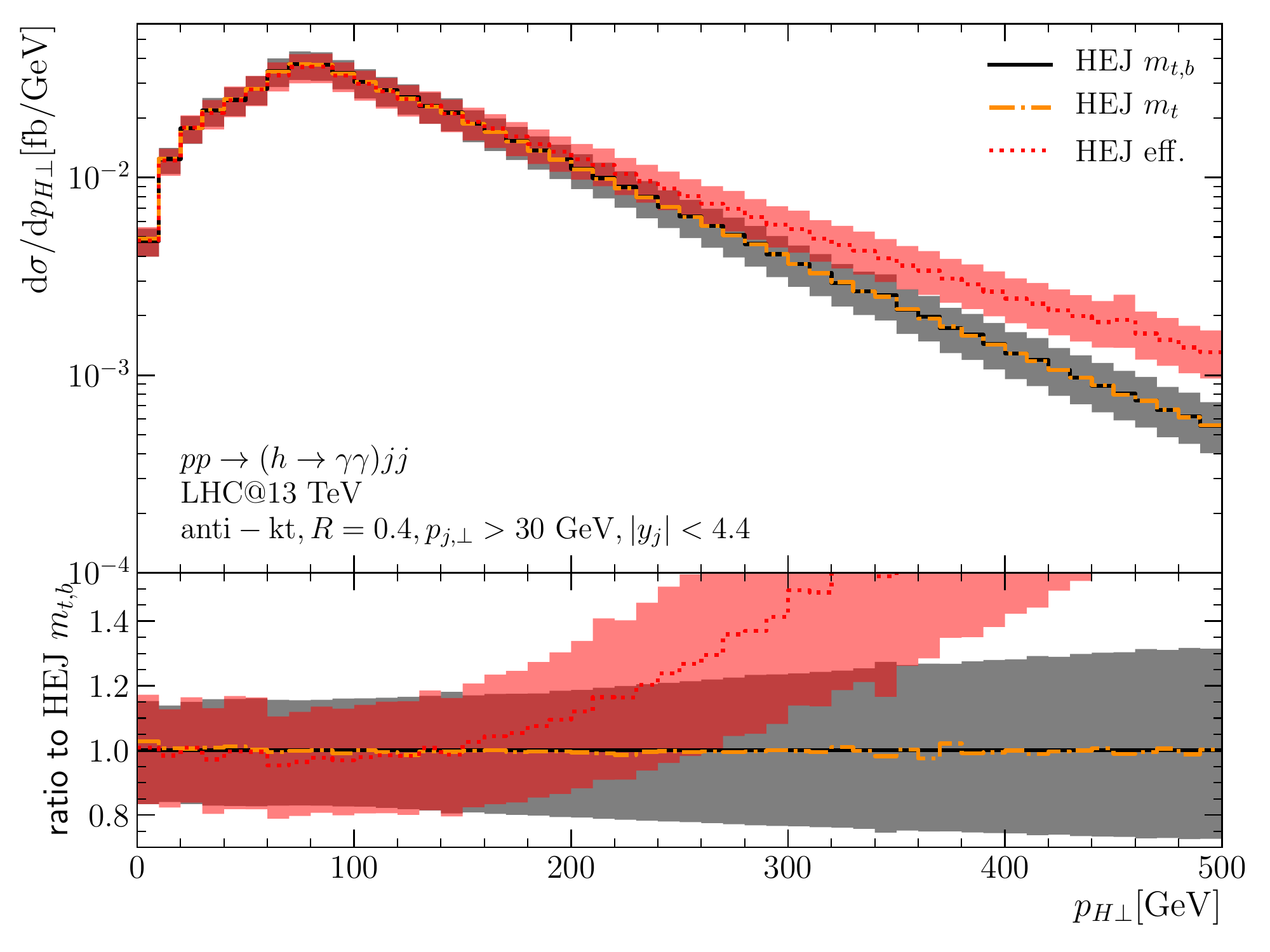}
  \caption{}
\label{fig:ResultsMassPHperp}
\end{subfigure}
\begin{subfigure}[t]{0.495\textwidth}
  \includegraphics[width=\linewidth]{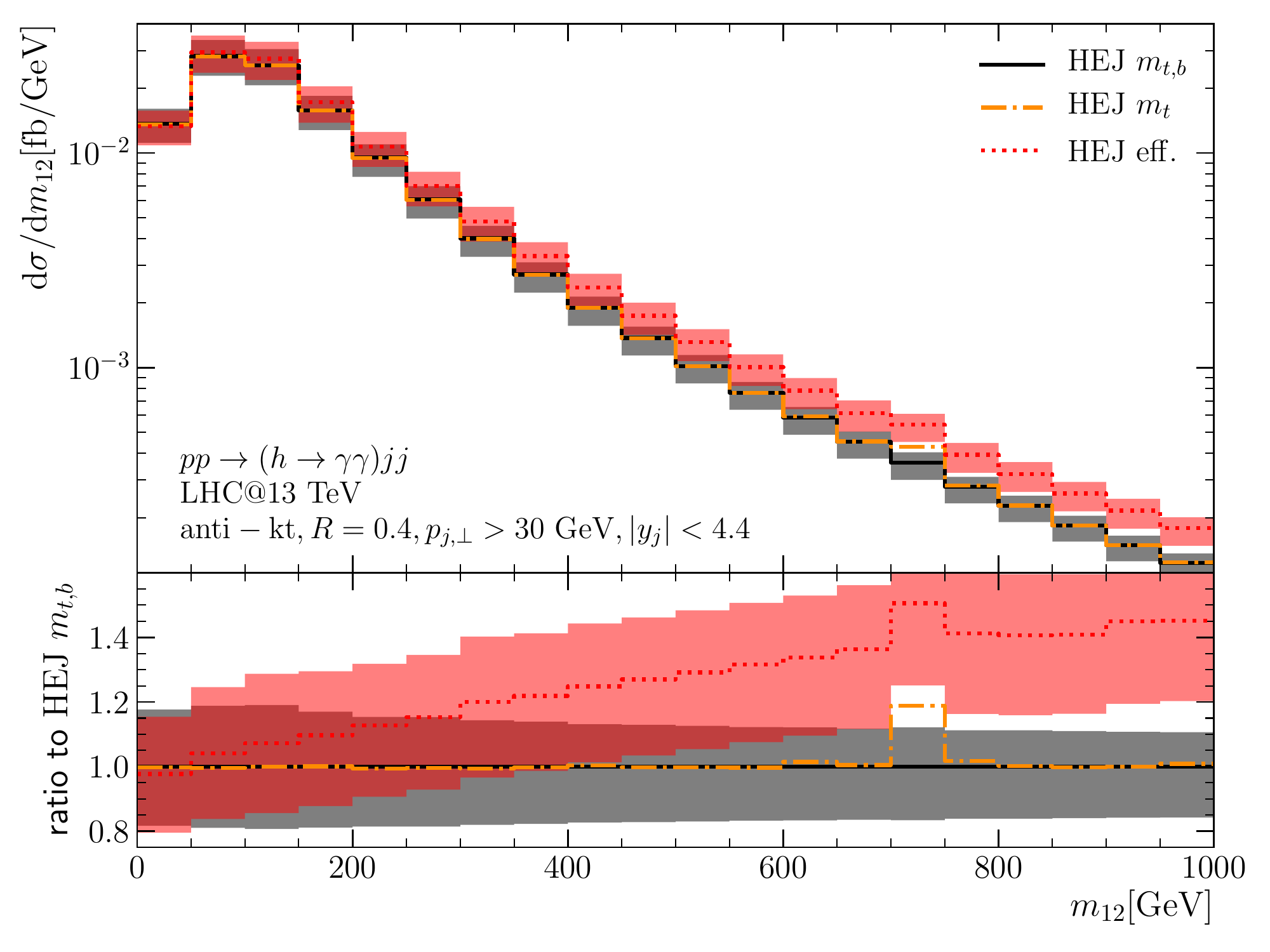}
  \caption{}
\label{fig:ResultsMassm12}
\end{subfigure}
\begin{subfigure}[t]{0.495\textwidth}
  \includegraphics[width=\linewidth]{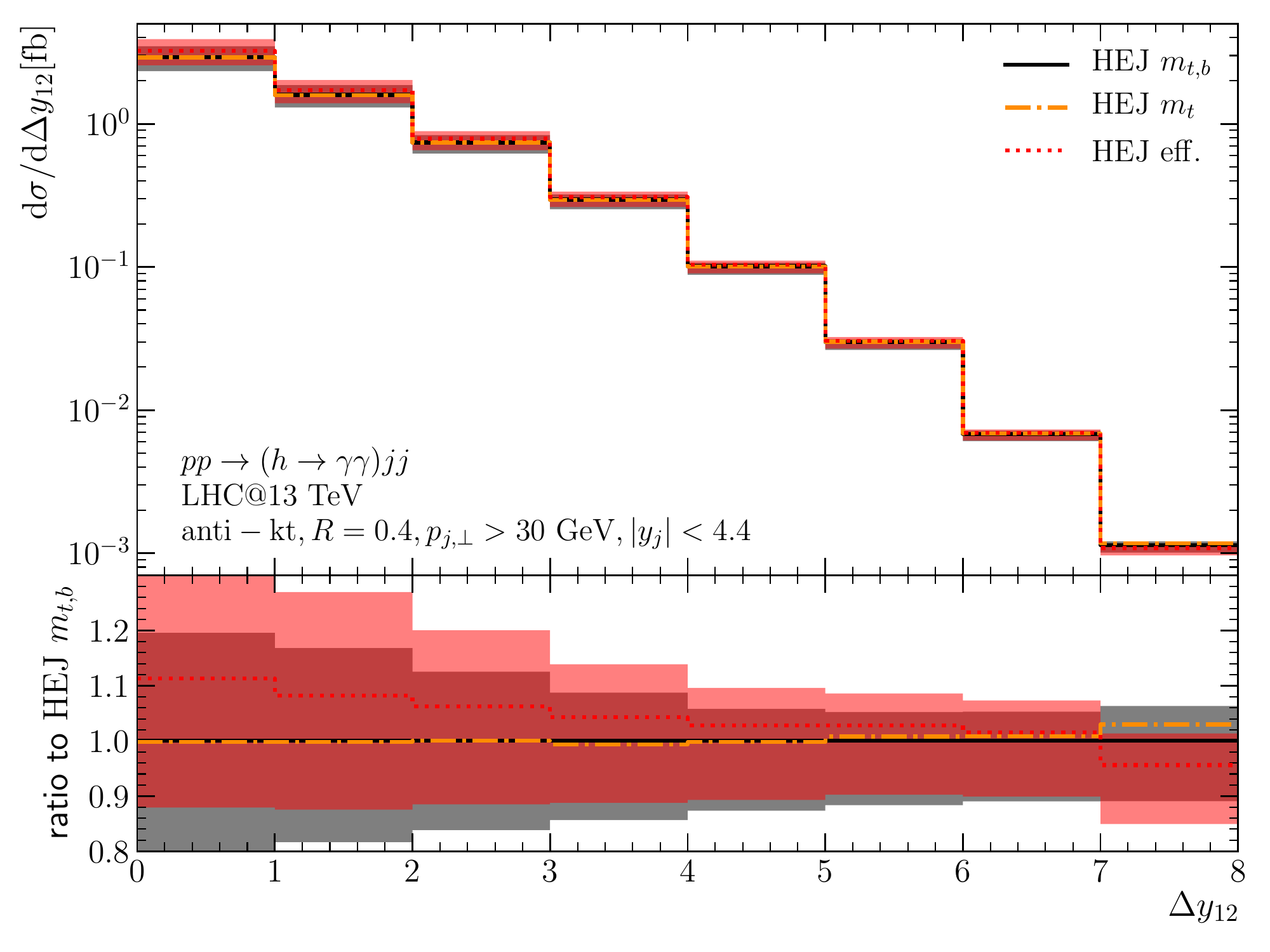}
  \caption{}
\label{fig:ResultsMassDY12}
\end{subfigure}
\begin{subfigure}[t]{0.495\textwidth}
  \includegraphics[width=\linewidth]{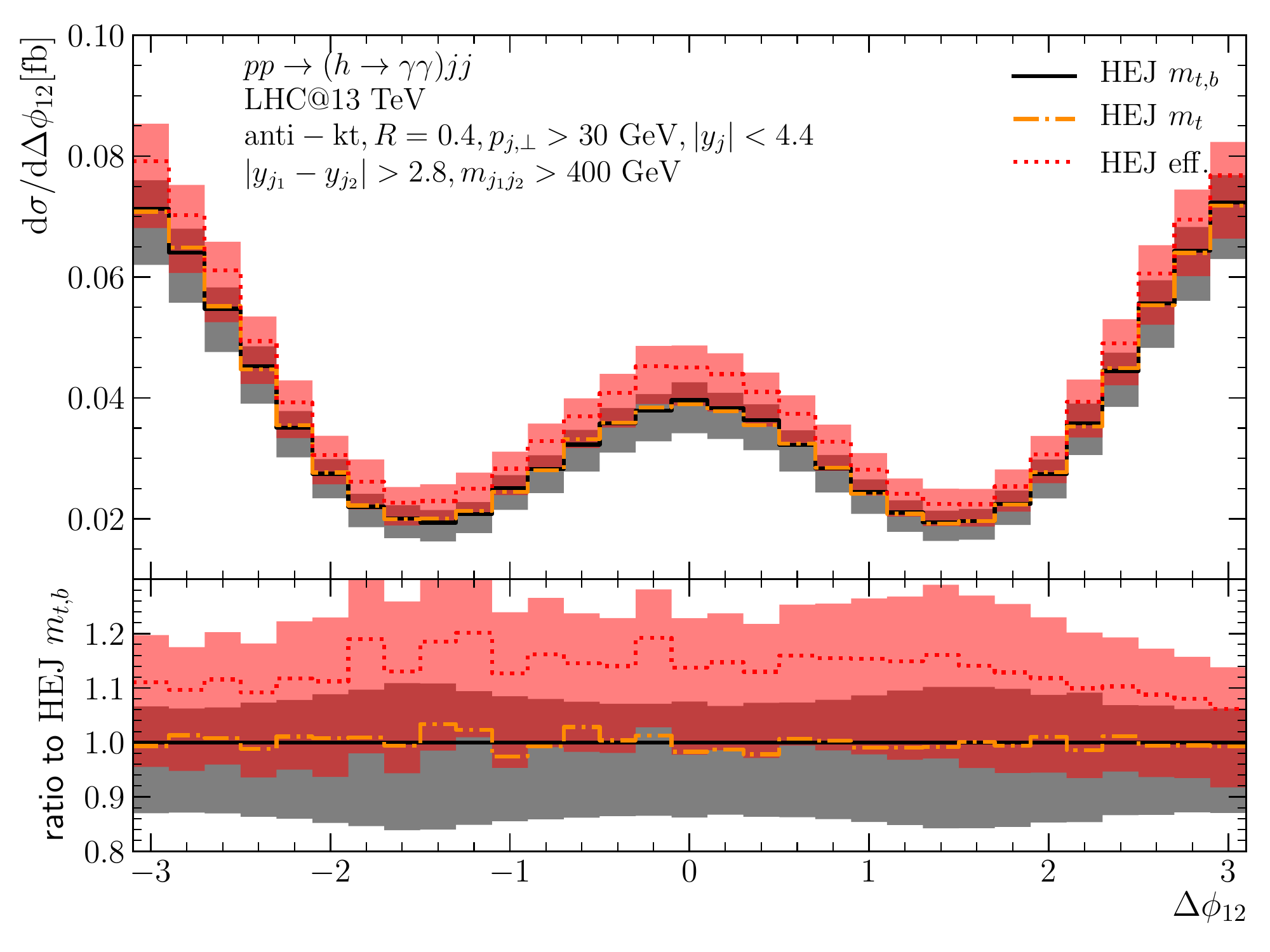}
  \caption{}
\label{fig:ResultsMassdphi12}
\end{subfigure}
\caption{The result of the all-order matched results of \HEJ with three
  descriptions of the quark masses: infinite top-quark mass (red, dotted), finite
  $m_t$ (orange, dot-dashed), and finite $m_t$ and $m_b$ (black/grey, solid). See text for
  further details.}
\label{fig:ResultsMass}
\end{figure}
Figure~\ref{fig:ResultsMass} compares the results obtained with \HEJ using
the three different descriptions of quark masses, namely infinite top-quark
mass, finite $m_t$ but $m_b=0$, and finite both $m_t$ and $m_b$. Evidently,
the effect of the finite $m_b$ is negligibly small and uniform in all the
distributions. As seen already in figures~\ref{fig:h2j} and
\ref{fig:ResultsHigherOrder}, the approximation of infinite top-mass fails
for transverse momenta significantly larger than the top-mass (illustrated
here by a plot of the distribution in the transverse momentum of the Higgs
boson in figure~\ref{fig:ResultsMassPHperp}). Similarly, the results using an
infinite top-mass overshoot the result of finite top-mass by 20\% at an
invariant mass between the two hardest jets of 400~GeV, increasing to 40\%
at 1~TeV. This is relevant for the description of the contribution from the
QCD process within the VBF-studies of $pp\to H+2j$. The corrections from
finite quark masses to the distributions in $\Delta y_{12}$
(figure~\ref{fig:ResultsMassDY12}) or $\Delta \phi_{12}$ with additional VBF cuts
(figure~\ref{fig:ResultsMassdphi12}) reach just 10\%.

\boldmath
\subsection{Final Results for \HEJ}
\label{sec:final-results-hej}
\begin{figure}
  \centering
  \begin{subfigure}[t]{0.495\textwidth}
    \includegraphics[width=\linewidth]{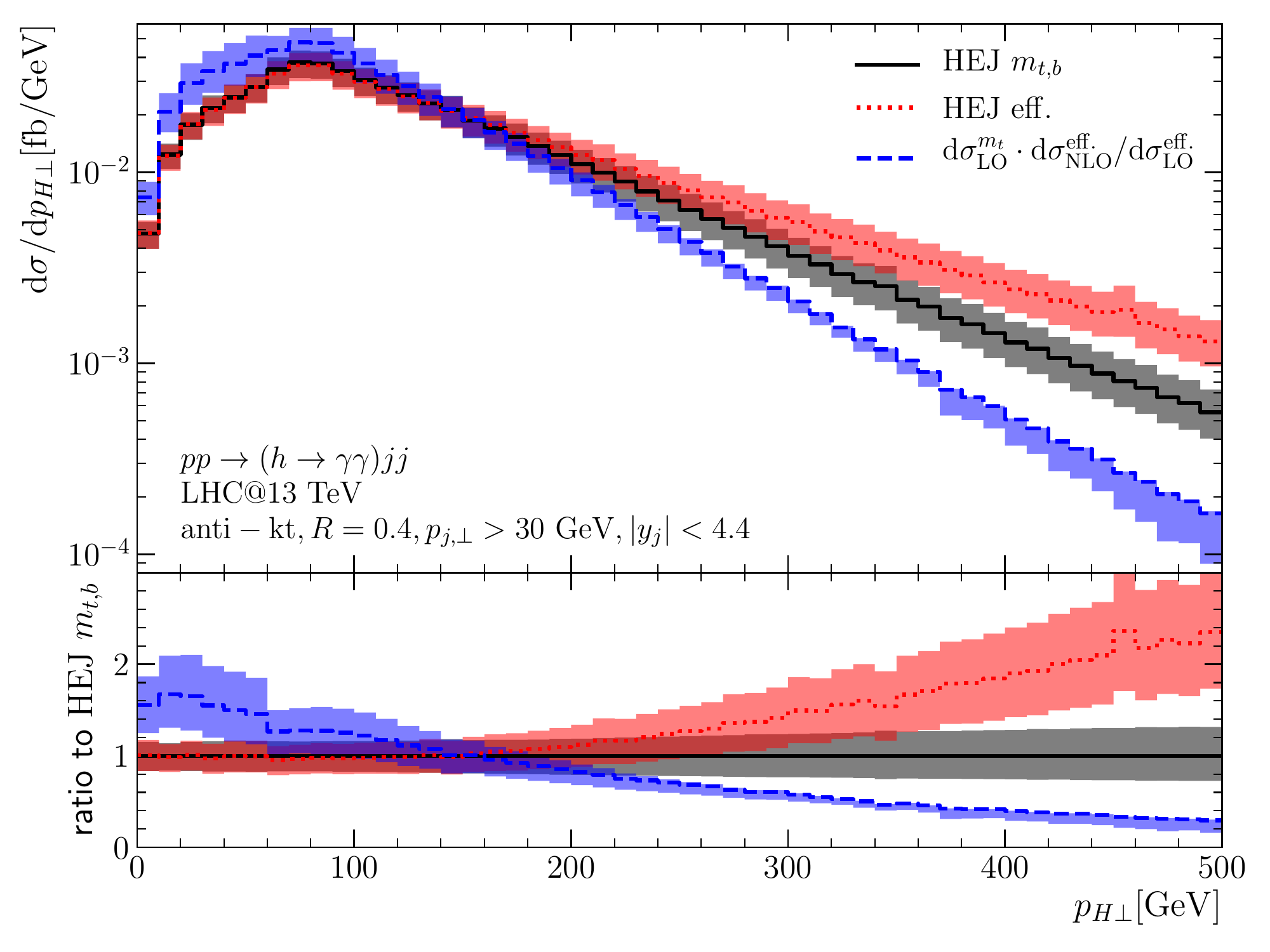}
    \caption{}
    \label{fig:finalHEJfixeda}
  \end{subfigure}
  \begin{subfigure}[t]{0.495\textwidth}
    \includegraphics[width=\linewidth]{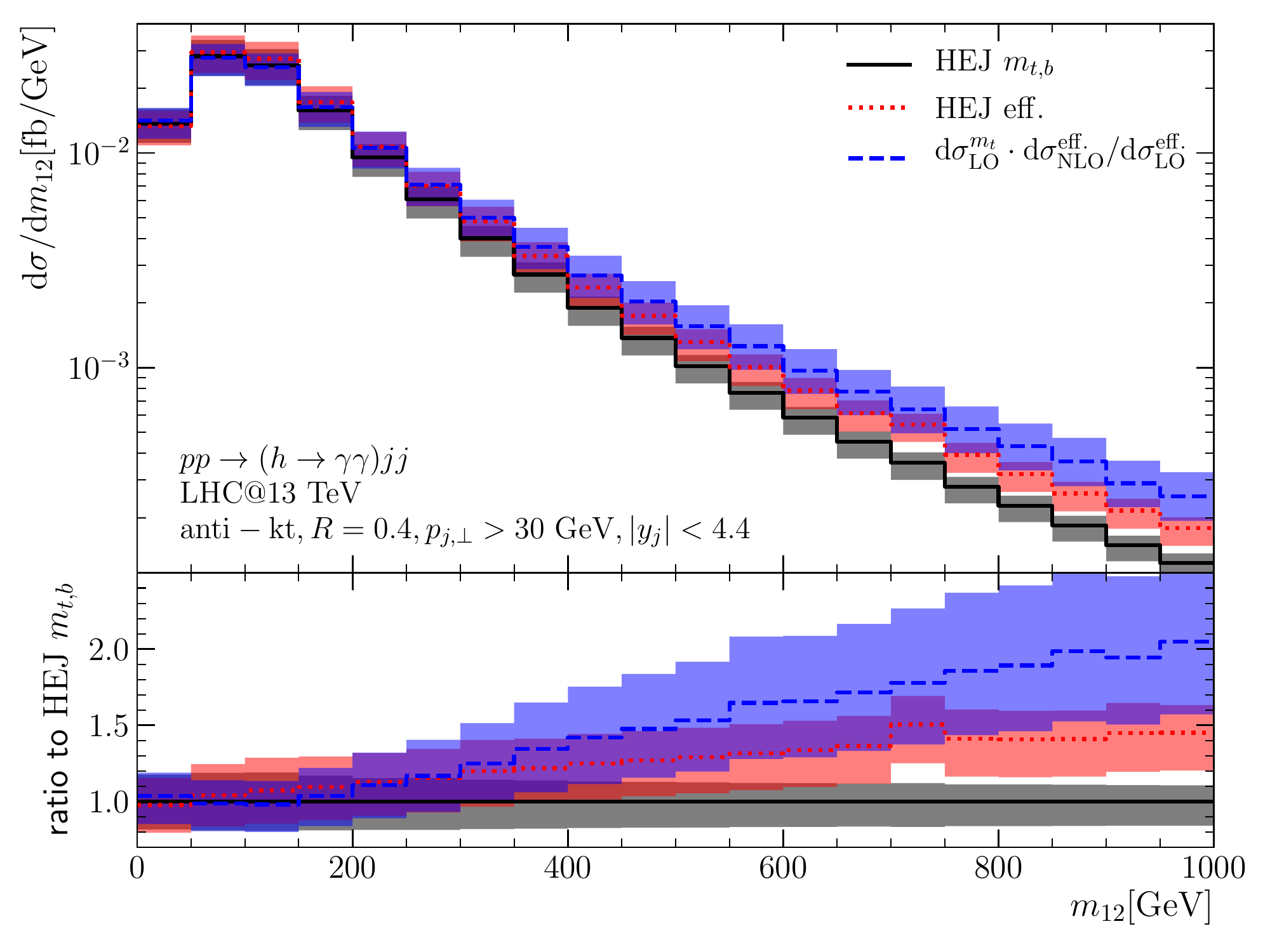}
    \caption{}
    \label{fig:finalHEJfixedb}
  \end{subfigure}
  \begin{subfigure}[t]{0.495\textwidth}
    \includegraphics[width=\linewidth]{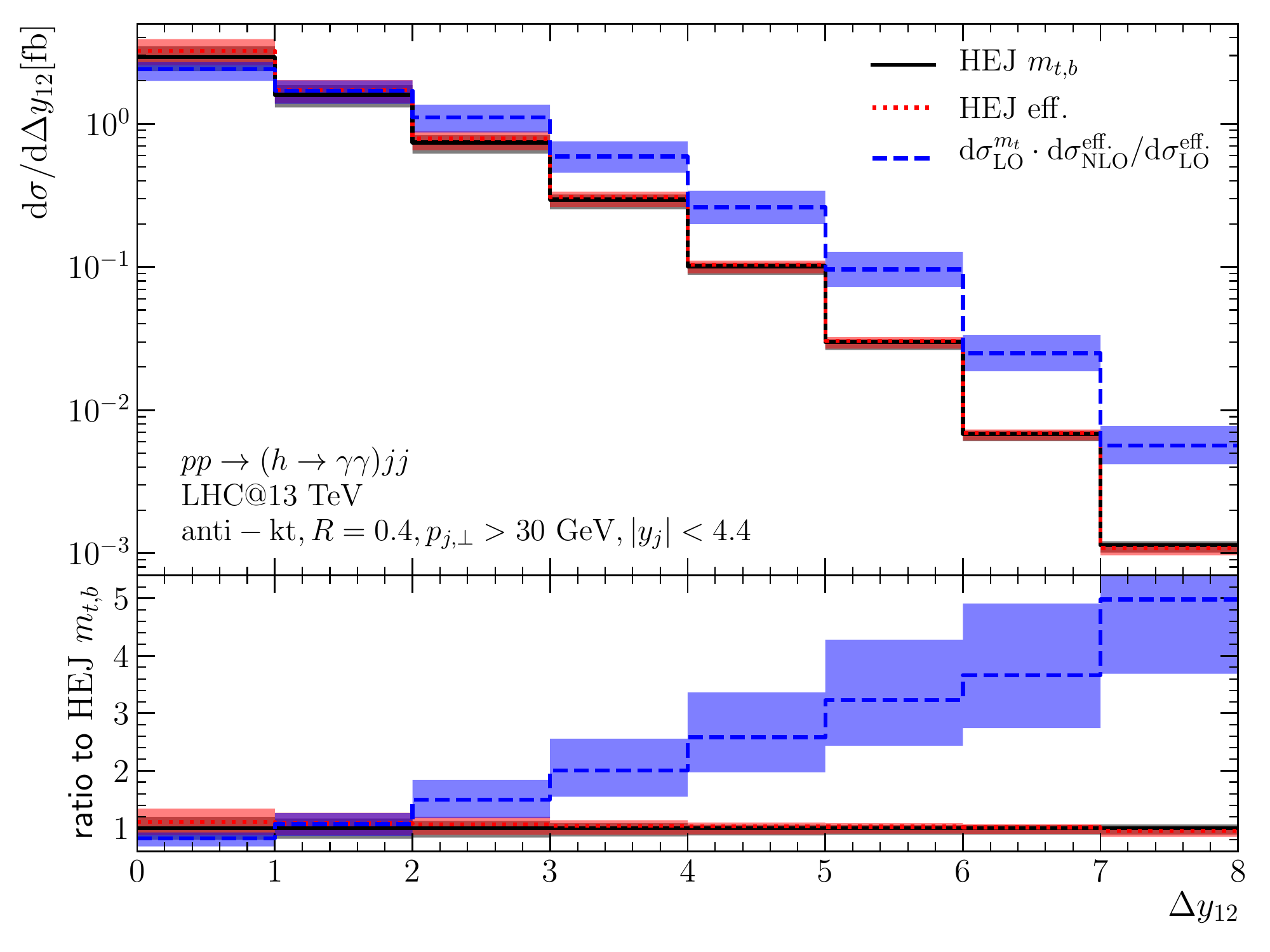}
    \caption{}
    \label{fig:finalHEJfixedc}
  \end{subfigure}
  \begin{subfigure}[t]{0.495\textwidth}
    \includegraphics[width=\linewidth]{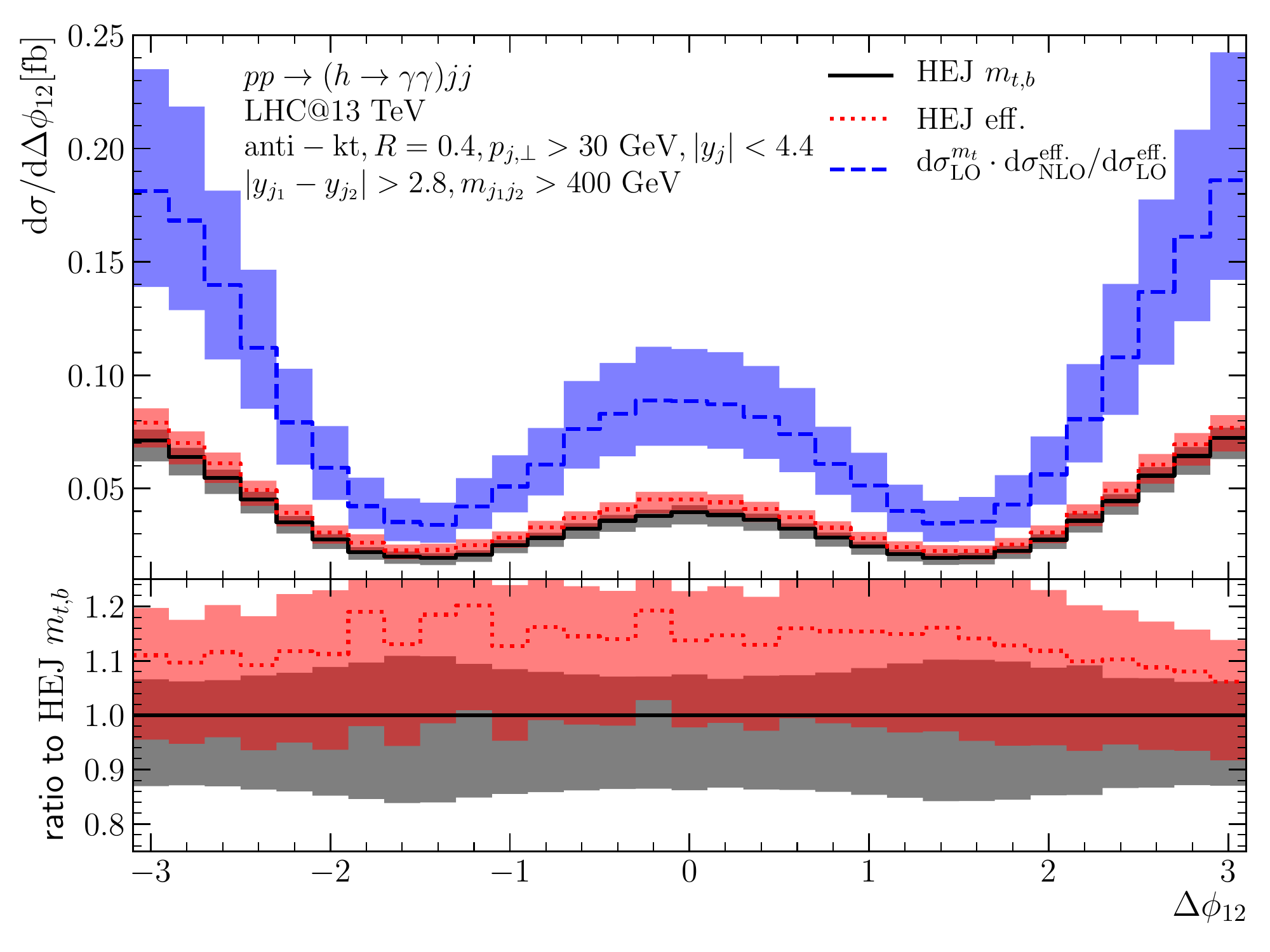}
    \caption{}
    \label{fig:finalHEJfixedd}
  \end{subfigure}
  \caption{The results obtained with \HEJ compared with fixed order for various
    key distributions. See text
  for further details.}
  \label{fig:finalHEJfixed}
\end{figure}

\unboldmath
In this section we compare the most accurate results obtained using the
methods described in this study in \HEJ to those obtained at fixed order. We start by
comparing the observables already investigated previously; as such, the red
and grey bands in figure~\ref{fig:finalHEJfixed} are
identical to those on figure~\ref{fig:ResultsMass}, but are here compared to
the results of using Born-level with finite top-quark mass, rescaled
bin-by-bin with the \NLO $K$-factor obtained using infinite top-quark mass.
We see in figure~\ref{fig:finalHEJfixeda} that the fixed-order result is
significantly softer in the transverse momentum of the Higgs boson than the
result obtained with \HEJ. We have already discussed how this leads to a
larger impact of the finite quark masses within \HEJ than at fixed order.

Figure~\ref{fig:finalHEJfixedb} illustrates that the distribution in the
invariant mass between the two hardest jets is increasingly suppressed for
increasing $m_{12}$ in \HEJ compared to fixed order. While the results are
similar for small $m_{12}$, the ratio of fixed order over \HEJ reaches 1.5 at
$m_{12}\approx 500$~GeV. Similarly, as illustrated in
figure~\ref{fig:finalHEJfixedc} the results of \HEJ are much suppressed
compared to \NLO at large $\Delta y_{12}$. The ratio of fixed order to \HEJ found here
increases in a straight line finally reaching 5 at $\Delta
y_{12}=8$. As discussed around figure~\ref{fig:NLOK-factor}, this is due to the
absence of a logarithmic suppression of the 3-jet component in the \NLO
prediction.
Although the \HEJ cross section is matched to the \NLO value, this does not
change the differences in the shapes of the distributions and the large $K$-factor at
large $m_{12}$ therefore persists.

Figure~\ref{fig:finalHEJfixedd} shows the distribution with respect to the
azimuthal angle between the two hardest jets, measured relative to the
positive rapidity direction, thus exploring
the full interval from $-\pi$ to $\pi$. VBF cuts have again been applied in
addition to the general cuts.  These require a significant invariant mass and
rapidity separation of the hardest two jets and hence the suppression in
figures~\ref{fig:finalHEJfixedb} and \ref{fig:finalHEJfixedc} translates into
a large difference (around a factor of 2) in the cross section between the
\HEJ and fixed order predictions (as also seen earlier in
table~\ref{tab:crosssections}).  The distinctive shape which arises as a
result of the CP structure of the $ggH$
vertex~\cite{Plehn:2001nj,Klamke:2007cu,Andersen:2010zx} is seen in all the
predictions.

We present the results of
figures~\ref{fig:ResultsHigherOrder}--\ref{fig:finalHEJfixed} for the
alternative central scale choice of $H_T/2$ in appendix~\ref{sec:HT2}.  The main
conclusions of the plots are unchanged: the impact of the higher-order
corrections in \HEJ lead to a harder distribution in $p_{H \perp}$, which
enhances the finite quark mass and loop propagator effects.  This in turn leads
to a suppression of the prediction at large $m_{12}$, and the predicted impact of a
VBF cut is more severe in the all-order calculations of \HEJ than that seen in
fixed-order predictions.

\begin{figure}
  \centering
  \begin{subfigure}[t]{0.495\textwidth}
    \includegraphics[width=\linewidth]{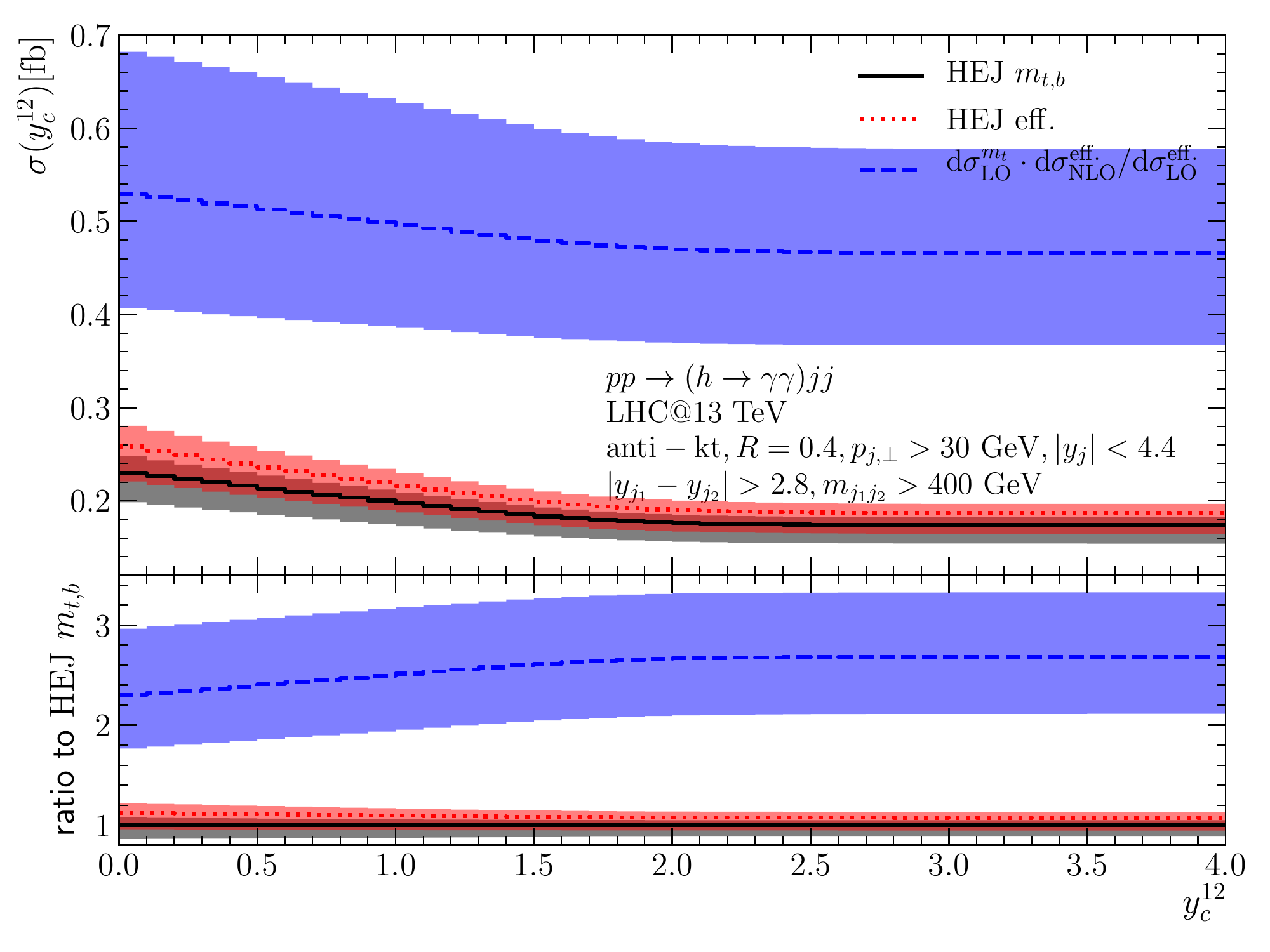}
    \caption{}
    \label{fig:yc12}
  \end{subfigure}
  \begin{subfigure}[t]{0.495\textwidth}
    \includegraphics[width=\linewidth]{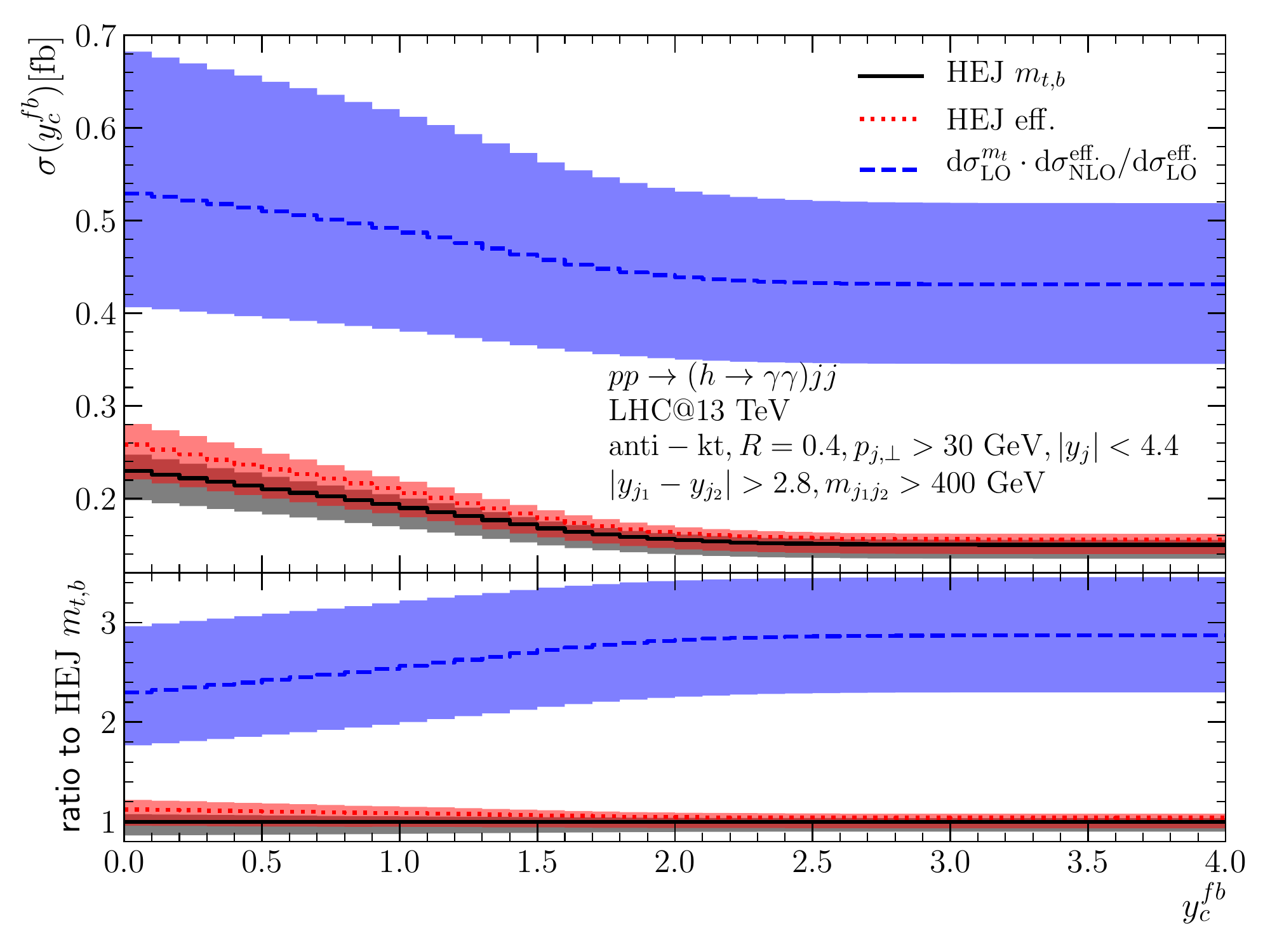}
    \caption{}
    \label{fig:ycfb}
  \end{subfigure}
  \caption{We compare the cross section from \HEJ and at \NLO as a function of a
  jet veto, $y_c$, defined in the text.  In (a) the tagging jets are the
  two hardest jets; in (b) the tagging jets are the most forward/backward jets.}
  \label{fig:yc}
\end{figure}

In the final figure in this section, figure~\ref{fig:yc}, we discuss an alternative
to a traditional jet veto~\cite{Dokshitzer:1991he,Rainwater:1996ud,Plehn:2001nj,Klamke:2007cu,Andersen:2010zx}.  We begin by defining two tagging
jets $t_1$ and $t_2$: firstly as the hardest two jets in the event
$t_{1,2}=j_{1,2}$ and secondly as the most forward/backward jets, $t_{1,2}=j_{f,b}$.  We may
then construct $y_0=(y_{t_1} + y_{t_2})/2$ for each event.  The event will then be vetoed if it
contains a further jet with transverse momentum above 30~GeV in-between the
two tagging jets which satisfies $|y_j-y_0|<y_c$.  This procedure applies to a
larger region in rapidity than a traditional jet veto which is only applied to jets
in-between the tagging jets.  This means that the same level of suppression can
be obtained with a higher (and therefore perturbatively safer) transverse
momentum cut.  In figure~\ref{fig:yc12}, we show the results when we choose the
two hardest jets as the tagging jets while in figure~\ref{fig:ycfb} the tagging
jets are
the most forward/backward jets.  In both cases, the cross section has reached a
plateau by about $y_c=2$.  The difference between the two choices is
relatively small but the cross section for a given value of $y_c$ is lower for the
forward/backward choice for the tagging jets than for the hardest jet choice.
As discussed in \cite{Bendavid:2018nar} this type of jet
veto has for $y_c$ up to 1.5 very little impact on the VBF process itself
(since most radiation is produced close in rapidity to the Born-level jets),
and is therefore an efficient tool in distinguishing the contribution from
the two processes for $pp\to H+2j$. We saw that the VBF cuts themselves have a
relatively larger impact on the cross sections of \HEJ than fixed order,
because of the steeper fall-off with $m_{12}$ and $\Delta
y_{12}$. Figure~\ref{fig:yc} shows that a further cut on jet activity will
have a yet larger effect on \HEJ compared to fixed order. This is all
expected since the fixed order results fail to reproduce the rise in jet
activity with increasing rapidity separation, which is observed in both data
and \HEJ~\cite{Aad:2011jz,Aad:2014pua}.

%%% Local Variables:
%%% mode: latex
%%% TeX-master: "main"
%%% End:

\newpage
\section{Conclusions}
\label{sec:summary}

% * Larger effect than for fixed-order results due to harder spectrum in
%   transverse momenta: 6.5% overall, 7.6% effect within VBF cuts
% * Fixed order rescaling procedure to obtain NLO in mt-> infty limit

We have calculated the gluon fusion contribution to $H+2j$ including both
\begin{itemize}
\item leading logarithmic corrections in $\hat s/p_t^2$ to all orders in
  $\alpha_s$, and
\item full dependence on top and bottom quark masses, including the
  loop-propagator kinematic effects absent in the $m_t\to \infty$ limit.
\end{itemize}
The components necessary for implementing the full quark-mass
dependence within the all-order resummation scheme of \HIGHEJ~(\HEJ) were
calculated, such that both the quark mass and the systematic logarithmic
corrections within the VBF cuts could be investigated.  This goes far beyond the
current state-of-the-art fixed order predictions.  The results thus
obtained have been compared to the fixed-order full top-mass-dependent
results evaluated at Born-level, but rescaled bin-by-bin with the \NLO
$K$-factor obtained in the limit of $m_t\to\infty$.

While the fixed-order results obtained with finite $m_t$ differ very little from
those obtained for $m_t\to\infty$ we find a much larger reduction of 9\% on the
inclusive cross section in \HEJ; this is because the transverse momentum of the
Higgs boson is found to be harder with \HEJ than at \LO, and that the finite
mass-corrections are larger at large transverse scales.  For the first time, our calculation allows
the computation of the interference between the top and bottom quark contributions
beyond leading order in $H+2j$.  We find that the interference is
extremely small for the running values of $m_b(m_H)$ and $m_t(m_H)$, less than a percent
for all observables.

We find that for a scale choice of $\scalemjj$ the \NLO $K$-factor increases systematically for both $m_{12}$,
$\Delta y_{12}$ and most dramatically for $\Delta y_{fb}$ (see
figure~\ref{fig:NLOK-factor}). With a scale choice of $\scaleht$
however, the $K$-factor \emph{decreases} with increasing $m_{12}$ or $\Delta
y_{12}$. The balance between the large virtual negative corrections and the real
positive corrections are clearly scale dependent. The large corrections illustrate a serious perturbative
instability of the fixed order expansion within the VBF-cuts. This
instability is specifically addressed
by \HEJ.

At
large $\Delta y_{12}$ and $m_{12}$, the all-order predictions from \HEJ are
systematically suppressed compared to fixed order. The discussion of scale
choice is independent of the discussion of the behaviour at large $m_{jj}$,
and so is the conclusion that a resummation of the leading terms at large
$m_{jj}$ leads to a reduction of the cross section within the VBF cuts. Our results show
that the gluon-fusion contamination in VBF studies is less severe than the
fixed-order estimate would imply. The finite-mass corrections to \HEJ within
the VBF-cuts lead to a further 11\% suppression compared to the results obtained with
infinite top-mass.

% Apart from the reduction in the cross section, the finite quark-mass
% corrections have no negative impact on the possibilities for using the
% gluon-fusion component in a measurement of possible $CP$-admixtures in the
% Higgs couplings.

%%% Local Variables:
%%% mode: latex
%%% TeX-master: "main"
%%% End:

\section*{Acknowledgements}
We are grateful to the following for helpful comments on the manuscript: Simon Badger,
Johannes Bl\"umlein, Luigi Del Debbio, Keith Ellis, Nigel Glover and Valya Khoze.
This work has received funding from the European Union's Horizon 2020
research and innovation programme as part of the Marie
Sk\l{}odowska-Curie Innovative Training Network MCnetITN3 (grant
agreement no.~722104), the Marie Sk\l{}odowska-Curie grant agreement No.
764850, SAGEX, and COST action CA16201: ``Unraveling new physics at the
LHC through the precision frontier'', and from the UK Science and Technology
Facilities Council (STFC). JMS is supported by a Royal Society
University Research Fellowship and the ERC Starting Grant 715049
``QCDforfuture''.

%%% Local Variables:
%%% mode: latex
%%% TeX-master: "main"
%%% End:

\appendix

\section{Form Factors for the Higgs-Boson Coupling to Gluons}
\label{sec:ggH_form_factors}

Quoting eq.~\eqref{eq:VH}, the coupling of the Higgs boson to gluons via
a virtual quark loop can be written as
\begin{equation}
  \label{eq:VH_appendix}
  V^{\mu\nu}_H(q_1, q_2) = \mathgraphics{V_H} =
  \frac{\alpha_s m^2}{\pi v}\big[
  g^{\mu\nu} T_1(q_1, q_2) - q_2^\mu q_1^\nu T_2(q_1, q_2)
  \big]\,.
\end{equation}
The outgoing momentum of the Higgs boson is $p_H = q_1 - q_2$. The form
factors $T_1$ and $T_2$ are then given by~\cite{DelDuca:2003ba}
\begin{align}
  \label{eq:T_1}
     T_1(q_1, q_2) ={}& -C_0(q_1, q_2)\*\left[2\*m^2+\frac{1}{2}\*\left(q_1^2+q_2^2-p_H^2\right)+\frac{2\*q_1^2\*q_2^2\*p_H^2}{\lambda}\right]\notag\\
      &-\left[B_0(q_2)-B_0(p_H)\right]\*\frac{q_2^2}{\lambda}\*\left(q_2^2-q_1^2-p_H^2\right)\notag\\
      &-\left[B_0(q_1)-B_0(p_H)\right]\*\frac{q_1^2}{\lambda}\*\left(q_1^2-q_2^2-p_H^2\right)-1\,,\displaybreak[0]\\
  \label{eq:T_2}
    T_2(q_1, q_2) ={}& C_0(q_1,
    q_2)\*\left[\frac{4\*m^2}{\lambda}\*\left(p_H^2-q_1^2-q_2^2\right)-1-\frac{4\*q_1^2\*q_2^2}{\lambda}
        -
        \frac{12\*q_1^2\*q_2^2\*p_H^2}{\lambda^2}\*\left(q_1^2+q_2^2-p_H^2\right)\right]\notag\\
      &-\left[B_0(q_2)-B_0(p_H)\right]\*\left[\frac{2\*q_2^2}{\lambda}+\frac{12\*q_1^2\*q_2^2}{\lambda^2}\*\left(q_2^2-q_1^2+p_H^2\right)\right]\notag\\
      &-\left[B_0(q_1)-B_0(p_H)\right]\*\left[\frac{2\*q_1^2}{\lambda}+\frac{12\*q_1^2\*q_2^2}{\lambda^2}\*\left(q_1^2-q_2^2+p_H^2\right)\right]\notag\\
      &-\frac{2}{\lambda}\*\left(q_1^2+q_2^2-p_H^2\right)\,,
\end{align}
where we have used the scalar bubble and triangle integrals
\begin{align}
  \label{eq:B0}
  B_0\left(p\right) ={}& \int \frac{d^dl}{i\pi^{\frac{d}{2}}}
                         \frac{1}{\left(l^2-m^2\right)\left((l+p)^2-m^2\right)}\,,\\
  \label{eq:C0}
  C_0\left(p,q\right) ={}& \int \frac{d^dl}{i\pi^{\frac{d}{2}}} \frac{1}{\left(l^2-m^2\right)\left((l+p)^2-m^2\right)\left((l+p-q)^2-m^2\right)}\,,
\end{align}
and the K\"{a}ll\'{e}n function
\begin{equation}
  \label{eq:lambda}
  \lambda = q_1^4 + q_2^4 + p_H^4 - 2\*q_1^2\*q_2^2 - 2\*q_1^2\*p_H^2- 2\*q_2^2\*p_H^2\,.
\end{equation}
The relation to the form factors $A_1, A_2$ given in \cite{DelDuca:2003ba} is
\begin{align}
  \label{eq:A_1}
  A_1(q_1, q_2) ={}& \frac{i}{16\pi^2}\*T_2(-q_1, q_2)\,,\\
  \label{eq:A_2}
  A_2(q_1, q_2) ={}& -\frac{i}{16\pi^2}\*T_1(-q_1, q_2)\,.
\end{align}

%%% Local Variables:
%%% mode: latex
%%% TeX-master: "main"
%%% End:

\section{Effective Current for Peripheral Emission}
\label{sec:jH}

We describe the emission of a peripheral Higgs boson close to a
scattering gluon with an effective current. In the following we consider
a lightcone decomposition of the gluon momenta, i.e. $p^\pm = E \pm p_z$
and $p_\perp = p_x + i p_y$. The incoming gluon momentum $p_a$ defines
the $-$ direction, so that $p_a^+ = p_{a\perp} = 0$. The outgoing
momenta are $p_1$ for the gluon and $p_H$ for the Higgs boson. We choose
the following polarisation vectors:
\begin{equation}
  \label{eq:pol_vectors}
  \epsilon_\mu^\pm(p_a) = \frac{j_\mu^\pm(p_1, p_a)}{\sqrt{2}
    \bar{u}^\pm(p_a)u^\mp(p_1)}\,, \quad \epsilon_\mu^{\pm,*}(p_1) = -\frac{j_\mu^\pm(p_1, p_a)}{\sqrt{2}
    \bar{u}^\mp(p_1)u^\pm(p_a)}\,.
\end{equation}
Following~\cite{DelDuca:2001fn}, we introduce effective polarisation
vectors to describe the contraction with the Higgs-boson production
vertex eq.~\eqref{eq:VH}:
\begin{align}
  \label{eq:eps_H}
  \epsilon_{H,\mu}(p_a) = \frac{T_2(p_a, p_a-p_H)}{(p_a-p_H)^2}\big[p_a\cdot
  p_H\epsilon_\mu(p_a) - p_H\cdot\epsilon(p_a) p_{a,\mu}\big]\,,\\
  \epsilon_{H,\mu}^*(p_1) = -\frac{T_2(p_1+p_H, p_1)}{(p_1+p_H)^2}\big[p_1\cdot
  p_H\epsilon_\mu^*(p_1) - p_H\cdot\epsilon^*(p_1) p_{1,\mu}\big]\,,
\end{align}
We also employ the usual short-hand notation
\begin{equation}
  \label{eq:spinor_helicity}
  \spa i.j = \bar{u}^-(p_i)u^+(p_j)\,,\qquad \spb i.j =
  \bar{u}^+(p_i)u^-(p_j)\,, \qquad[ i | H | j\rangle = j_\mu^+(p_i, p_j)p_H^\mu\,.
\end{equation}
Without loss of generality, we consider only the case where the incoming
gluon has positive helicity. The remaining helicity configurations can
be obtained through parity transformation.

Labeling the effective current by the helicities of the gluons we obtain
for the same-helicity case
\begin{equation}
  \label{eq:jH_same_helicity}
  \begin{split}
    j_{H,\mu}^{++}{}&(p_1,p_a,p_H) =
    \frac{m^2}{\pi v}\bigg[\\
      &-\sqrt{\frac{2p_1^-}{p_a^-}}\frac{p_{1\perp}^*}{|p_{1\perp}|}\frac{t_2}{\spb a.1}\epsilon^{+,*}_{H,\mu}(p_1)
      +\sqrt{\frac{2p_a^-}{p_1^-}}\frac{p_{1\perp}^*}{|p_{1\perp}|}\frac{t_2}{\spa 1.a}\epsilon^{+}_{H,\mu}(p_a)\\
      &+ [1|H|a\rangle \bigg(
        \frac{\sqrt{2}}{\spa 1.a}\epsilon^{+}_{H,\mu}(p_a)
        +         \frac{\sqrt{2}}{\spb a.1}\epsilon^{+,*}_{H,\mu}(p_1)-\frac{\spa 1.a T_2(p_a, p_a-p_H)}{\sqrt{2}(p_a-p_H)^2}\epsilon^{+,*}_{\mu}(p_1)\\
        &
        \qquad
        -\frac{\spb a.1 T_2(p_1+p_H,
          p_1)}{\sqrt{2}(p_1+p_H)^2}\epsilon^{+}_{\mu}(p_a)-\frac{RH_4}{\sqrt{2}\spb a.1}\epsilon^{+,*}_{\mu}(p_1)+\frac{RH_5}{\sqrt{2}\spa 1.a}\epsilon^{+}_{\mu}(p_a)
        \bigg)\\
        &
        - \frac{[1|H|a\rangle^2}{2 t_1}(p_{a,\mu} RH_{10} - p_{1,\mu} RH_{12})\bigg]
  \end{split}
\end{equation}
with $t_1 = (p_a-p_1)^2$, $t_2 = (p_a-p_1-p_H)^2$ and $R = 8 \pi^2$. The form factors $H_4,
H_5, H_{10}, H_{12}$ are given in~\cite{DelDuca:2003ba}.

The current with a flip in the gluon helicity reads
\begin{equation}
  \label{eq:jH_helicity_flip}
  \begin{split}
    j_{H,\mu}^{+-}{}&(p_1,p_a,p_H) =
    \frac{m^2}{\pi v}\bigg[\\
    &-\sqrt{\frac{2p_1^-}{p_a^-}}\frac{p_{1\perp}^*}{|p_{1\perp}|}\frac{t_2}{\spb
      a.1}\epsilon^{-,*}_{H,\mu}(p_1)
    +\sqrt{\frac{2p_a^-}{p_1^-}}\frac{p_{1\perp}}{|p_{1\perp}|}\frac{t_2}{\spb a.1}\epsilon^{+}_{H,\mu}(p_a)\\
    &+ [1|H|a\rangle \left(
    \frac{\sqrt{2}}{\spb a.1} \epsilon^{-,*}_{H,\mu}(p_1)
    -\frac{\spa 1.a T_2(p_a, p_a-p_H)}{\sqrt{2}(p_a-p_H)^2}\epsilon^{-,*}_{\mu}(p_1)
    - \frac{RH_4}{\sqrt{2}\spb a.1}\epsilon^{-,*}_{\mu}(p_1)\right)
    \\
    &+ [a|H|1\rangle \left(
      \frac{\sqrt{2}}{\spb a.1}\epsilon^{+}_{H,\mu}(p_a)
      -\frac{\spa 1.a
        T_2(p_1+p_H,p_1)}{\sqrt{2}(p_1+p_H)^2}\epsilon^{+}_{\mu}(p_a)
      +\frac{RH_5}{\sqrt{2}\spb a.1}\epsilon^{+}_{\mu}(p_a)
    \right)\\
    & - \frac{[1|H|a\rangle [a|H|1\rangle}{2 \spb a.1 ^2}(p_{a,\mu} RH_{10} - p_{1,\mu}
    RH_{12})\\
    &+ \frac{\spa 1.a}{\spb a.1}\bigg(RH_1p_{1,\mu}-RH_2p_{a,\mu}+2
    p_1\cdot p_H \frac{T_2(p_1+p_H, p_1)}{(p_1+p_H)^2} p_{a,\mu}
    \\
    &
    \qquad- 2p_a \cdot p_H \frac{T_2(p_a, p_a-p_H)}{(p_a-p_H)^2} p_{1,\mu}+ T_1(p_a-p_1, p_a-p_1-p_H)\frac{(p_1 + p_a)_\mu}{t_1}\\
    &\qquad-\frac{(p_1+p_a)\cdot p_H}{t_1} T_2(p_a-p_1, p_a-p_1-p_H)(p_1 - p_a)_\mu
    \bigg)
    \bigg]\,.
  \end{split}
\end{equation}
If we instead choose the gluon momentum in the $+$ direction, so that
$p_a^- = p_{a\perp} = 0$, the corresponding currents are obtained by
replacing $p_1^- \to p_1^+, p_a^- \to p_a^+,
\frac{p_{1\perp}}{|p_{1\perp}|} \to -1$ in the second line of
eq.~\eqref{eq:jH_same_helicity} and eq.~\eqref{eq:jH_helicity_flip}.

%%% Local Variables:
%%% TeX-master: "main"
%%% End:

\section{The Current for a Single Unordered Gluon Emission}
\label{sec:curr-single-unord}

In section~\ref{sec:first-set-next}, we use an effective current, $ j_\mu^{{\rm
    uno\; cd}}(p_2,p_1,p_a)$, to describe the emission of an unordered gluon (one
additional gluon outside in rapidity of an FKL configuration).    The current
for
%\begin{align}
  $q(p_a) \to g(p_1) q(p_2) g^*(\tilde{q}_2)$
%\end{align}
was derived in~\cite{Andersen:2017kfc} to be:
\begin{align}
  \label{eq:juno}
  \begin{split}
    &j^{{\rm uno}\; \mu\ cd}(p_2,p_1,p_a) = i \varepsilon_{1\nu} \left(  T_{2i}^{c}T_{ia}^d\
      \left(U_1^{\mu\nu}-L^{\mu\nu} \right) + T_{2i}^{d}T_{ia}^c\ \left(U_2^{\mu\nu} +
        L^{\mu\nu} \right) \right). \\
      U_1^{\mu\nu} &= \frac1{s_{21}} \left( j_{21}^\nu j_{1a}^\mu + 2 p_2^\nu
      j_{2a}^\mu \right) \qquad  \qquad U_2^{\mu\nu} = \frac1{t_{a1}} \left( 2
      j_{2a}^\mu p_a^\nu - j_{21}^\mu  j_{1a}^\nu \right) \\
    L^{\mu\nu} &= \frac1{t_{a2}} \left(-2p_1^\mu j_{2a}^\nu+2p_1.j_{2a}
      g^{\mu\nu} + (\tilde{q}_1+\tilde{q}_2)^\nu j_{2a}^\mu + \frac{t_{b2}}{2} j_{2a}^\mu \left(
      \frac{p_2^\nu}{p_1.p_2} + \frac{p_b^\nu}{p_1.p_b} \right) \right) ,
  \end{split}
\end{align}
where $\tilde{q}_1=p_a-p_1$ and $\tilde{q}_2=\tilde{q}_1-p_2$. This differs from
our other currents as there is no longer a single overall colour factor, and
hence colour factors (with free indices $c$ and $d$) must be included.  Upon
contracting with another current squaring, this leads to terms with
different colour factors.  For example, for $q(p_a)Q(p_b)\to
g(p_1)q(p_2)Q(p_3)$, we find~\cite{Andersen:2017kfc}
\begin{align}
  \label{eq:coloursquare}
    \begin{split}
    \left| \overline{\mathcal{M}_{{\rm tree}\ qQ\to gqQ}^{HEJ}} \right|^2
    =& -\frac{g_s^6}{16 t_{b3}^2}\ \sum_{h_a,h_1,h_b,h_2}\ \bigg[ C_F \big( 2
      \mathrm{Re}\big( [(L^{\mu\nu}-U_1^{\mu\nu})\cdot j_{3b\, \mu}]\ [
          (L^{\rho}_{\;\,\nu}+U_{2\;\, \nu}^{\;\rho})\cdot j_{3b\, \rho} ]^* \big)
      \big)  \\
    & \hspace{4cm} + 2\frac{C_F^2}{C_A} \left| (U_1^{\mu\nu}+U_2^{\mu\nu}) \cdot
        j_{3b_\mu} \right|^2 \bigg] \\
      \equiv& -\frac{g_s^6}{16 t_{b3}^2} C_F \left\|S_{f_1 f_2\to gf_1 f_2}^{\rm uno}\right\|^2 .
  \end{split}
\end{align}
The factor we require in eq.~\eqref{eq:MunocentralH} is therefore given by
\begin{align}
  \label{eq:SunoH}
  \begin{split}
  &||S^{\rm uno}_{qf_2 \to gqHf_2}(p_1,p_2,p_3,p_a,p_b,q_1,q_2)||^2 \\ =& \sum_{h_a,h_1,h_b,h_2}\ \bigg[ \big( 2
      \mathrm{Re}\big( [(L^{\mu\nu}-U_1^{\mu\nu})\cdot J_{\mu}]\ [
          (L^{\rho}_{\;\,\nu}+U_{2\;\, \nu}^{\;\rho})\cdot J_{\rho} ]^* \big)
      \big)   \\ & \hspace{4cm} + 2\frac{C_F}{C_A} \left| (U_1^{\mu\nu}+U_2^{\mu\nu}) \cdot
        J_\mu \right|^2 \bigg] \,,
      \end{split}
\end{align}
where we use the shorthand $J^\mu = V_H^{\mu\nu}(q_1,q_2)j_\nu(p_3,p_b)$.

%%% Local Variables:
%%% mode: latex
%%% TeX-master: "main"
%%% End:

\section{Results with $\scaleht$}
\label{sec:HT2}

\begin{figure}
  \centering
  \begin{subfigure}[t]{0.495\textwidth}
    \includegraphics[width=\linewidth]{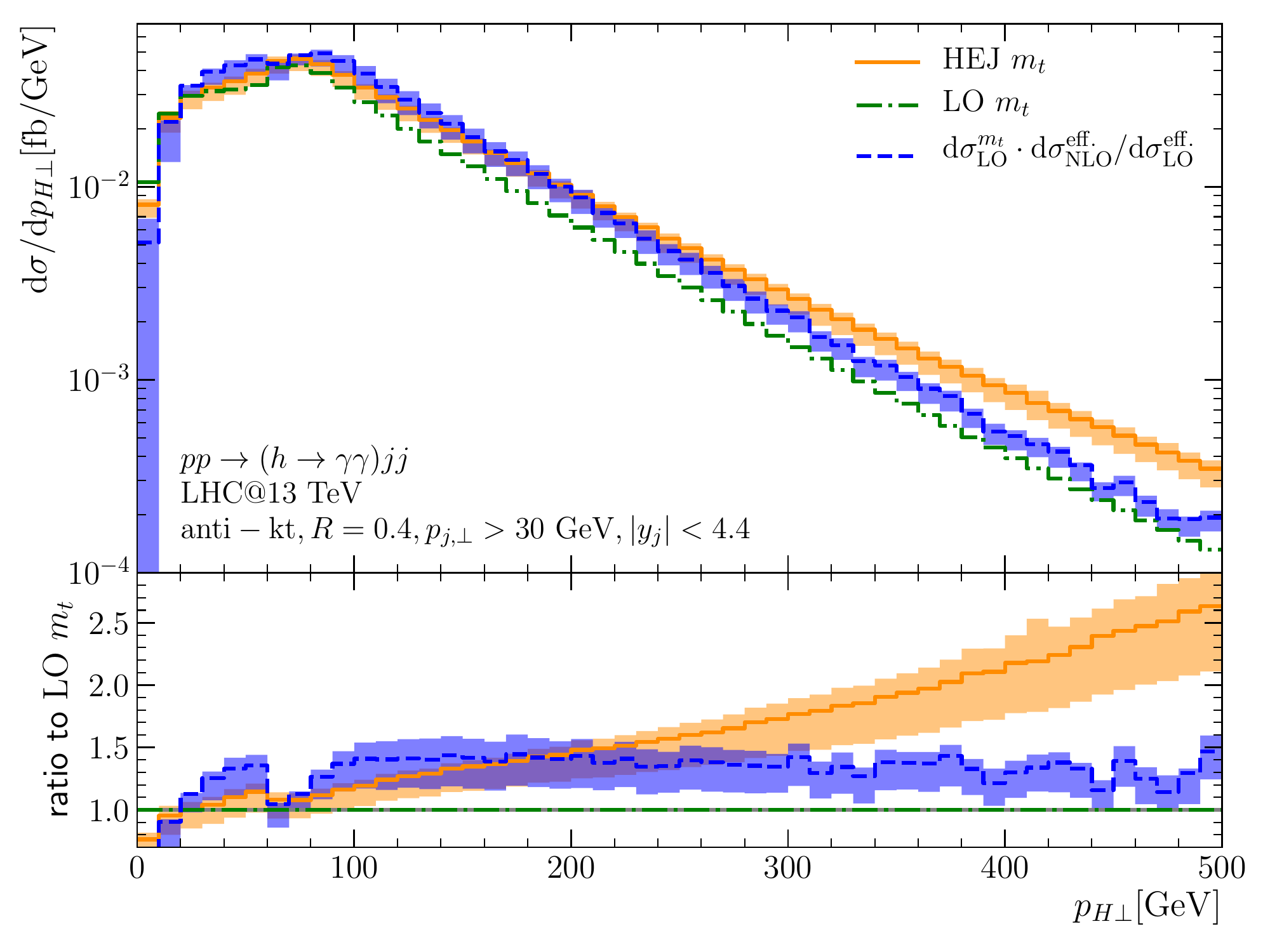}
    \caption{}
    \label{fig:HT2:ResultsHigherOrdera}
  \end{subfigure}
  \begin{subfigure}[t]{0.495\textwidth}
    \includegraphics[width=\linewidth]{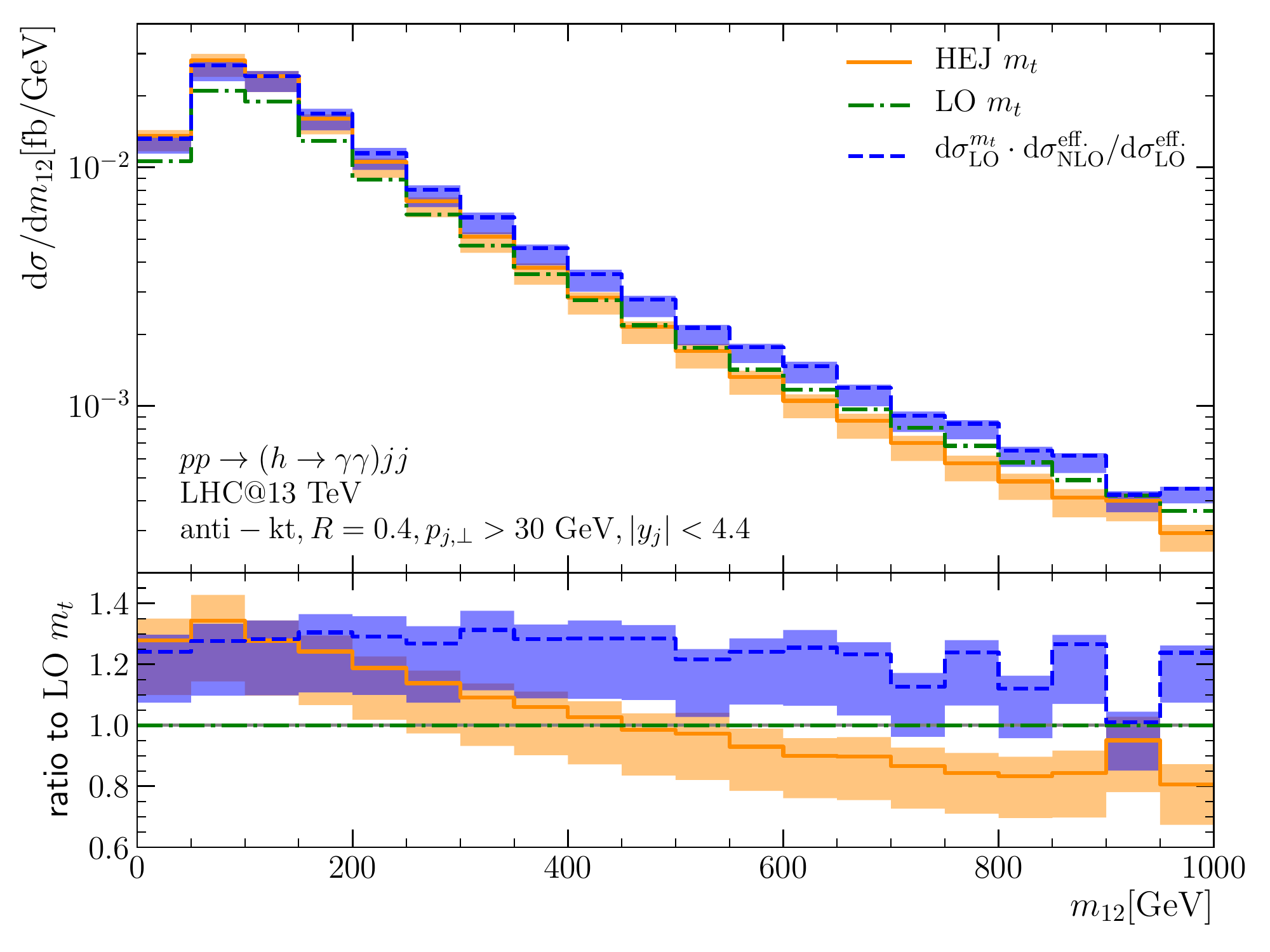}
    \caption{}
    \label{fig:HT2:ResultsHigherOrderb}
  \end{subfigure}
  \begin{subfigure}[t]{0.495\textwidth}
    \includegraphics[width=\linewidth]{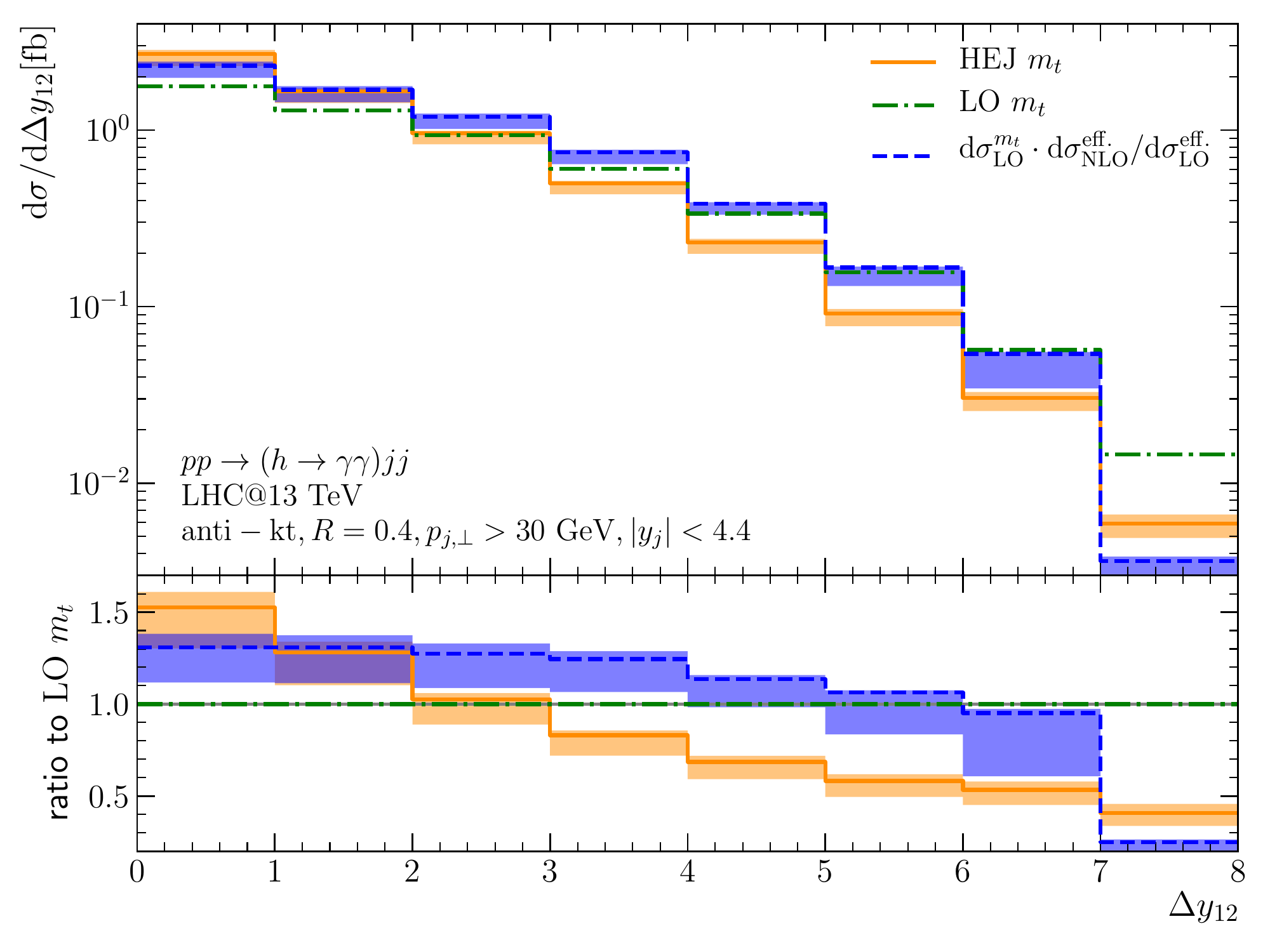}
    \caption{}
    \label{fig:HT2:ResultsHigherOrderc}
  \end{subfigure}
  \begin{subfigure}[t]{0.495\textwidth}
    \includegraphics[width=\linewidth]{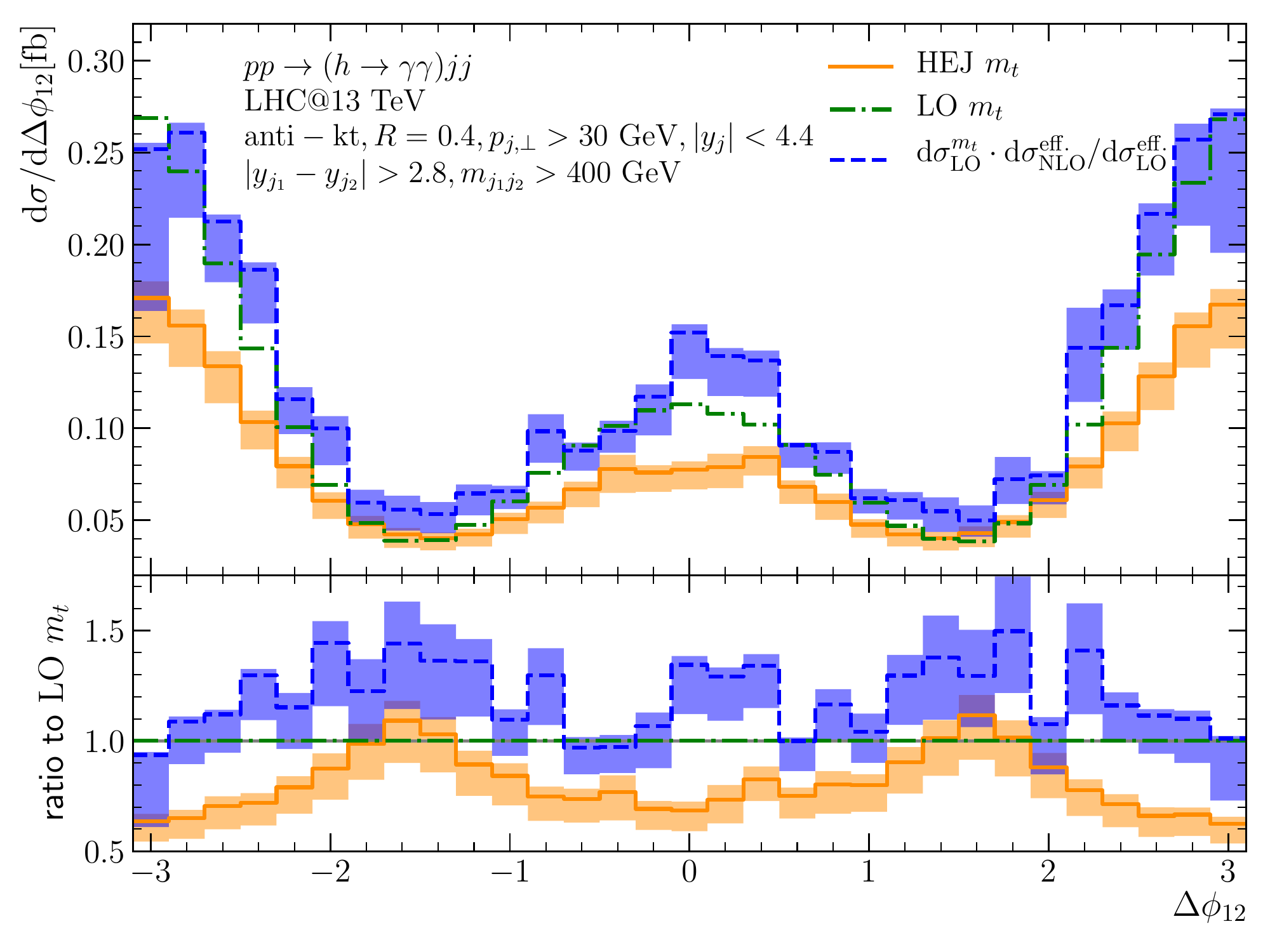}
    \caption{}
    \label{fig:HT2:ResultsHigherOrderd}
  \end{subfigure}
  \caption{Predictions for various distributions obtained with \HEJ, pure
    leading order, and leading order rescaled with differential $K$ factors
    for the central scale choice $\scaleht$. See
    figure~\ref{fig:ResultsHigherOrder} for the corresponding plots with
    $\scalemjj$.}
  \label{fig:HT2:ResultsHigherOrder}
\end{figure}

\begin{figure}
\centering
\begin{subfigure}[t]{0.495\textwidth}
  \includegraphics[width=\linewidth]{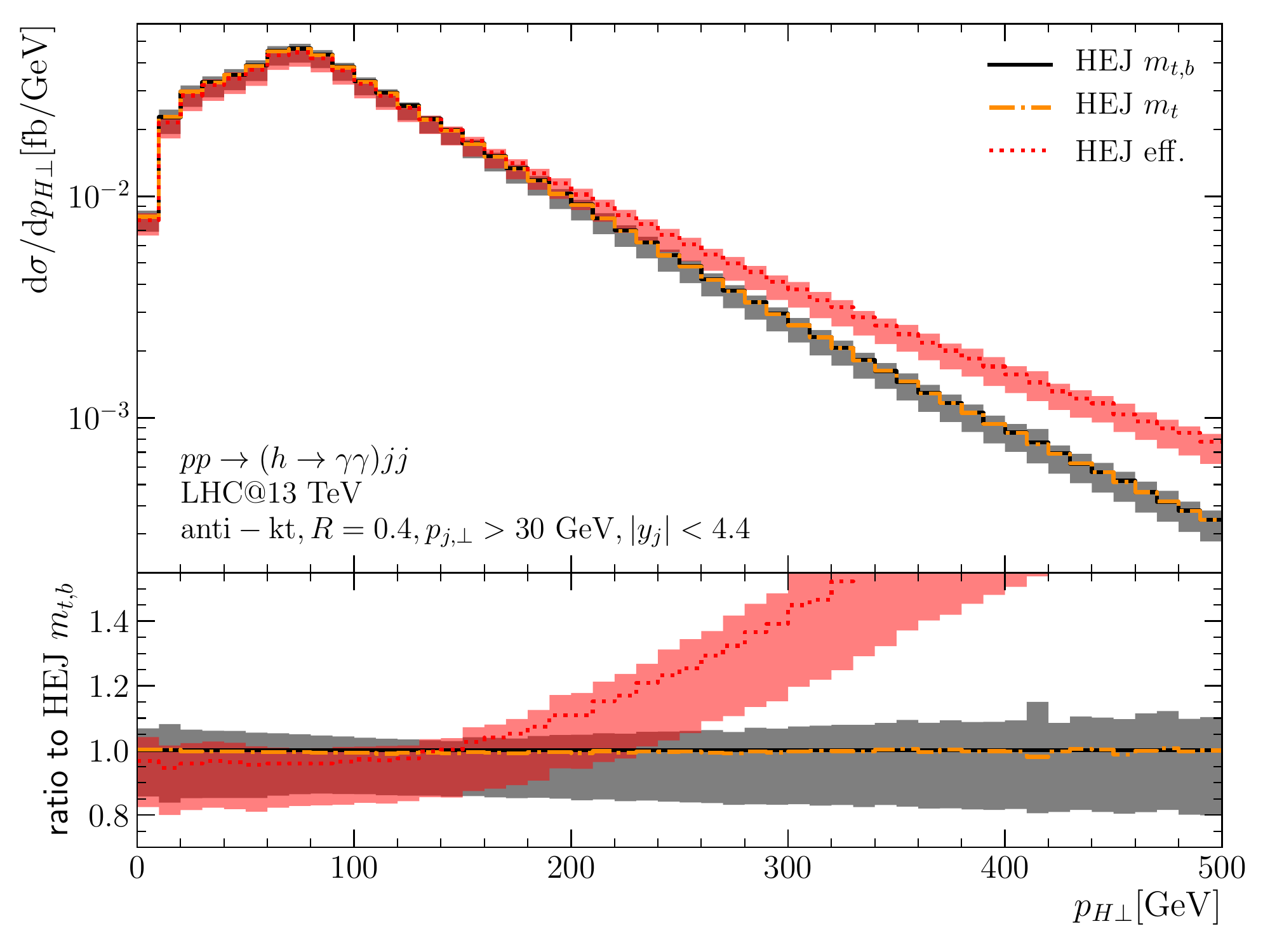}
  \caption{}
\label{fig:HT2:ResultsMassPHperp}
\end{subfigure}
\begin{subfigure}[t]{0.495\textwidth}
  \includegraphics[width=\linewidth]{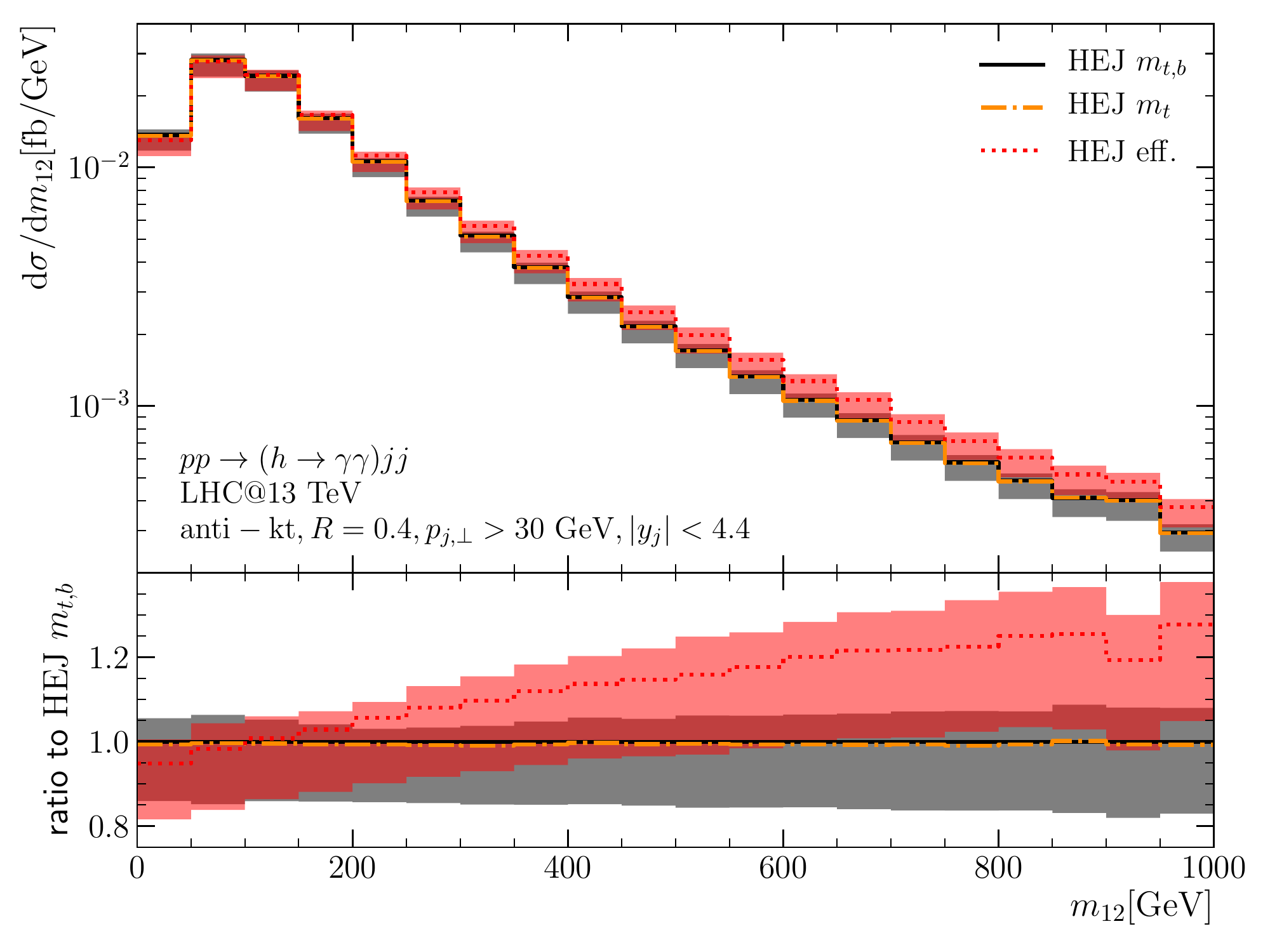}
  \caption{}
\label{fig:HT2:ResultsMassm12}
\end{subfigure}
\begin{subfigure}[t]{0.495\textwidth}
  \includegraphics[width=\linewidth]{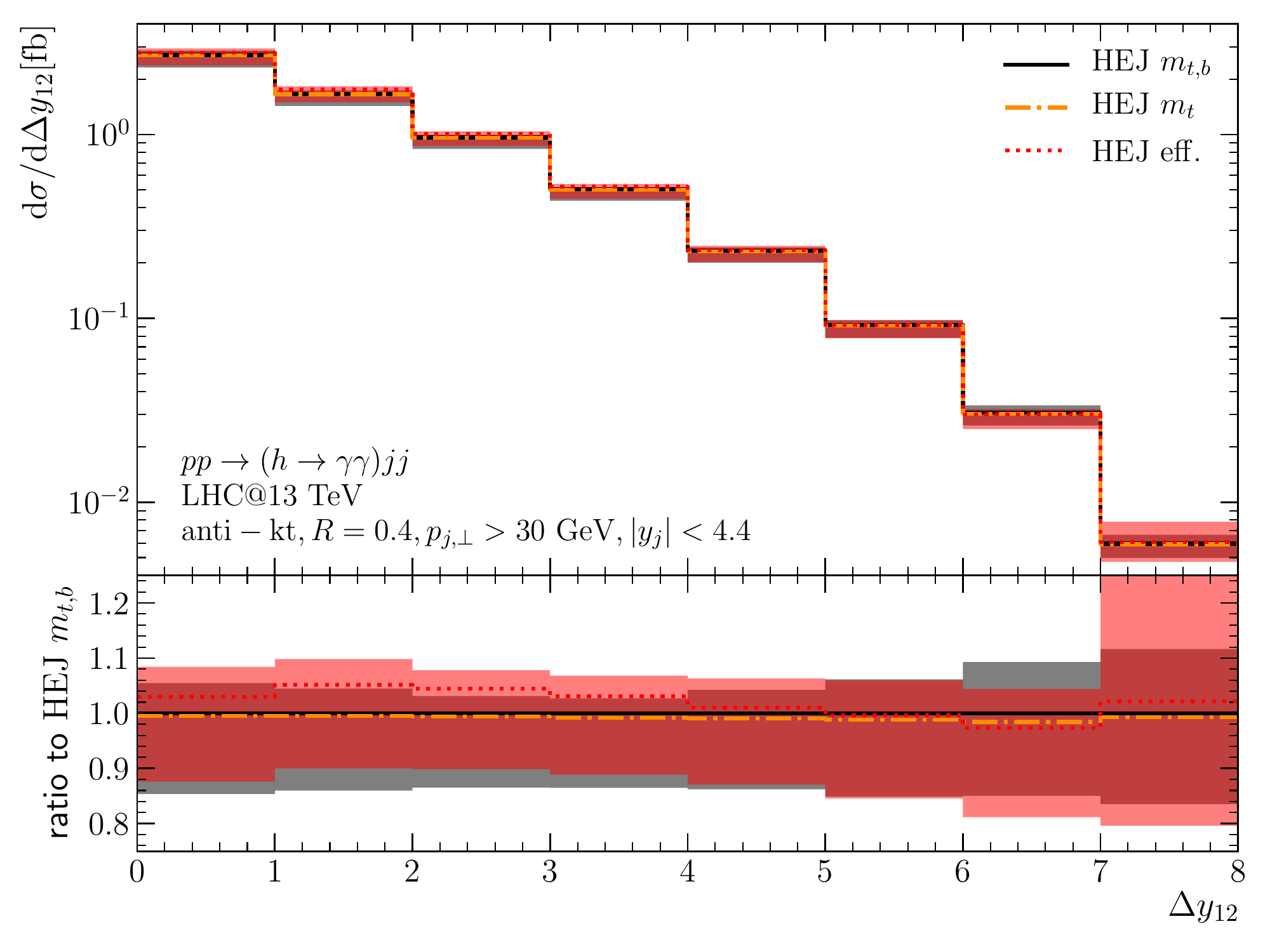}
  \caption{}
\label{fig:HT2:ResultsMassDY12}
\end{subfigure}
\begin{subfigure}[t]{0.495\textwidth}
  \includegraphics[width=\linewidth]{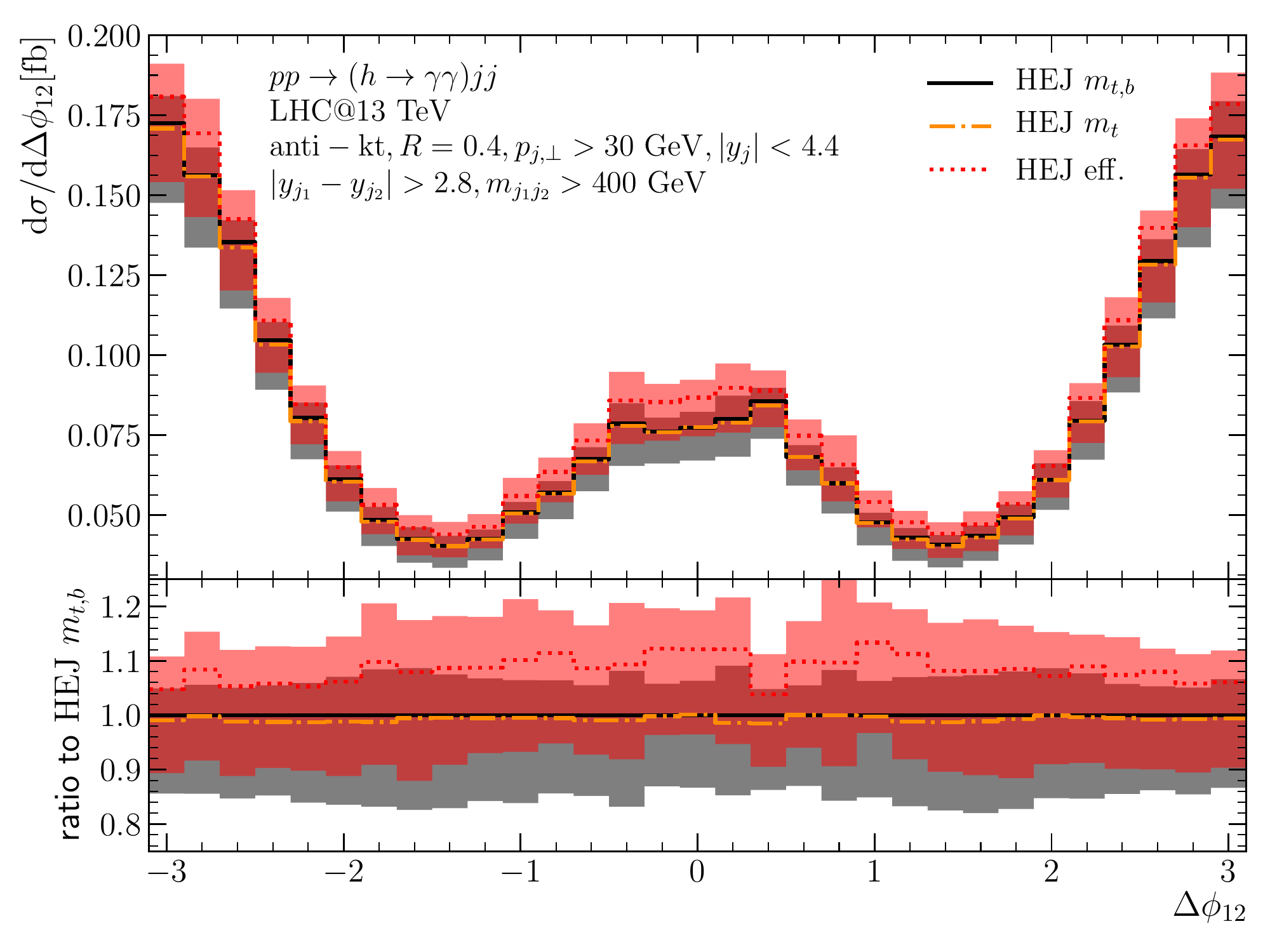}
  \caption{}
\label{fig:HT2:ResultsMassdphi12}
\end{subfigure}
\caption{\HEJ predictions for various distributions and different
  choices for the heavy-quark mass with the central scale choice
  $\scaleht$. See figure~\ref{fig:ResultsMass} for the corresponding plots
  with $\scalemjj$.}
\label{fig:HT2:ResultsMass}
\end{figure}

\begin{figure}
  \centering
  \begin{subfigure}[t]{0.495\textwidth}
    \includegraphics[width=\linewidth]{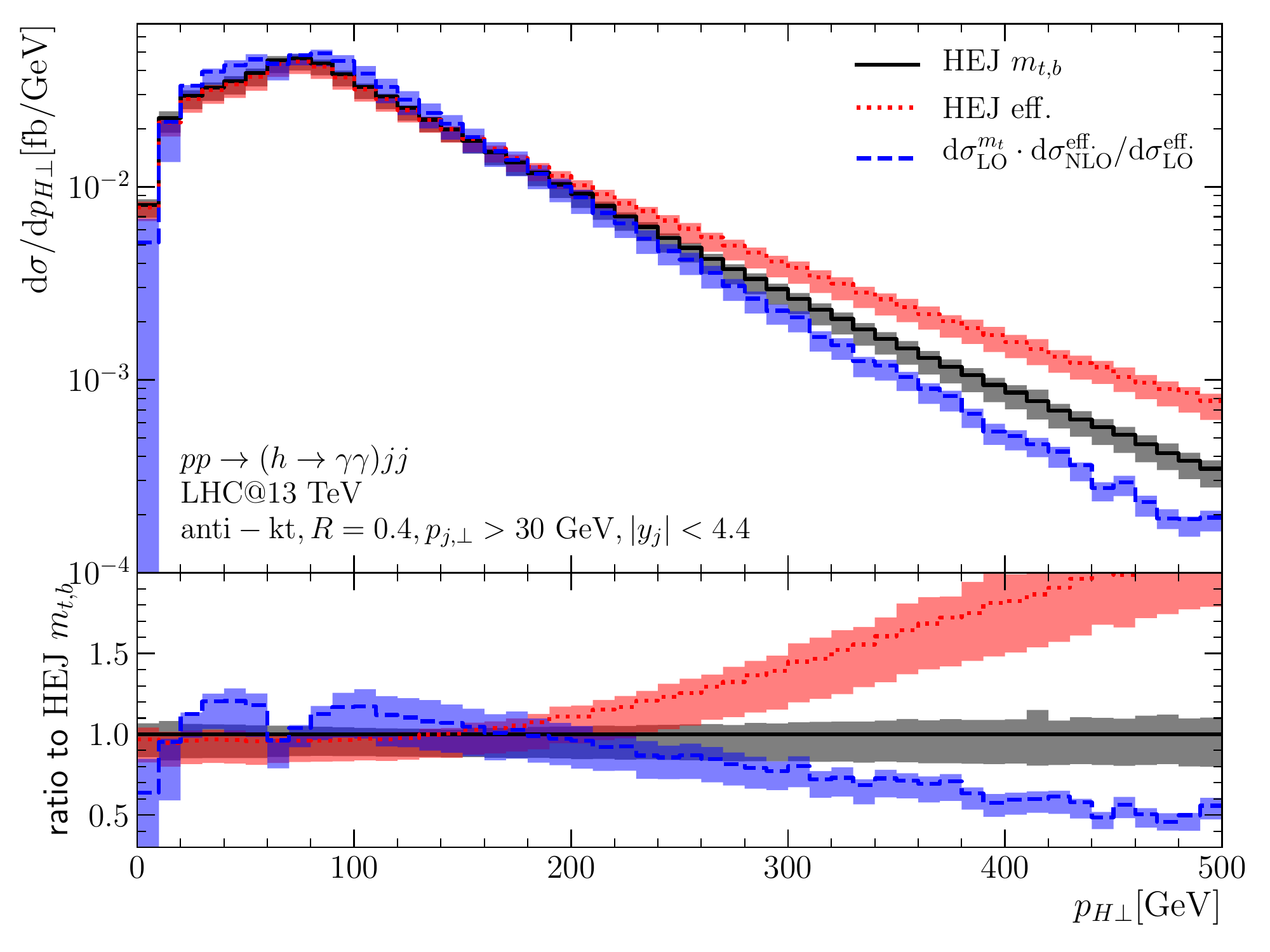}
    \caption{}
    \label{fig:HT2:finalHEJfixeda}
  \end{subfigure}
  \begin{subfigure}[t]{0.495\textwidth}
    \includegraphics[width=\linewidth]{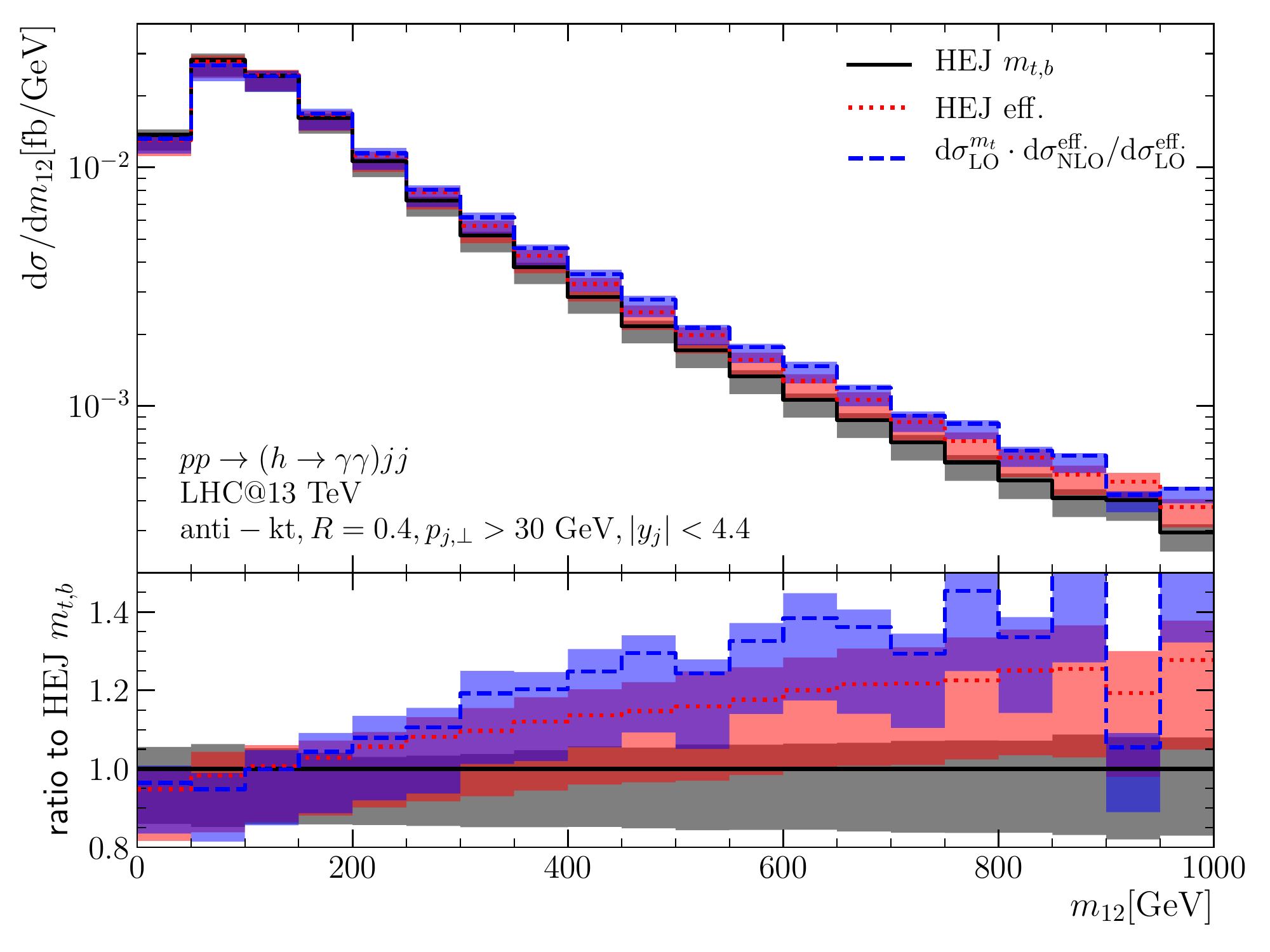}
    \caption{}
    \label{fig:HT2:finalHEJfixedb}
  \end{subfigure}
  \begin{subfigure}[t]{0.495\textwidth}
    \includegraphics[width=\linewidth]{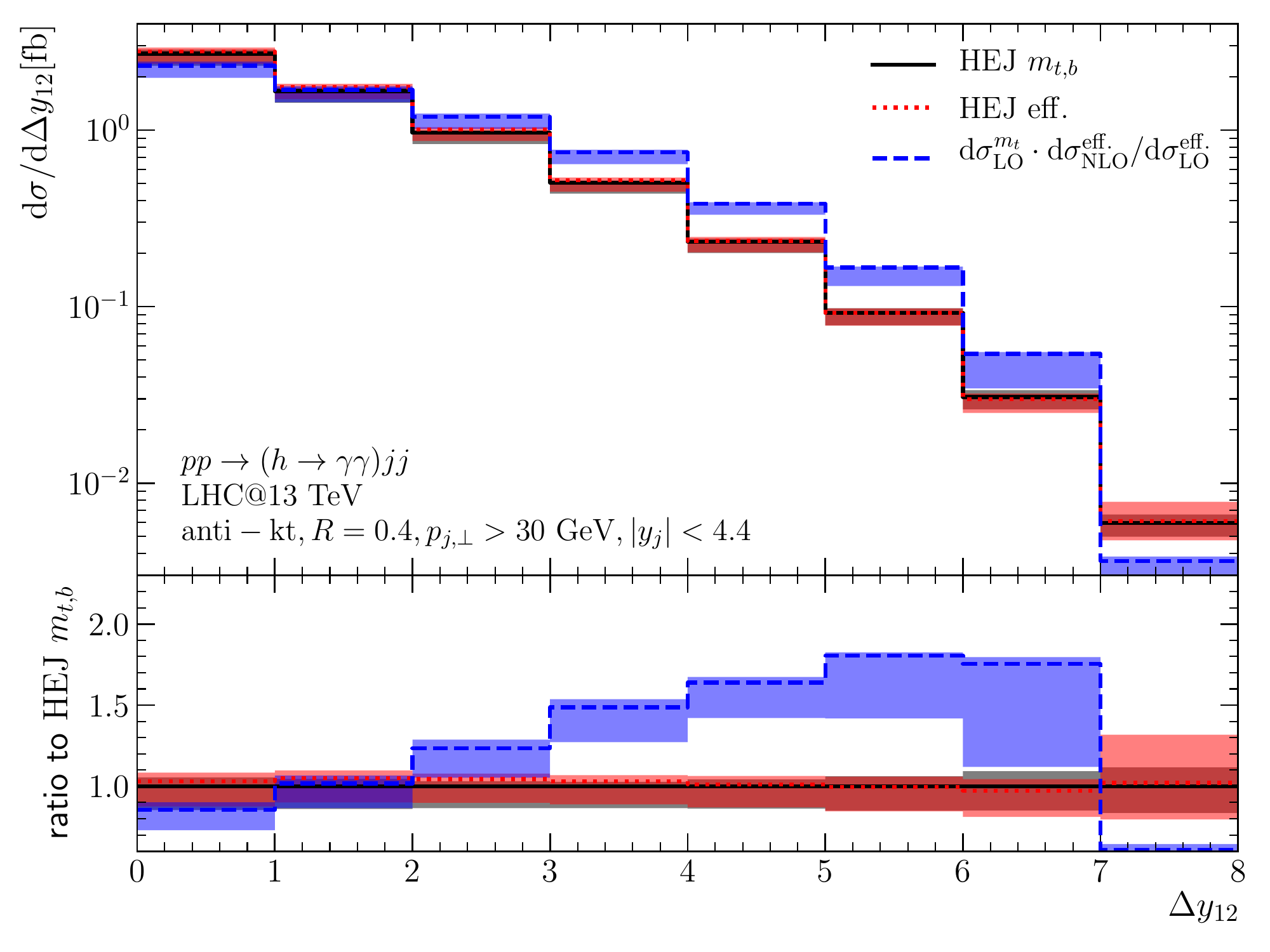}
    \caption{}
    \label{fig:HT2:finalHEJfixedc}
  \end{subfigure}
  \begin{subfigure}[t]{0.495\textwidth}
    \includegraphics[width=\linewidth]{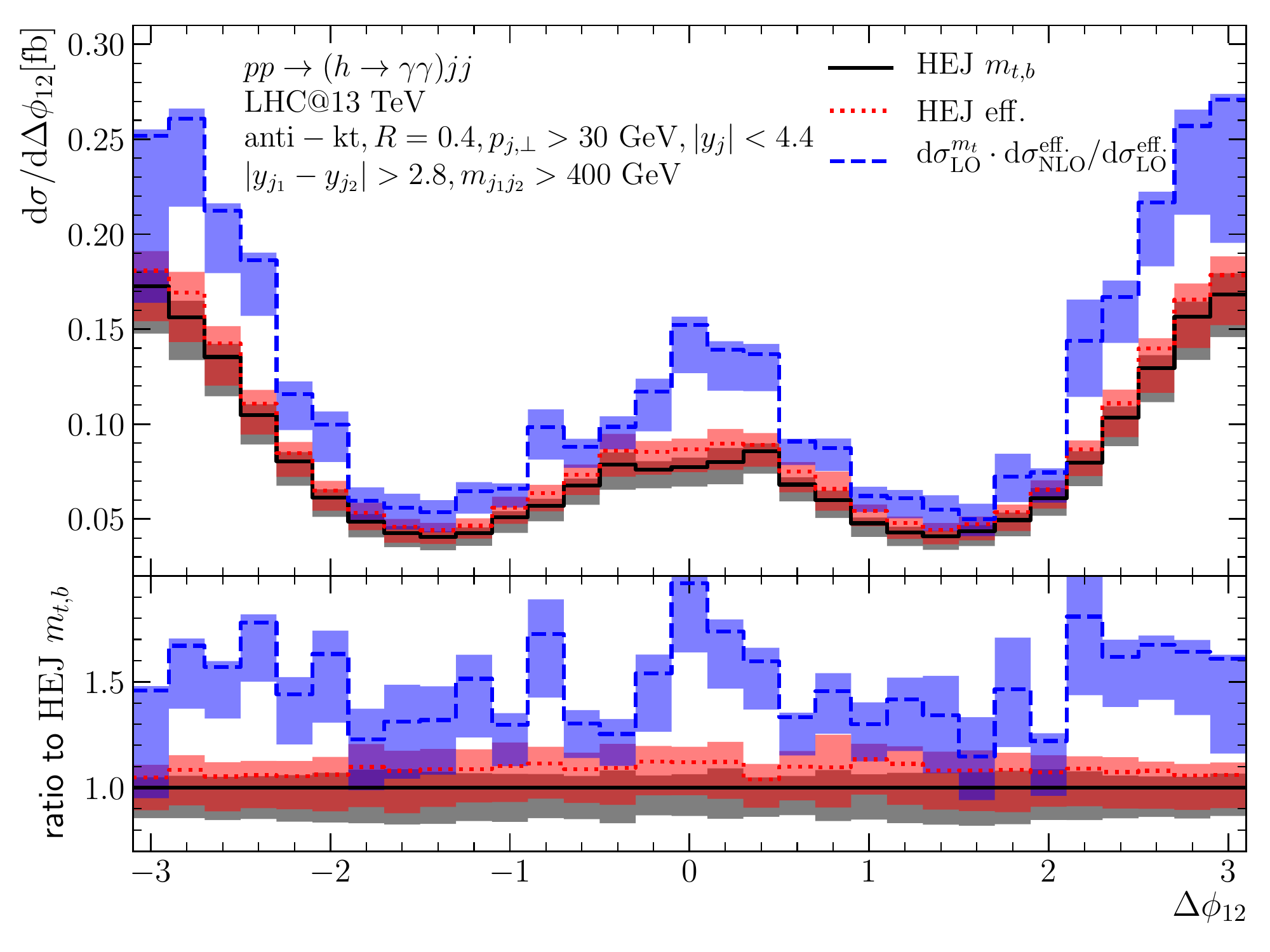}
    \caption{}
    \label{fig:HT2:finalHEJfixedd}
  \end{subfigure}
\caption{Comparison between \HEJ and the rescaled leading-order
prediction for various distributions with the central scale choice
$\scaleht$. See figure~\ref{fig:finalHEJfixed} for the corresponding
plots with $\scalemjj$}
  \label{fig:HT2:finalHEJfixed}
\end{figure}

In this appendix we study the effect of using a central
scale of \scaleht instead of the choice
\scalemjj used in the main text.  In table~\ref{tab:HT2:crosssections} we
present the cross section results for a central scale choice of \scaleht.  These
correspond to the results in table~\ref{tab:crosssections} in
section~\ref{sec:effects-fin-top-mass}.
We continue in
figures~\ref{fig:HT2:ResultsHigherOrder}--\ref{fig:HT2:finalHEJfixed} by repeating the comparisons of
figures~\ref{fig:ResultsHigherOrder}--\ref{fig:finalHEJfixed}.  While there are
variations in numerical values, we find that the conclusions of the impact of
the higher-order corrections in \HEJ and of the finite quark mass and loop
propagator effects are unchanged.

% We begin by studying the impact of the effect of higher perturbative orders in
% figure~\ref{fig:HT2:ResultsHigherOrder}, which should be compared to
% figure~\ref{fig:ResultsHigherOrder} in the main text.  We see with this scale
% choice that the NLO $K$-factors (seen in the blue bands in the ratio panels) do
% not exhibit the same large and increasing values.  This is the same effect seen
% between figure~\ref{fig:NLOK-factor} and figure~\ref{fig:HT2:NLOK-factor} and
% discussed in section~\ref{sec:match-lead-order}.

% As in figure~\ref{fig:ResultsHigherOrdera}, we see in
% figure~\ref{fig:HT2:ResultsHigherOrdera} that the all-order corrections in \HEJ
% still lead to a harder spectrum in $p_{H\perp}$ when compared to the fixed-order
% predictions.  This leads to a greater sensitivity to the effect of the finite
% quark mass and propagator effects.  Turning our attention to
% figure~\ref{fig:ResultsHigherOrderb}, we see that \HEJ still predicts a lower
% cross section at large $m_{12}$ than at NLO (despite the $K$-factor being reduced).

\begin{figure}
  \centering
  \begin{subfigure}[t]{0.495\textwidth}
    \includegraphics[width=\linewidth]{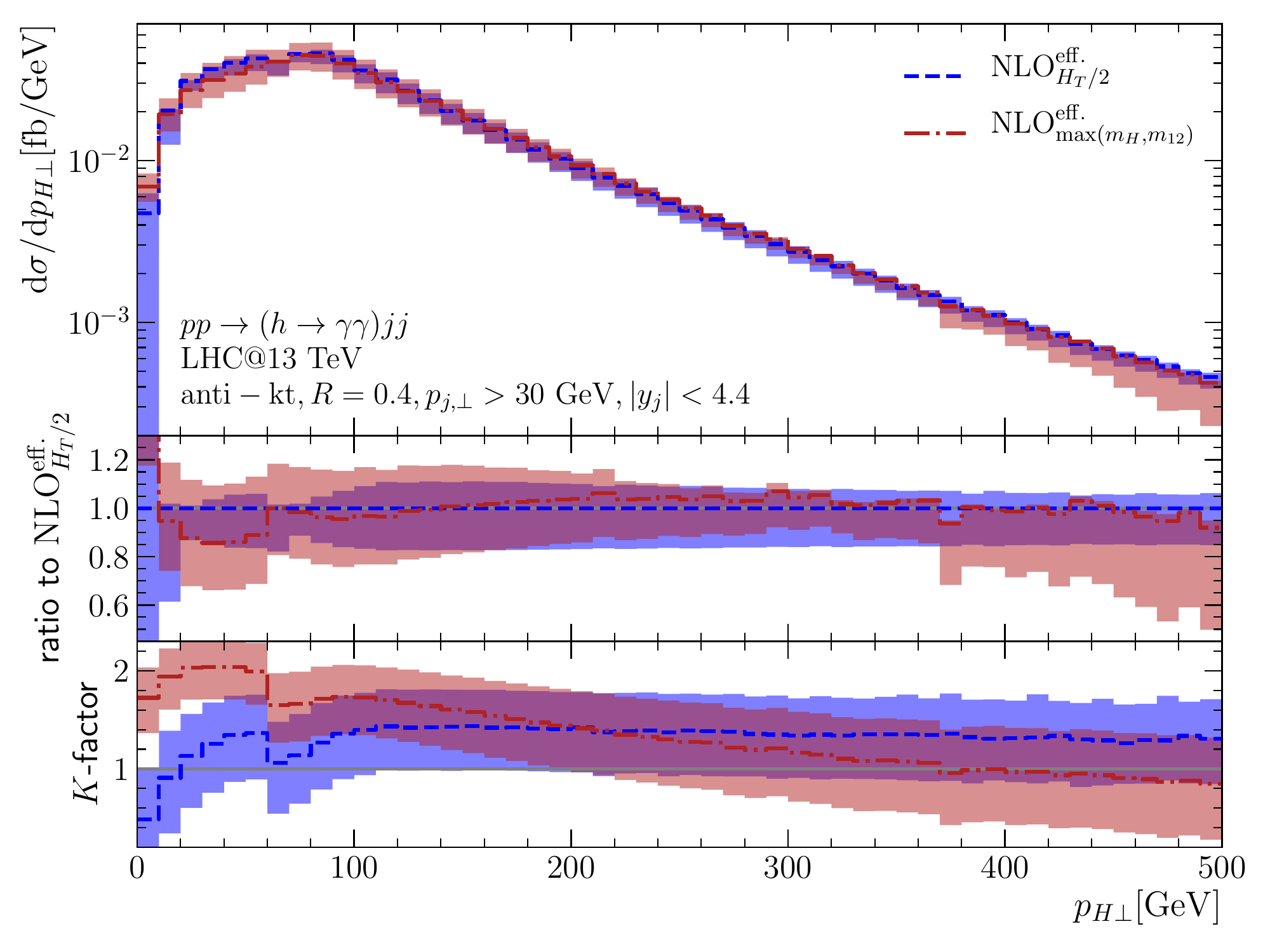}
    \caption{}
    \label{fig:CompareScales_NLOa}
  \end{subfigure}
  \begin{subfigure}[t]{0.495\textwidth}
    \includegraphics[width=\linewidth]{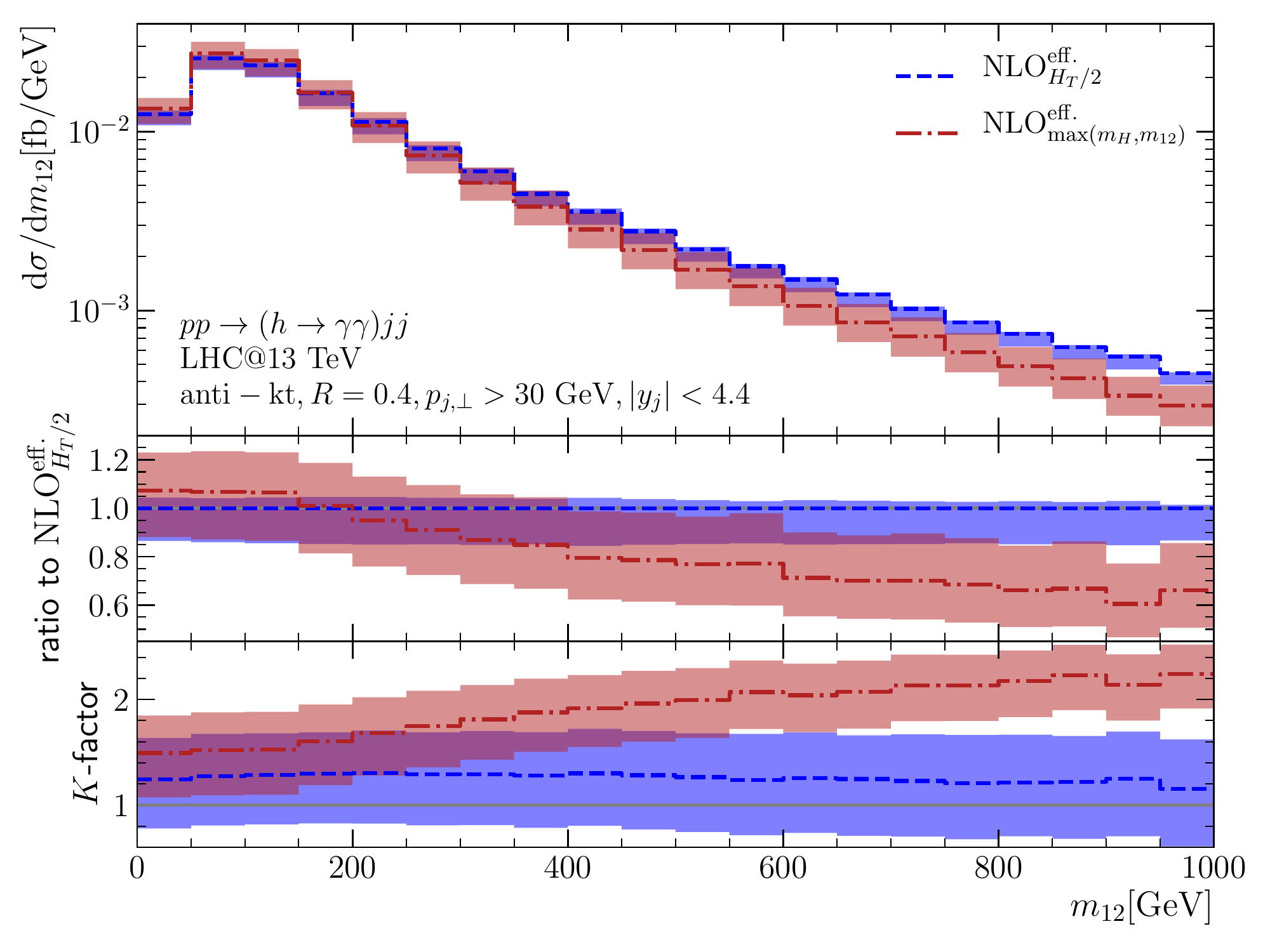}
    \caption{}
    \label{fig:CompareScales_NLOb}
  \end{subfigure}
  \begin{subfigure}[t]{0.495\textwidth}
    \includegraphics[width=\linewidth]{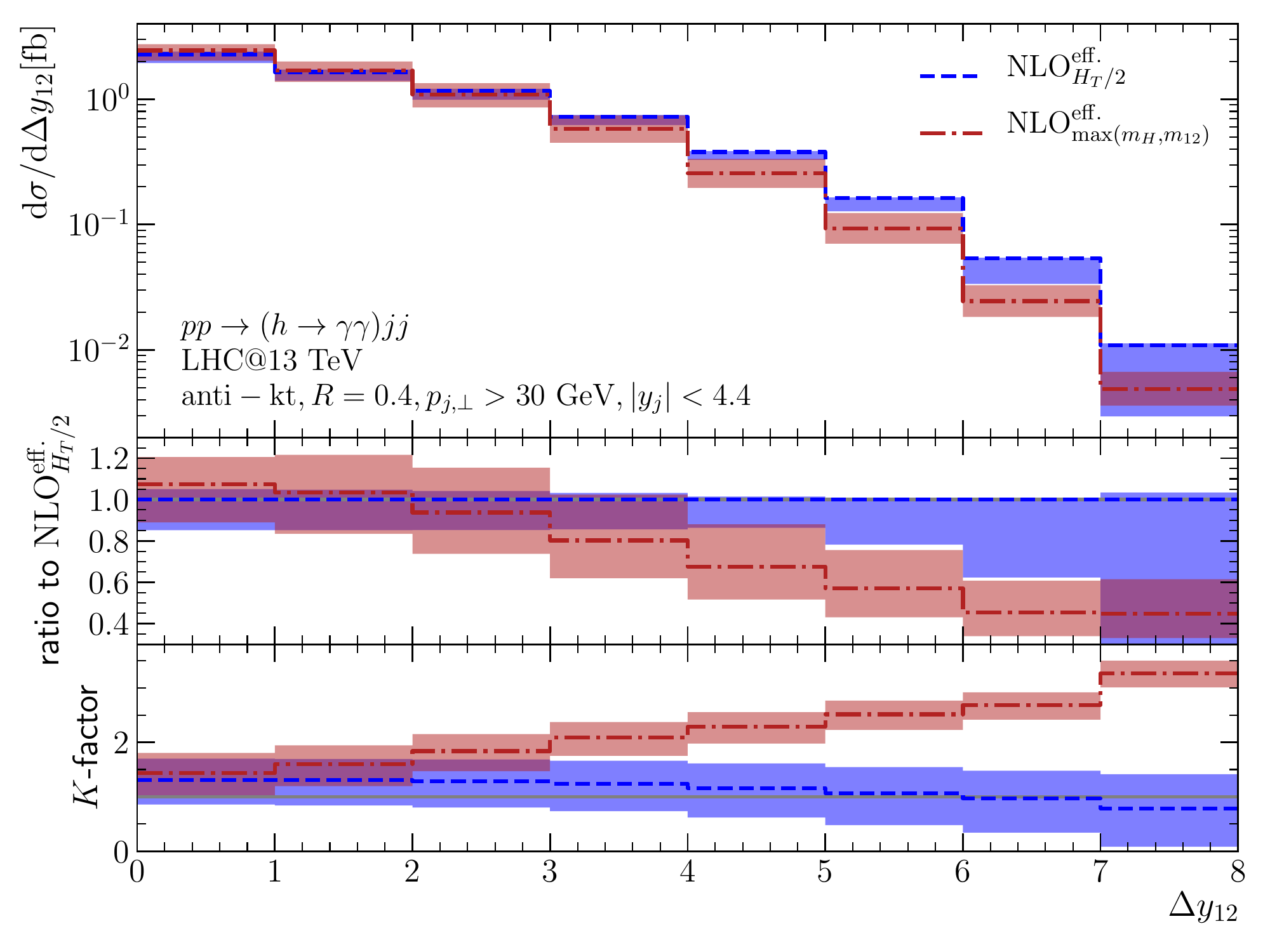}
    \caption{}
    \label{fig:CompareScales_NLOc}
  \end{subfigure}
  \begin{subfigure}[t]{0.495\textwidth}
    \includegraphics[width=\linewidth]{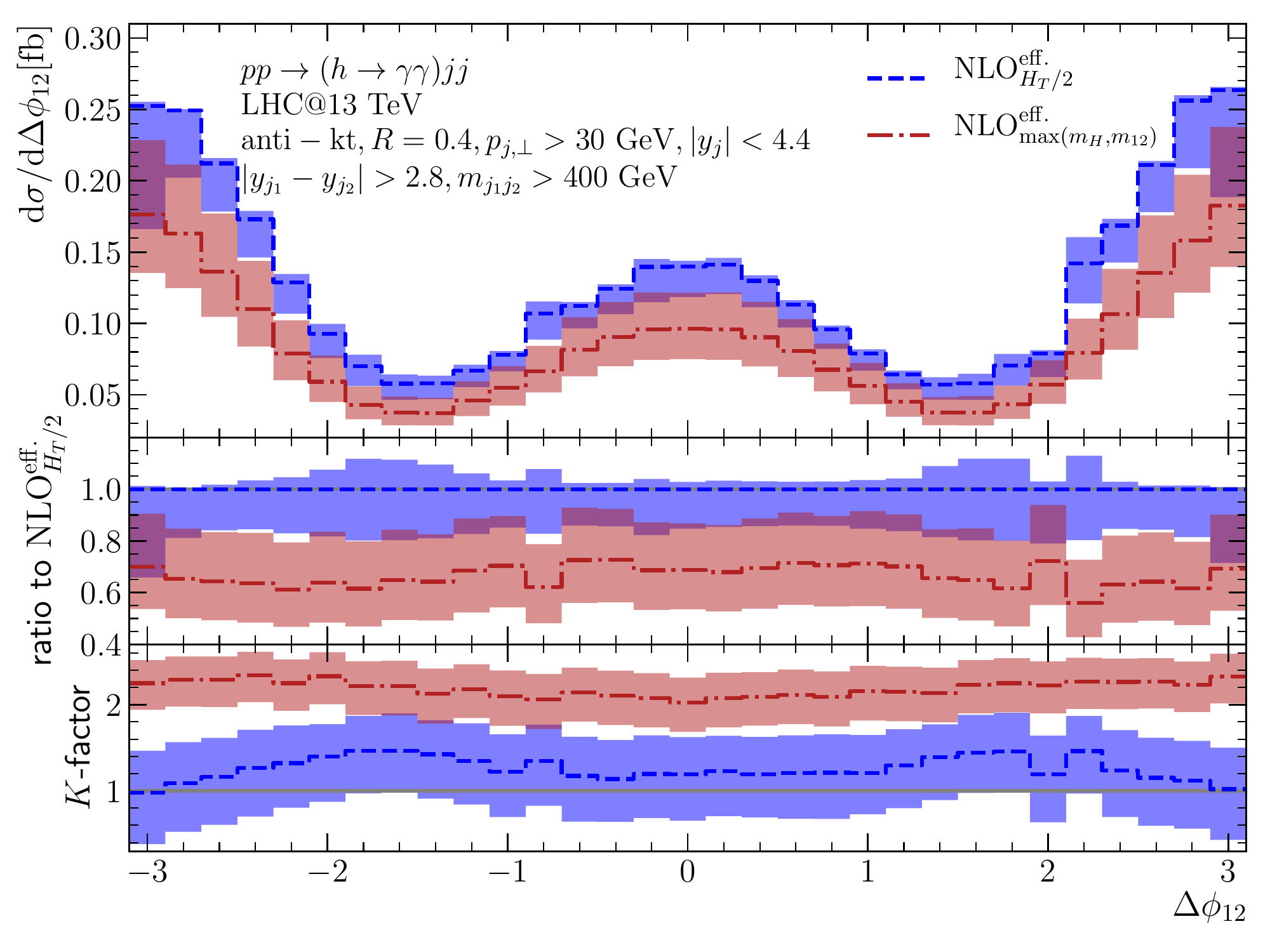}
    \caption{}
    \label{fig:CompareScales_NLOd}
  \end{subfigure}
  \caption{Comparison between pure \NLO results with central scale
    choices $\scaleht$ and  $\scalemjj$}
  \label{fig:CompareScales_NLO}
\end{figure}

The results obtained at NLO for the two central scale choices \scalemjj and
\scaleht are compared in figure~\ref{fig:CompareScales_NLO}. It is noteworthy
that the difference in the results in figure~\ref{fig:CompareScales_NLOd} for
the cross section within the VBF-cuts is similar to the difference between
the results of \NLO and \HEJ obtained with the same scale.

\begin{figure}
  \centering
  \begin{subfigure}[t]{0.495\textwidth}
    \includegraphics[width=\linewidth]{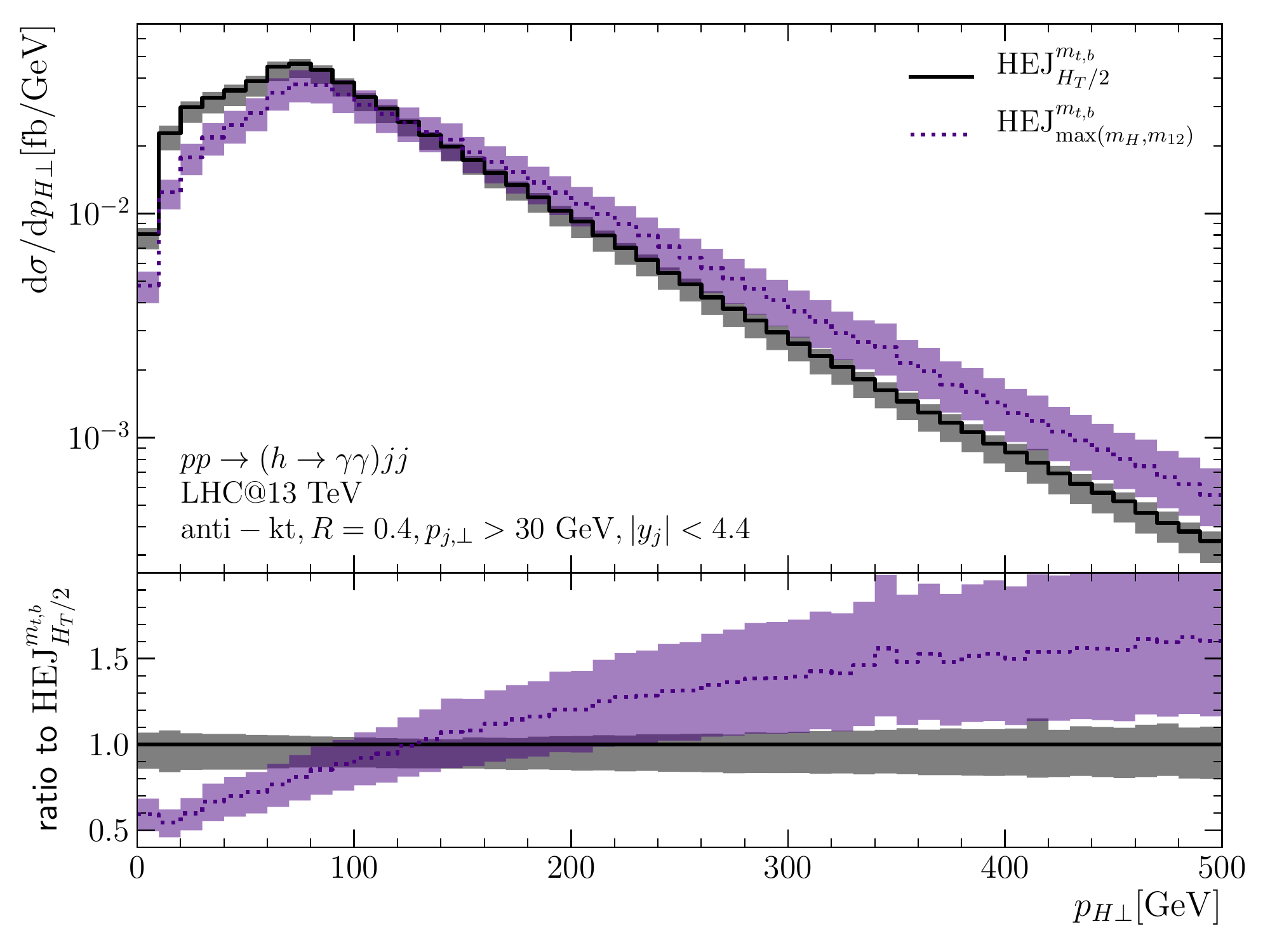}
    \caption{}
    \label{fig:CompareScales_HEJa}
  \end{subfigure}
  \begin{subfigure}[t]{0.495\textwidth}
    \includegraphics[width=\linewidth]{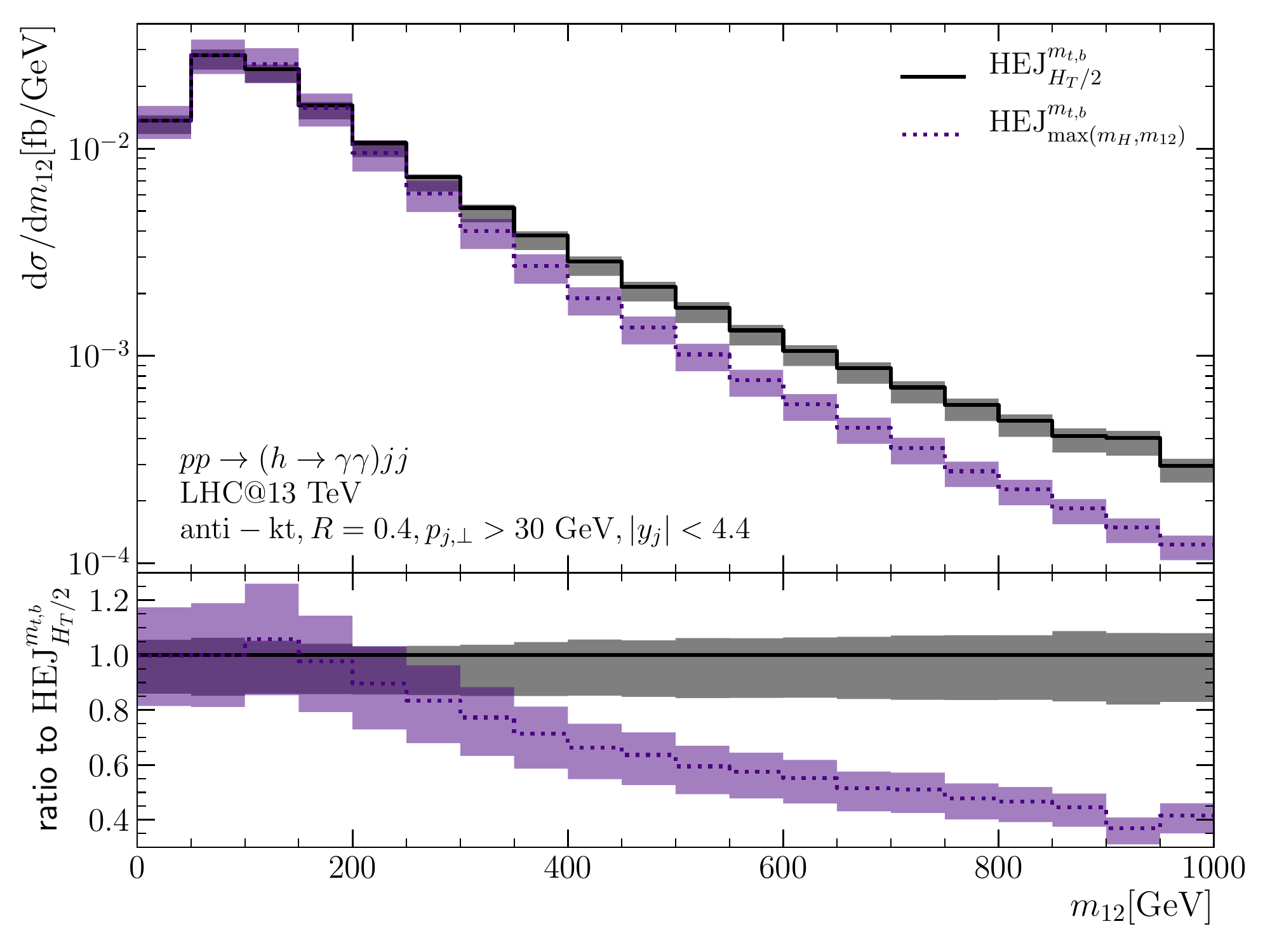}
    \caption{}
    \label{fig:CompareScales_HEJb}
  \end{subfigure}
  \begin{subfigure}[t]{0.495\textwidth}
    \includegraphics[width=\linewidth]{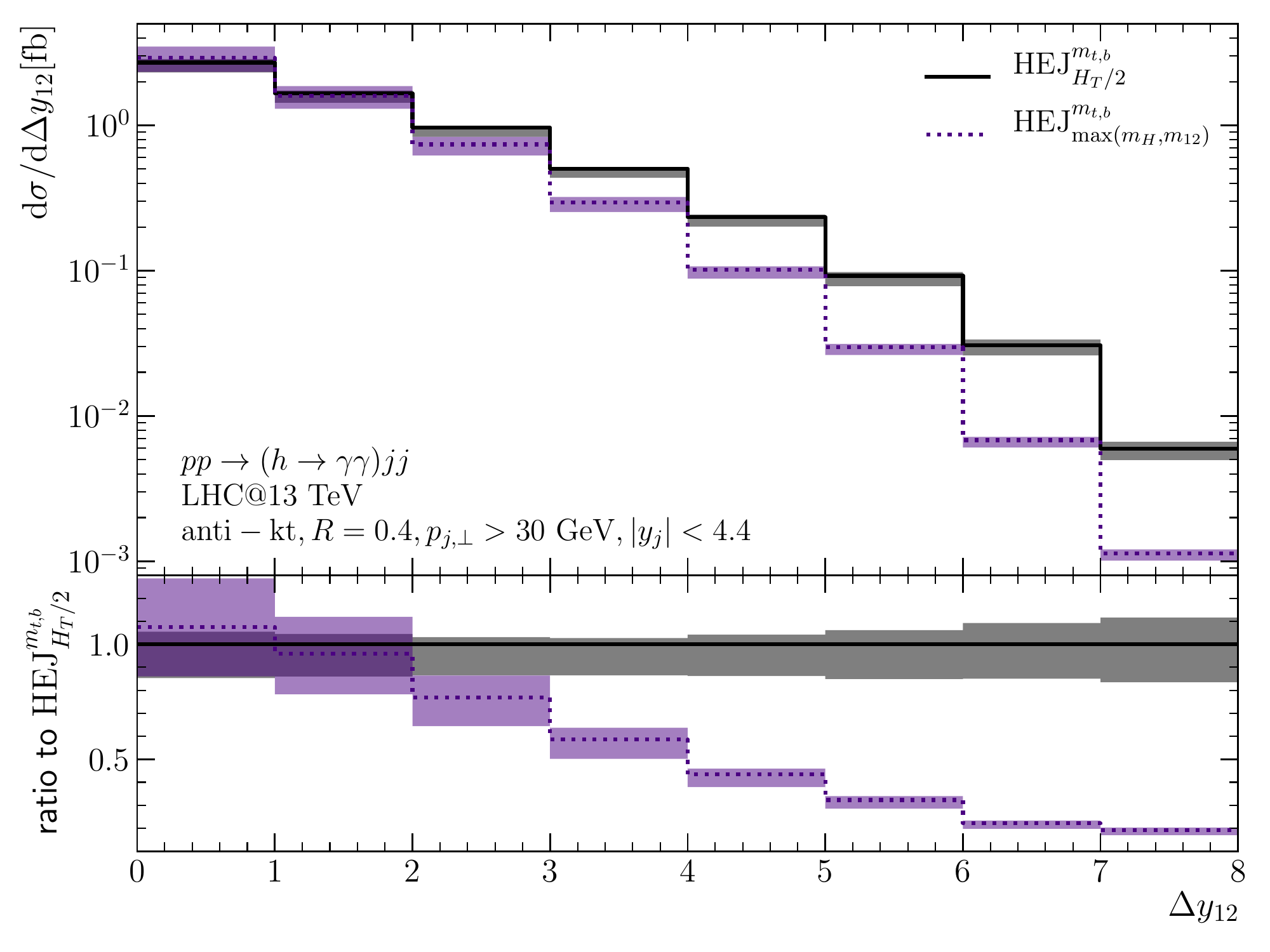}
    \caption{}
    \label{fig:CompareScales_HEJc}
  \end{subfigure}
  \begin{subfigure}[t]{0.495\textwidth}
    \includegraphics[width=\linewidth]{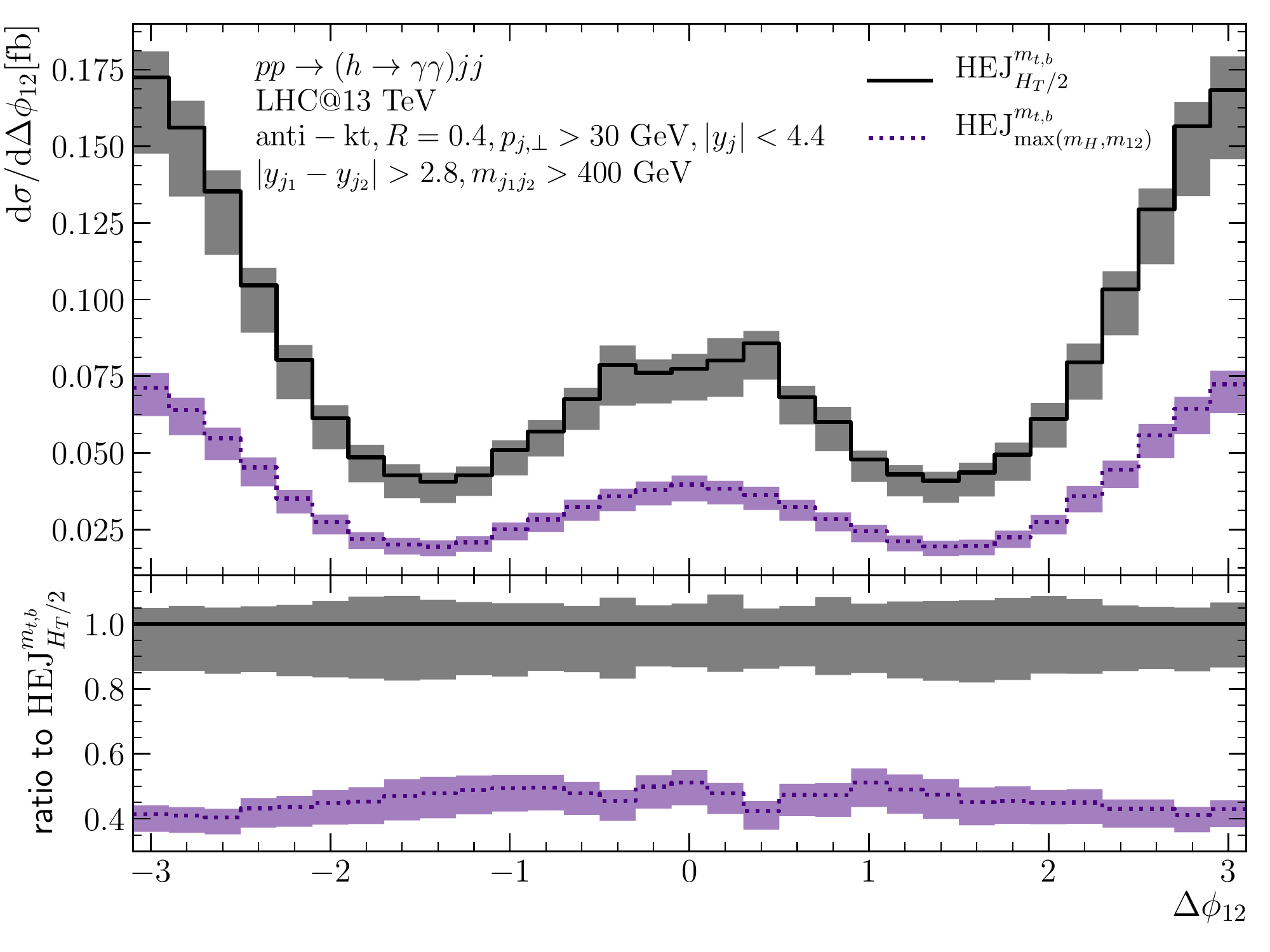}
    \caption{}
    \label{fig:CompareScales_HEJd}
  \end{subfigure}
  \caption{Comparison between \HEJ results with central scale
    choices $\scaleht$ and $\scalemjj$}
  \label{fig:CompareScales_HEJ}
\end{figure}

Finally, figure~\ref{fig:CompareScales_HEJ} compares the results obtained for
\HEJ with the two central scale choices. The differences in the results for
the distributions are larger than indicated by the scale variation. This is
not surprising, since the leading logarithmic behaviour at large $m_{jj}$ is
unrelated to $\beta_0$-terms from the running of the coupling. As stated
earlier, comparisons with data for other processes can determine which of
these scale choices obtains the best description. The discussion of scale
choice is independent of the discussion of the behaviour at large $m_{jj}$,
and so is the conclusion that a resummation of the leading terms at large
$m_{jj}$ leads to a reduction of the cross section within the VBF cuts.

\begin{table}
  \centering
  \begin{tabular}{lrrrr}
    \toprule
    &\multicolumn{2}{c}{Fixed Order}&\multicolumn{2}{c}{\HEJ}\\
    \cmidrule(r){2-3}\cmidrule(l){4-5}
    &Inclusive $H+2j$&VBF cuts&Inclusive $H+2j$&VBF cuts\\\midrule
    $m_t\to\infty$   & $6.4^{+0.3}_{-0.9}$ fb& $0.82^{+0.02}_{-0.11}$ fb& $6.4^{+0.3}_{-0.9}$ fb& $0.56^{+0.04}_{-0.09}$ fb\\[.5em]
    $m_t=163$ GeV     & $6.6^{+0.3}_{-1.0}$ fb & $0.82^{+0.02}_{-0.11}$ fb& $6.2^{+0.3}_{-0.9}$ fb& $0.51^{+0.03}_{-0.08}$ fb\\[.5em]

    $m_t=163$ GeV & \multirow{2}{*}{-} & \multirow{2}{*}{-}
        &\multirow{2}{*}{ $6.2^{+0.3}_{-0.9}$ fb}&
            \multirow{2}{*}{ $0.52^{+0.03}_{-0.08}$ fb}\\
    $m_b = 2.8$ GeV\\
\bottomrule
  \end{tabular}
  \caption{Total cross section predictions for the central scale choice
$\scaleht$ and different values of the heavy-quark masses. See
table~\ref{tab:crosssections} for the corresponding predictions with
$\scalemjj$. }
  \label{tab:HT2:crosssections}
\end{table}

%%% Local Variables:
%%% mode: latex
%%% TeX-master: "main"
%%% End:

\clearpage

\bibliographystyle{JHEP}
\bibliography{papers}

\end{document}